\documentclass{amsart}
\usepackage[english]{babel}
\usepackage{graphicx}
\usepackage{amssymb}
\usepackage{amsmath}
\usepackage[foot]{amsaddr}
\usepackage{amsfonts}
\usepackage{bbm}
\usepackage{xcolor}
\usepackage{float}
\usepackage{algorithm}
\usepackage{algpseudocode}
\usepackage{rotating}

\usepackage{amsmath}
\usepackage{amssymb}
\usepackage{mathrsfs}
\usepackage{color}
\usepackage{graphicx}
\usepackage[percent]{overpic}
\usepackage{url}
\usepackage{subfig}
\usepackage{enumerate}
\usepackage{overpic}
\usepackage{amsthm}
\usepackage{moreverb}
\usepackage[pdftex,colorlinks,bookmarksopen,bookmarksnumbered,citecolor=red,urlcolor=red]{hyperref}

\def\f{{\bm f}}
\def\g{{\bm g}}
\def\n{{\bm n}}
\def\u{{\bm u}}

\def\0{\boldsymbol{0}}
\def\ss{\boldsymbol{\sigma}}

\def\dt{\partial_t}

\def\cl {\nonumber \\}
\def\el {\nonumber }

\newtheorem{rem}{Remark}[section]

\newcommand{\bm}[1]{\mbox{\boldmath{$#1$}}}
\def\div{\nabla\cdot}
\def\grad{\nabla}

\usepackage{tikz}
\usepackage[top=1in, bottom=1.25in, left=1.25in, right=1.25in]{geometry}

\usepackage{xcolor}
\usepackage{hyperref}

\newcommand{\bb}{\mathbf b}

\newcommand{\bk}{\mathbf k}

\newcommand{\bu}{\mathbf u}

\def\cl {\nonumber \\}
\def\el {\nonumber }

\graphicspath{{images/}}

\begin{document}

\title[]{{Validation of an OpenFOAM\textsuperscript{\textregistered}-based solver for the Euler equations with benchmarks for mesoscale atmospheric modeling}} % in a Finite Volume environment

\author{Michele Girfoglio$^1$, Annalisa Quaini$^2$ and Gianluigi Rozza$^1$}
\address{$^1$ mathLab, Mathematics Area, SISSA, via Bonomea 265, I-34136 Trieste, Italy}
\address{$^2$ Department of Mathematics, University of Houston, Houston TX 77204, USA}
%\maketitle

%\affil[1]{SISSA, International School for Advanced Studies, Mathematics Area, mathLab, via Bonomea, Trieste 265 34136, Italy}
%\affil[2]{Department of Mathematics, University of Houston, Houston TX 77204, USA}

%\maketitle

\begin{abstract}
Within OpenFOAM, we develop a pressure-based solver for the Euler equations written in conservative form using density, momentum, and total energy as variables. Under simplifying assumptions, these equations are used to describe non-hydrostatic atmospheric flow. For the stabilization of the Euler equations and to capture sub-grid processes, we consider two Large Eddy Simulation models: the classical Smagorinsky model and the one equation eddy-viscosity model.
To achieve high computational efficiency, 
our solver uses a splitting scheme that decouples the computation of each variable. The numerical results obtained with our solver are validated against numerical data available in the literature for two classical benchmarks: the rising thermal bubble and the density current. Through qualitative and quantitative comparisons, we show that our approach is accurate. This work is meant to lay the foundation for a new open
source package specifically created for quick assessment of new computational approaches for the simulation of atmospheric flows at the mesoscale level.
\end{abstract}

\maketitle


%\anna{Ho commentato le affiliazioni perche' non riconosce il comando. Inoltre, mi pare che non riporti gli indirizzi mail da nessuna parte.}
 
\section{Introduction}

Forecasting the rapid changes in the Earth's climate is one of the biggest challenges of our times. Since the complexity of the problem requires inputs and close collaboration from scientists across various disciplines, ranging from physics of the atmosphere to computer science, open source software has emerged as the obvious choice. In fact, the availability of open source software for use, modification, and distribution makes it ideal for collaborative development. Some examples of open source software for climate simulations are the Community Earth System Model \cite{CESM}, the Energy Exascale Earth System Model \cite{E3SM}, the MIT General Circulation Model \cite{MITgcm} and the Climate Machine \cite{clima}. All of these examples have large communities of active users and developers that have contributed over the course of several years. Since climate simulations are an extension of weather forecasting, we note that open source software has been a tool of choice for weather prediction as well. Examples include WRF \cite{WRF} and Atlas \cite{atlas}. 

Fast and accurate weather/climate forecasts need state-of-the-art numerical and computational methodologies. While the above mentioned software packages are great tools for realistic simulations, testing and assessing new numerical approaches within them is non-trivial. In fact, the very complexity that allows for realistic simulations requires a considerable amount of time to familiarize with, making such software impractical as testbed. This paper lays the foundation for a new open source package, called GEA (Geophysical and Environmental Applications) \cite{GEA}, specifically created for assessment of new computational approaches for the simulation of mesoscale atmospheric flows and ocean flows \cite{GIRFOGLIO2023114656,GQR_ROM_QGE22}. That idea is that if a new approach fails to meet desired accuracy or efficiency standards in our simplified software, it would not be considered for implementation in more advanced software. To maximize the reach and impact, we choose to build our software package on OpenFOAM\textsuperscript{\textregistered} \cite{Weller1998}, an open source, freely available C++ finite volume library that has become widely used in Computational Fluid Dynamics (CFD). 
%In part, its success is due to the fact that is facilitates the development of custom-built open-source software, such as, e.g.,  ITHACA-FV \cite{RoSta17}. 

In recent years, OpenFOAM has been adopted to study a large variety of computational approaches for different CFD applications, including plasma cutting \cite{GODINAUD2022105479}, fire plumes \cite{MARAGKOS2022105208}, and 
metal forming processes \cite{SKURIC2018226}. Thanks to parallel computing support, multiphase modeling capabilities, a wide range of existing turbulence models and ease of implementation for new ones, OpenFOAM represents a great tool for the development and assessment of computational strategies for atmospheric modeling. 
In this context, researchers have chosen OpenFOAM to perform atmospheric boundary layer simulations (see, e.g., \cite{Kristof2009,BALOGH2012360,FLORES20131}), 
multi-fluid atmospheric convection
(see, e.g., \cite{Weller2020}), mesh $r$-adaptivity for weather prediction over steep terrain (see, e.g., \cite{YAMAZAKI2022111217}). The results in \cite{Weller2020,YAMAZAKI2022111217} have been obtained with AtmosFOAM \cite{atmosfoam_software}, an OpenFOAM-based model of the global atmosphere using arbitrary shaped cells. Our solver starts with a focus on local atmospheric problem. However, we do not exclude interfacing it with AtmosFOAM in the future. 

As the core of our open source package,
we present an OpenFOAM-based solver for the Euler equations for stratified fluid flow, which are of importance in non-hydrostatic mesoscale atmospheric modeling, and assess it through two well-known test cases. We consider the Euler equations written in conservative form using density, momentum, and total energy as variables. It is shown in \cite{giraldo_2008} that this set of equations yields less
dissipative results than other forms of the Euler equations for relevant test problems. Hence, it was recommended to pursue this formulation for the development of mesoscale models despite the fact that other formulations (i.e., non-conservative forms and the conservative form using density, momentum, and potential temperature as variables)
have received more attention for
atmospheric studies. 
For the stabilization of the Euler equations and to capture sub-grid processes, we will assess and compare two Large Eddy Simulation (LES) models built in OpenFOAM: the classical Smagorinsky model \cite{smagorinsky1963} and the one equation eddy-viscosity model
\cite{Yoshizawa1985}. We will show that both LES models yield numerical results that agree very well with data published in the literature. 

While nowadays both pressure-based and density-based approaches for the Euler equations are applicable to a wide range of flows, from incompressible to highly compressible, historically pressure-based solvers have been used for mildly compressible flows (as is the case for atmospheric flows). 
Despite this, the majority of the software for weather prediction adopts density-based approaches, which were originally designed for high-speed compressible flows. In this paper, we opt for a pressure-based solver and show that it yields accurate results when compared to data in the literature. In the spirit of OpenFOAM, our solver uses a splitting scheme that decouples the computation of each variable in order to achieve high computational efficiency. 

The rest of the paper is organized as follows. In Sec.~\ref{sec:pbd}, we briefly describe the formulation of the Euler equations and the LES models under consideration. Sec.~\ref{sec:splitting} presents our pressure-based approach and provides the details of space and time discretization. Numerical results for the two benchmark tests are discussed in Sec.~\ref{sec:num_res}. Conclusions are drawn in Sec.~\ref{sec:concl}.

%\textcolor{red}{Enfatizzare il fatto che tutti (to the best of our knowledge) usano un density based solver (Godunov-like schemes), mentre noi proponiamo per questa classe di problemi un pressure-based che storicamente, in altri ambiti, tipo in aerodinamica, sono usati in maniera preferenziale rispetto ai density based quando parliamo di Mach bassi. Al riguardo Anna dai un'occhiata a questa overview in ambito Fluent: \url{https://www.afs.enea.it/project/neptunius/docs/fluent/html/ug/node776.htm}. Ovviamente i software commerciali a sorgente chiusa noi li odiamo per natura ahaha :D pero' secondo me questa contrapposizione pressure based vs density based viene resa bene}.

\section{Problem definition}
\label{sec:pbd}

\subsection{The compressible Euler equations}
\label{sec:NS Equations}

We consider the dynamics of dry atmosphere in a spatial domain of interest $\Omega$ by neglecting the effects of moisture, solar radiation, and heat flux from the ground.
We assume that dry air behaves like an ideal gas.
%Before stating the equations that govern dry air motion, we need
%to introduce some notation. We denote by $c_{\alpha}$ the specific heat capacities at constant $\alpha = p, v$ (pressure, volume). The 
%specific gas constants of dry air is denoted by $R$.

Let $\rho$ be the air density, $\bu = (u, v, w)$  
the wind velocity, and $e$ the total energy density. Note that 
%\textcolor{red}{(qui avevamo fatto un errore, è $c_v T$ e non $c_p T$ perchè il primo addendo in $e$ è l'energia interna. L'entalpia invece può essere espressa come $c_p T$ come riporto nel Remark 2.1. Ho corretto comunque :))} 
$e = c_v T + |\bu|^2/2 + g z$, where $c_{v}$ the specific heat capacity at constant volume, $T$ is the absolute temperature, $g$ is the gravitational constant, and $z$ is the vertical coordinate.
The equations stating conservation of mass, momentum, and energy for the dry atmosphere written in terms of $\rho$, $\bu$, and $e$ over a time interval of interest $(0,t_f]$ read: 
\begin{align}
&\frac{\partial \rho}{\partial t} + \nabla \cdot (\rho \bu) = 0 &&\text{in } \Omega \times (0,t_f], \label{eq:mass}  \\
&\frac{\partial (\rho \bu)}{\partial t} +  \nabla \cdot (\rho \bu \otimes \bu) + \nabla p   + \rho g \widehat{\bk} = \boldsymbol{0} &&\text{in } \Omega \times (0,t_f],  \label{eq:mom} \\
&\frac{\partial (\rho e)}{\partial t} +  \nabla \cdot (\rho \bu e) + \nabla \cdot (p \bu) = 0 &&\text{in } \Omega \times (0,t_f],
\label{eq:ent}
\end{align}
%\sm{[SM] The gravity term in the energy equation must be removed.}\\
%\sm{[SM] I thought Michele changed the code to solve the theta equation, not the total energy equation. As I mentioned before, we will have major problems in including moisture if we use this equation. That's why nobody uses total energy. Not even ClimaMachine does anymore.}
where $\widehat{\bk}$ is the unit vector aligned with the vertical axis $z$ and $p$ is pressure. To close system \eqref{eq:mass}-\eqref{eq:ent}, we need a 
thermodynamics equation of state for $p$. Following the assumption that dry air behaves like an ideal gas, we have: %\textcolor{red}{(in realtà l'ipotesi di gas ideale la facciamo fin da quando assumiamo che l'energia interna possa scriversi come $C_v T$, vogliamo spostare questo commento prima?)}:
\begin{align}
p = \rho R T, %p_0 \left( \frac{\rho R \theta}{p_0} \right)^\frac{c_p}{c_v}. 
\label{eq:p}
\end{align}
where $R$ is the specific gas constant of dry air.

%Concerning the hydrostatic term, it is numerically convenient to solve for an alternative pressure defined by

Let us write the pressure as the sum of a fluctuation $p'$ with respect to a background state
\begin{align}
p = p' + \rho g z. \label{eq:p_split}
\end{align}
%\anna{Dove scriviamo come cambiare lo splitting di $p$, qui o in Sec.~\ref{sec:HB}?
%Se non vogliamo che un reviewer di problemi atmosferici si arrabbi subito, possiamo dire che presentiamo l'approccio per \eqref{eq:p_split} anche se su terreni non pianeggianti e' problematico e poi in Sec.~\ref{sec:HB} proponiamo come cambiarla se ci sono montagne.} \textcolor{red}{mmm bella domanda. Forse meglio presentarla direttamente in Sec. 4.3? Se gli diciamo fin da qui che sulla montagna cambiamo, usando l'approccio piu' famoso e riconosciuto, non vorrei che si incattivisse chiedendoci: ma perchè non lo avete fatto subito? Non so, ci pensiamo un altro po'}
By plugging \eqref{eq:p_split} into \eqref{eq:mom}, we obtain:
\begin{align}
\frac{\partial (\rho \bu)}{\partial t} +  \nabla \cdot (\rho \bu \otimes \bu) + \nabla p' + gz \nabla \rho =0\quad \text{in } \Omega \times (0,t_f].  \label{eq:mom_split}
\end{align}
%while by plugging \eqref{eq:p_split} into \eqref{eq:p}, we get
%\begin{align}
%p' = \rho (R T - g z).  \label{eq:state_split}
%\end{align}

%The pressure gradient and gravity force terms are rearranged in the following form:
%\begin{align}
%\nabla p + \rho g \widehat{\bk}  = \nabla p' + gz %\nabla \rho, %p_0 \left( \frac{\rho R \theta}{p_0} %\right)^\frac{c_p}{c_v}.
%\label{eq:pPrime}
%\end{align}
%where $p' = p - \rho g z$.
%\textcolor{red}{Possiamo riscrivere il sistema con la posizione (5)}

%\textcolor{red}{forse questo remark è da commentare perchè non ci consente di definire bene lo splitting che applichiamo. Qui potremmo ad esempio scrivere l'equazione in termini di entalpia}

%Let $h = c_p T + |\bu|^2/2$. Notice that, using eq.~\eqref{eq:mass}, eq.~\eqref{eq:ent} can be rewritten as:
%\begin{align}
%\frac{\partial (\rho h)}{\partial t} +  \nabla \cdot (\rho h \bu) + \rho g \bu \cdot \widehat{\bk} + \nabla \cdot (p \bu) = 0.
%\label{eq:over_ent}
%\end{align}

Let $c_{p}$ be the specific heat capacity at constant pressure for dry air and let
\begin{equation}\label{eq:K_h}
K = |\bu|^2/2, \quad h = c_v T + p/\rho = c_p T,    
\end{equation}
be the kinetic energy density and the specific enthalpy, respectively. The total energy density can be written as $e = h - p/\rho + K + gz$. Then, eq.~\eqref{eq:ent} can be rewritten as:
\begin{align}
\frac{\partial (\rho h)}{\partial t} +  \nabla \cdot (\rho \bu h) + 
\frac{\partial (\rho K)}{\partial t} +  \nabla \cdot (\rho \bu K) - \dfrac{\partial p}{\partial t}  +  
\rho g \bu \cdot \widehat{\bk} = 0,
\label{eq:over_ent}
\end{align}
where we have used eq.~\eqref{eq:mass} for further simplification.  We will devise a splitting approach for problem \eqref{eq:mass},\eqref{eq:p}-\eqref{eq:over_ent} because this formulation of the Euler equation facilitates the decoupling of all variables. In addition, it allows for an explicit treatment of the kinetic and potential energies.

\begin{rem}
A quantity of interest for atmospheric problems is the potential temperature
\begin{align}
\theta = \frac{T}{\pi}, \quad \pi = \left( \frac{p}{p_0} \right)^{\frac{R}{c_{p}}}, \label{eq:theta}
\end{align}
i.e., the temperature that a parcel of dry air would have if it were expanded or compressed
adiabatically to standard pressure $p_0 = 10^5$ Pa, which is the atmospheric pressure at the ground. Additionally, 
we define the potential temperature fluctuation $\theta'$ as the difference between $\theta$ and its mean hydrostatic value $\theta_0$:
\begin{align}
\theta'(x,y,z,t) = 
\theta(x,y,z,t) - \theta_0(z) . \label{eq:theta_split} 
\end{align}
Note that the hydrostatic reference state is a  function
of the vertical coordinate $z$ only. See, e.g., \cite{kellyGiraldo2012} for more details. 
\end{rem}

\subsection{LES models}\label{sec:LES}

The numerical discretization of the model presented in the previous subsection would lead to an accurate description of atmospheric flow if one could afford a discretization mesh able to capture all the scales of turbulent structures.
Since the Kolmogorov scale of a typical atmospheric problem is about $10^{-4}$ m, a Direct Numerical Simulation is currently beyond reach. One way to keep the computational cost affordable is to solve for the flow average by using a coarser mesh and model the effects of the small scales that are not directly solved. This can be done via Large Eddy Simulation. 

We will focus on two LES models: the classical Smagorinsky model \cite{smagorinsky1963} and the one equation eddy-viscosity model
\cite{Yoshizawa1985}.
Both approaches are equivalent to introducing additional terms in eq.~\eqref{eq:mom_split} and \eqref{eq:over_ent} of the form: %\textcolor{red}{Anna, secondo te si possono allineare meglio le eq. qui sotto o vanno bene cosi'? Inoltre ho cambiato il termine diffusivo, ho preferito metterlo in termini di $h$ rendendolo esattamente coerente con OF :) Prima era $\nabla \left(\dfrac{\mu_a c_p}{Pr}\nabla T \right)$. Alla fine non cambia nulla, giusto per coerenza. Inoltre ho messo un termine di viscosità in più nella quantità di moto che nel comprimibile ci deve essere a rigore}: %\textcolor{red}{(se togliamo da mezzo il remark ricordiamoci di modificare di conseguenza anche la seconda equazione qui sotto)}:
\begin{align}
&\frac{\partial (\rho \bu)}{\partial t} +  \nabla \cdot (\rho \bu \otimes \bu) + \nabla p' + gz \nabla \rho -  \nabla \cdot (2 \mu_a \boldsymbol{\epsilon}(\bu)) + \nabla \left(\frac{2}{3}\mu_a \nabla \cdot \bu \right)= 0 &&\text{in } \Omega \times (0,t_f],  \label{eq:mom_LES} \\
&\frac{\partial (\rho h)}{\partial t} +  \nabla \cdot (\rho \bu h) + 
\frac{\partial (\rho K)}{\partial t} +  \nabla \cdot (\rho \bu K) - \dfrac{\partial p}{\partial t}  +  
\rho g \bu \cdot \widehat{\bk}  - \nabla \cdot \left(\frac{\mu_a}{Pr} \nabla h \right) = 0 &&\text{in } \Omega \times (0,t_f],
%&\frac{\partial (\rho h)}{\partial t} +  \nabla \cdot (\rho h \bu) + \nabla \cdot (p \bu) + \rho g \bu \cdot \widehat{\bk}  - \nabla \cdot \left(\frac{\mu_a c_p}{Pr} \nabla T \right) = 0 &&\text{in } \Omega \times (0,t_f],
\label{eq:ent_LES}
\end{align}
where $\mu_a$ is an artificial viscosity, $\boldsymbol{\epsilon}(\bu) = (\nabla \bu + (\nabla \bu)^T)/2$ is the strain-rate tensor, and $Pr$ is the Prandtl number, i.e., the dimensionless number defined as the ratio of momentum diffusivity to thermal diffusivity.
The definition of $\mu_a$ is what distinguishes the different LES models. 

The artificial dynamics viscosity, also called subgrid scale eddy viscosity, introduced by the Smagorinsky model is given by:
\begin{align}
\mu_a = \rho (C_s \delta)^2 \sqrt{ 2 \boldsymbol{\epsilon} : \boldsymbol{\epsilon}}, \quad C_s^2 = C_k \sqrt{\dfrac{C_k}{C_{\epsilon}}} \label{eq:smago}
\end{align}
where $\delta$ is the filter width, and $C_k$ and $C_{\epsilon}$ are model parameters. For the results in Sec.~\ref{sec:num_res},
we compute $\delta$ for each cell by taking the maximum distance between the cell center and a face center multiplied by 2. The main limitation of the Smagorinsky model lies in the assumption of local balance
between the subgrid scale energy production and dissipation. The one equation
eddy-viscosity model (kEqn model, for short) in \cite{Yoshizawa1985} was developed to overcome this limitation. The artificial dynamics viscosity introduced by this model is given by:
\begin{align}
    \mu_a = \rho C_k \sqrt{k_\text{sgs}} \delta,
    \label{eq:1eq}
\end{align}
where $C_k$ and $\delta$ represent the same quantities as in the Smagorinsky model and $k_\text{sgs}$ is the subgrid scale kinetic energy computed from a transport equation that accounts for production, dissipation, and diffusion. See \cite{Yoshizawa1985} for more details.

\section{A splitting approach}\label{sec:splitting}

This section presents a space and time discretization for the generic LES model
\eqref{eq:mass},\eqref{eq:p},\eqref{eq:p_split},\eqref{eq:K_h},\eqref{eq:mom_LES},
\eqref{eq:ent_LES}.

Let $\Delta t \in \mathbb{R}$, $t^n = n \Delta t$, with $n = 0, ..., N_f$ and $t_f = N_f \Delta t$. Moreover, we denote by $y^n$ the approximation of a generic quantity $y$ at the time $t^n$. We adopt a Backward Differentiation Formula of order 1 (BDF1) for the discretization of the Eulerian time derivatives. Problem \eqref{eq:mass},\eqref{eq:p},\eqref{eq:p_split},\eqref{eq:K_h},\eqref{eq:mom_LES},\eqref{eq:ent_LES}
discretized in time reads: given $\rho^0$, $\bu^0$, $h^0$,  $p^0$, and $T^0$, set $K^0 = |\bu^0|^2/2$ and for $n \geq 0$ find solution $(\rho^{n+1}, \bu^{n+1},h^{n+1},K^{n+1},p^{n+1}, p'^{,n+1}, T^{n+1})$ of system: %\textcolor{red}{forse cambiare l'eq. 16}:
\begin{align}
& \frac{\rho^{n+1}}{\Delta t} + \nabla \cdot (\rho^{n+1} \bu^{n+1}) = b^{n+1}_\rho, \label{eq:mass_td}  \\
&\frac{\rho^{n+1} \bu^{n+1}}{\Delta t} +  \nabla \cdot (\rho^{n+1} \bu^{n+1} \otimes \bu^{n+1}) + \nabla p'^{,n+1} + gz \nabla \rho^{n+1} \cl 
&\quad -  \nabla \cdot (2 \mu_a^{n+1} \boldsymbol{\epsilon} ( \bu^{n+1})) +  \nabla \left(\frac{2}{3}\mu_a^{n+1} \nabla \cdot \bu^{n+1} \right) =\bb^{n+1}_\bu,  \label{eq:mom_td} \\
& \frac{\rho^{n+1} h^{n+1}}{\Delta t} +  \nabla \cdot (\rho^{n+1} \bu^{n+1} h^{n+1}) + \frac{\rho^{n+1} K^{n+1}}{\Delta t} +  \nabla \cdot (\rho^{n+1} \bu^{n+1} K^{n+1})  \cl
&\quad- \frac{p^{n+1}}{\Delta t} + \rho^{n+1} g \bu^{n+1}  \cdot \widehat{\bk}  - \nabla \cdot \left(\frac{\mu_a^{n+1} }{Pr} \nabla h^{n+1} \right) = b_e^{n+1}, 
\label{eq:ent_td} \\
&p^{n+1} = p'^{,n+1} + \rho^{n+1} g z, \label{eq:p_td} \\
& p^{n+1} = \rho^{n+1} R T^{n+1},\label{eq:p_td2}  \\
&h^{n+1} - h^{n} = c_p (T^{n+1} - T^n) \label{eq:T_td}, \\
& K^{n+1} = \frac{|\bu^{n+1}|^2}{2}, \label{eq:K_td}
\end{align}
where $b^{n+1}_\rho = \rho^n/\Delta t$, $\bb^{n+1}_\bu = \rho^{n}\bu^n/\Delta t$, and
$b_e^{n+1} = (\rho^nh^n + \rho^n K^n - p^n)/\Delta t$. Notice that in \eqref{eq:T_td} we have chosen to update the value of the specific enthalpy in an incremental fashion. 

%\anna{Should I use $\mu_a^{n+1}$ instead, since it might be a quantity dependent on certain variables?} \textcolor{red}{Direi di si! L'ho predisposto :)}

%\begin{rem}
%We adopt a first order extrapolation for the convective terms in \eqref{eq:mass_td}-\eqref{eq:ent_td} although we use a second order approximation for the time derivatives because this is what  
%OpenFOAM solvers typically do \cite{Weller1998}
%\anna{Michele, see if you want to add anything. Magari aggiungiamo anche una (o piu') referenze.} \textcolor{red}{ho inserito una refernza generale di OF.  Non aggiungerei altro, tutto sommato è solo un dettaglio implementativo di OF (anche se ovviamente influenza l'ordine di accuratezza dello schema).}
%\end{rem}

A monolithic approach for coupled problem \eqref{eq:mass_td}-\eqref{eq:K_td} would lead to high computational costs. 
Thus, to save computational time we adopt the following splitting approach: given $\rho^0$, $\bu^0$, $h^0$, $p^0$ and $T^0$, set $K^0 = |\bu^0|^2/2$ and for $n \geq 0$ perform
\begin{itemize}
\item[-] Step 1: find first intermediate density ${\rho}^{n+\frac{1}{3}}$, intermediate velocity ${\bu}^{n+\frac{1}{3}}$ and associated kinetic energy density $K^{n+\frac{1}{3}}$ such that 
\begin{align}
& \frac{{\rho}^{n+\frac{1}{3}}}{\Delta t}  = b^{n+1}_\rho - \nabla \cdot (\rho^{n} \bu^{n}),\label{eq:step1} \\
& \frac{{\rho}^{n+\frac{1}{3}} {\bu}^{n+\frac{1}{3}}}{\Delta t} +  \nabla \cdot (\rho^{n} \bu^{n}\otimes {\bu}^{n+
\frac{1}{3}}) + \nabla p^{n}  -  \nabla \cdot (2 \mu_a^n \boldsymbol{\epsilon}
({\bu}^{n+\frac{1}{3}})) +\nabla \left(\frac{2}{3}\mu_a^{n} \nabla \cdot \bu^{n} \right)  =\bb^{n+1}_\bu, \label{eq:step1_2} \\
& K^{n+\frac{1}{3}} = \frac{|\bu^{n+\frac{1}{3}}|^2}{2} \label{eq:step1_3}.
\end{align}
\item[-] Step 2: find specific enthalpy $h^{n+1}$, temperature $T^{n+1}$, and second intermediate density ${\rho}^{n+\frac{2}{3}}$ such that
%\anna{(stessa cosa per il termine convettivo e non sono sicura del termine con $\widehat{\bk}$)}
\begin{align}
& \frac{{\rho}^{n+\frac{1}{3}} h^{n+1}}{\Delta t} +  \nabla \cdot (\rho^{n} \bu^{n} h^{n+1}) - \nabla \cdot \left(\frac{\mu_a^n}{Pr} \nabla h^{n+1} \right) = \tilde{b}_e^{n} - \frac{{\rho}^{n+\frac{1}{3}} K^{n+\frac{1}{3}}}{\Delta t}\cl
& \quad  -  \nabla \cdot (\rho^{n} \bu^{n} K^{n+\frac{1}{3}}) + \frac{p^{n}}{\Delta t} - {\rho}^{n+\frac{1}{3}} g {\bu}^{n+\frac{1}{3}}  \cdot \widehat{\bk}, \label{eq:step2_1} \\
&h^{n+1} - c_p T^{n+1} =  h^{n} - c_p T^{n},
%|\tilde{\bu}^{n+1}|^2. 
\label{eq:step2_2} \\
& {\rho}^{n+\frac{2}{3}} R T^{n+1} = p^{n},  \label{eq:step2_3} %\\
\end{align}
where $\tilde{b}_e^{n} = (\rho^n h^n + \rho^n K^{n-1} - p^{n-1}) / \Delta t$.  
%where
%$\tilde{b}_e^{n+1} = (\rho^nh^n + \rho^n K^n - p^n)/\Delta t$.
\item[-] Step 3: find end-of-step velocity $\bu^{n+1}$ and associated kinetic energy density $K^{n+1}$, pressure $p^{n+1}$ and pressure fluctuation $p'^{,n+1}$, and end-of-step density $\rho^{n+1}$ 
such that %\anna{Michele, per favore controlla il termine convettivo, e' giusto che sia $\rho^{n} \bu^n$ o dovrebbe essere 
%${\rho}^{n+\frac{1}{3}} \tilde{\bu}^{n+1}$?}
%\anna{Michele, nel primo termine di \eqref{eq:step3} non e' ${\rho}^{n+\frac{2}{3}}$ invece di ${\rho}^{n+\frac{1}{3}}$? E nell'ultimo termine prima dell'uguale non e' ${\bu}^{n+1}$? Manca un'eq.}
\begin{align}
& \frac{{\rho}^{n+\frac{1}{3}} {\bu}^{n+1}}{\Delta t} +  \nabla \cdot (\rho^{n} \bu^n \otimes {\bu}^{n+1}) + \nabla p^{n+1}  -  \nabla \cdot (2 \mu_a^n \boldsymbol{\epsilon}
({\bu}^{n+1})) +\nabla \left(\frac{2}{3}\mu_a^{n} \nabla \cdot \bu^{n} \right)  =\bb^{n+1}_\bu, \label{eq:step3} \\  
%\end{align}
%\item[-] Step 4: find pressure $p^{n+1}$, and density $\rho^{n+1}$, together with pressure fluctuation $p'^{,n+1}$, such that %\anna{(Michele, I changed the convective term in eq.~\eqref{eq:step4_1} because I need it like this at p.~5, please correct if it's wrong. If it is correct, we may want to change \eqref{eq:mass_td}.)} \textcolor{red}{Anna ho cambiato un po' di cose per rendere il tutto il piu' possibile fedele al codice}
%\begin{align}
%& p^{n} = {\rho}^{n+\frac{2}{3}} R T^{n+1} \label{eq:step3_1}\\
%&\frac{3}{2 \Delta t} {\rho}^{n+\frac{1}{3}} \bu^{n+1} +  \nabla \cdot (\rho^{n} \bu^n \otimes \bu^{n+1}) + \nabla p^{n+1}
%-  \nabla \cdot (\mu_a(\nabla \bu^{n+1} + (\nabla \bu^{n+1})^T))  =\bb^{n+1}_\bu,  \label{eq:step4_2} \\
& p^{n+1} - p'^{,n+1} = {\rho}^{n+\frac{2}{3}} g z, \label{eq:step3_2} \\
%& p^{n+1} = \rho^{n+1} R T^{n+1}\label{eq:step4_4}, \\
%& \frac{\tilde{\tilde{\tilde{\rho}}}^{n+1}}{\Delta t} + \nabla \cdot ({\rho}^{n+\frac{2}{3}} {\bu}^{n+1}) = b^{n+1}_\rho, \label{eq:step3_3}  \\
& p^{n+1} - \rho^{n+1} R T^{n+1} = 0, \label{eq:step3_3} \\
& \frac{\rho^{n+\frac{2}{3}}}{\Delta t} + \nabla \cdot ({\rho}^{n+\frac{2}{3}} {\bu}^{n+1}) = b^{n+1}_\rho \label{eq:step3_4}, \\
& K^{n+1} = \frac{|\bu^{n+1}|^2}{2} \label{eq:step3_5}.
\end{align}
\end{itemize}
Notice that in eq.~\eqref{eq:step3} one diffusive term is kept implicit, while the other is treated explicitly. This is customary in OpenFOAM.

%Notice that by starting from \eqref{eq:over_ent}, instead of \eqref{eq:ent}, we can write \eqref{eq:step2_1}-\eqref{eq:step2_2} as a problem with $T^{n+1}$ as the only effective unknown. 

The next section will describe the space descretization of the equations at each step and explain how to further reduce the computational cost associated with Step 3.

\subsection{Space discretization}
For the space discretization of \eqref{eq:step1}-\eqref{eq:step3_5},
we adopt a finite volume method. This requires to partition the computational domain $\Omega$ into cells or control volumes $\Omega_i$, with $i = 1, \dots, N_{c}$, where $N_{c}$ is the total number of cells in the mesh. 
%\textcolor{blue}{Ho introdotto il numero di volumi finiti $N$ in cui discretizziamo il dominio computazionale} 
Let  \textbf{A}$_j$ be the surface vector of each face of the control volume, 
with $j = 1, \dots, M$.  
%We chose to implement the proposed discretization within
%the finite volume C++ library OpenFOAM\textsuperscript{\textregistered} \cite{Weller1998}.

Let us start with Step 1.
After applying the Gauss-divergence theorem,
the integral form of eq.~\eqref{eq:step1} for each volume $\Omega_i$ is given by:
% \begin{align}
% \frac{3}{2 \Delta t}\int_{\Omega_i} {\rho}^{n+\frac{1}{3}} d\Omega +  \int_{\Omega_i}  \nabla \cdot (\rho^n \bu^n) d\Omega = \int_{\Omega_i} b_\rho^{n+1} d\Omega. \el
% \end{align}
%By applying the Gauss-divergence theorem, the equation above becomes:
\begin{align}
\frac{1}{\Delta t}\int_{\Omega_i} {\rho}^{n+\frac{1}{3}} d\Omega  = \int_{\Omega_i} b_\rho^{n+1} d\Omega - \int_{\partial \Omega_i} \rho^n \bu^n \cdot d\textbf{A}. \label{eq:e1_l_gauss}
\end{align}
Let us denote with $(\rho^n\bu^{n})_i$ and ${\rho}^{n+\frac{1}{3}}_i$ the average flux 
and intermediate density in control volume $\Omega_i$, respectively. Then, eq.~\eqref{eq:e1_l_gauss} is approximated as follows:
\begin{align}
\frac{1}{\Delta t} {\rho}^{n+\frac{1}{3}}_i + \sum_j  \varphi^n_j = b_{\rho,i}^{n+1}, \quad \varphi^n_j =(\rho^n\bu^{n})_{i,j}  \cdot  \textbf{A}_j,\label{eq:e1_d} 
\end{align}
where $ \varphi^n_j$ denotes 
the convective
flux
through face $j$ of $\Omega_i$
and $b_{\rho,i}^{n+1}$ is the average right-hand side term in $\Omega_i$. The convective flux at the cell faces is computed by a linear interpolation of the values from
the adjacent cells. %\anna{E' per la divergenza dei flussi che usiamo il 
%terz'ordine? Se si', possiamo dirlo qui e dire che vale per tutte le altre eq.}

Next, we consider eq.~\eqref{eq:step1_2}. Its integral form for each volume $\Omega_i$, it is given by: 
% \begin{align}%\label{eq:mom_if}
% \frac{3}{2\Delta t} \int_{\Omega_i} {\rho}^{n+\frac{1}{3}} \tilde{\bu}^{n+1} d\Omega &+ \int_{\Omega_i} \div \left(\rho^n \bu^n \otimes \tilde{\bu}can be written as^{n+1}\right) d\Omega  + \int_{\Omega_i}\nabla p^{n} d\Omega 
% \cl
% & - \int_{\Omega_i} \nabla \cdot (\mu_a(\nabla \tilde{\bu}^{n+1} + (\nabla \tilde{\bu}^{n+1})^T)) d\Omega
%  = \int_{\Omega_i}\bb^{n+1}_{\bu} d\Omega. \el
% \end{align}
%By applying the Gauss-divergence theorem, the equation above becomes:
\begin{align}
\frac{1}{\Delta t} \int_{\Omega_i} {\rho}^{n+\frac{1}{3}} \bu^{n+\frac{1}{3}} d\Omega &+\int_{\partial \Omega_i} \left(\rho^n \bu^n \otimes \bu^{n+\frac{1}{3}}\right) \cdot d\textbf{A}  
+ \int_{\Omega_i}\nabla p^{n} d\Omega \cl
&  - \int_{\partial \Omega_i} 2 \mu_a^n \boldsymbol{\epsilon} (\bu^{n+\frac{1}{3}}) \cdot d\textbf{A}  + \int_{\partial \Omega_i} \left(\frac{2}{3}\mu_a^{n} \nabla \cdot \bu^{n}\right)   d\textbf{A} = \int_{\Omega_i}{\bb}_\bu^{n+1} d\Omega, \label{eq:e2_gauss}
\end{align}
after the application of the Gauss-divergence theorem.
For the approximation of most of the terms in \eqref{eq:e2_gauss}, we follow \cite{Girfoglio2019}. So, 
we write the discretized form of \eqref{eq:e2_gauss} divided by the control volume $\Omega_i$ as:
\begin{align}
\frac{1}{\Delta t} {\rho}^{n+\frac{1}{3}}_i \bu^{n+\frac{1}{3}}_i &+\sum_j^{} \varphi^n_j \bu^{n+\frac{1}{3}}_{i,j} + \nabla p^{n}_{i} - \sum_j^{} 2 \mu_a^n \boldsymbol{\epsilon}(\bu^{n+\frac{1}{3}}_i)_j \cdot \textbf{A}_j   \cl
&+ \sum_j \left(\frac{2}{3}\mu_a^{n} (\nabla \cdot \bu^{n})_j\right) \textbf{A}_j = {\bm b}^{n+1}_{\bu, i}, \label{eq:step2_sd}
\end{align}
where $\bu^{n+\frac{1}{3}}_i$, $p^{n}_{i}$, and ${\bm b}^{n+1}_{\bu, i}$ are the average intermediate velocity, pressure, and source term in control volume $\Omega_i$, while
$\bu^{n+\frac{1}{3}}_{i,j}$ denotes the intermediate velocity associated to the centroid of face $j$ normalized by the volume of $\Omega_i$. This approximation of $\bu^{n+\frac{1}{3}}$ at cell face $j$ is obtained with a third order interpolation scheme \cite{Weller1998}. We will use the same scheme for all the flux terms in the equations below.
For term $\boldsymbol{\epsilon}(\bu^{n+\frac{1}{3}}_i)_j$, we need to approximate 
the gradient of $\bu_i^{n+\frac{1}{3}}$ at face $j$. We choose to do it with second order accuracy. 
%Such gradient is approximated with second order \anna{(or third order?)} \textcolor{red}{second order!} accuracy.
See \cite{jasakphd} for more details. 
The pressure term is treated with a second-order face flux reconstruction in order to suppress spurious oscillations \cite{gradrho}. We discretize all the pressure terms in the following equations in the same way. 

To complete Step 1, we need the discretized form of eq.~\eqref{eq:step1_3}, which is given by
\begin{align}
K^{n+\frac{1}{3}}_i = \frac{|\bu_i^{n+\frac{1}{3}}|^2}{2}. \label{eq:step1_3sd}
\end{align}

Let us now consider Step 2. After the application of the Gauss-divergence theorem, the integral form of eq.~\eqref{eq:step2_1} for each volume $\Omega_i$ becomes:
\begin{align}
& \frac{1}{\Delta t}\int_{\Omega_i} {\rho}^{n+\frac{1}{3}} h^{n+1} d\Omega + \int_{\partial \Omega_i} (\rho^{n} \bu^{n} h^{n+1}) \cdot d\textbf{A}  %+ \int_{\partial \Omega_i} (p^{n} \bu^{n}) \cdot d\textbf{A}  
 - \int_{\partial \Omega_i} \left(\frac{\mu_a^n }{Pr} \nabla h^{n+1} \right) \cdot d\textbf{A}  = \int_{\Omega_i} \tilde{b}_e^{n} d\Omega 
 \cl 
& \quad  -  \frac{1}{\Delta t}\int_{\Omega_i} {\rho}^{n+\frac{1}{3}} K^{n+\frac{1}{3}} d\Omega - \int_{\partial \Omega_i} (\rho^{n} \bu^{n} K^{n+\frac{1}{3}}) \cdot d\textbf{A} +\frac{1}{\Delta t}\int_{\Omega_i} p^{n} d\Omega - \int_{\Omega_i} {\rho}^{n+\frac{1}{3}} g \bu^{n+\frac{1}{3}} \cdot \widehat{\bk} d\Omega , \label{eq:step3_1if}
\end{align}
The discretized form of eq.~\eqref{eq:step3_1if} can be written as:
\begin{align}
&\frac{1}{\Delta t} {\rho}^{n+\frac{1}{3}}_i h^{n+1}_i + \sum_j^{} \varphi^n_j h^{n+1}_{i,j} - \sum_j^{} \frac{\mu_a}{Pr} (\nabla h^{n+1}_i)_j \cdot \textbf{A}_j  = \tilde{b}_{e,i}^{n} \cl
&\quad - 
\frac{1}{\Delta t} {\rho}^{n+\frac{1}{3}}_i K^{n+\frac{1}{3}}_i - \sum_j^{} \varphi^n_j K^{n+\frac{1}{3}}_{i,j} + \frac{1}{\Delta t} p_i^n - {\rho}^{n+\frac{1}{3}}_i g \bu^{n+\frac{1}{3}}_i \cdot \widehat{\bk}, \label{eq:step3_sd}
\end{align}
where $h^{n+1}_i$ and $b^{n}_{e, i}$ are the average specific enthalpy
and source term in control volume $\Omega_i$, while
$h^{n+1}_{i,j}$ and $K^{n+\frac{1}{3}}_{i,j}$ denote the specific enthalpy and kinetic energy density
associated to the centroid of face $j$ normalized by the volume of $\Omega_i$. Finally,  
$(\nabla h_i^{n+1})_j$ is the gradient of $h_i^{n+1}$ at face $j$. %\textcolor{red}{which is approximated in the same way as $(\nabla\bu^{n+1})_j$}.

To complete Step 2, we need the discretized form of eqs.~\eqref{eq:step2_2} and~\eqref{eq:step2_3}, which are given by
\begin{align}
h^{n+1}_i - c_p T^{n+1}_i =  h^{n}_i - c_p T^{n}_i,
%h^{n+1}_i = c_v T^{n+1}_i + \frac{p_i^n}{\tilde{\rho}_i^{n+1}},  %\frac{1}{2} |\tilde{\bu}^{n+1}_i|^2.
\label{eq:step3_2sd} \\
{\rho}^{n+\frac{2}{3}}_i R T^{n+1}_i = p^{n}_i. \label{eq:step3_3sd}
\end{align}
Note that the specific enthalpy is computed from eq.~\eqref{eq:step3_sd}, so one computes the end-of-step temperature from \eqref{eq:step3_2sd} and
the second intermediate density from \eqref{eq:step3_3sd} in a completely decoupled fashion. 

%The treatment of Step 4 requires careful attention in order to contain the computational cost. We will compute first the pressure fluctuation as follows.
%Let us start by writing eq.~\eqref{eq:step2_sd} with the pressure term evaluated at time $t^{n+1}$ as in \eqref{eq:step4_3} and in semi-discretized form, i.e.~with 
%the pressure term in continuous form while all the other terms are in discrete form: 
%\begin{align}
%{\rho}^{n+\frac{1}{3}}_i \tilde{\bu}^{n+1}_i &= \frac{2\Delta t}{3} \left(\mathbf{H}(\tilde{\bu}^{n+1}) - gz \nabla {\rho}^{n+\frac{1}{3}} - \nabla p'^{,n+1} \right), \label{g} \\
%\mathbf{H}(\tilde{\bu}^{n+1}) &= - \sum_j^{} \varphi^n_j \tilde{\bu}^{n+1}_{i,j}
%+ \sum_j^{} \mu_a ((\nabla \tilde{\bu}^{n+1}_i)_j+ (\tilde{\bu}^{n+1}_i)^T_j) \cdot \textbf{A}_j + {\bm b}^{n+1}_{\bu, i}.
%\el
%\end{align}
%We approximate eq.~\eqref{eq:step4_1} in the same way as eq.~\eqref{eq:step1} and obtain:
%\begin{align}
%\frac{3}{2 \Delta t} {\rho}^{n+\frac{1}{3}}_i + \sum_j ({\rho}^{n+\frac{1}{3}}\tilde{\bu}^{n+1})_{i,j}  \cdot  \textbf{A}_j = b_{\rho,i}^{n+1}, \label{eq:step4_1sd} 
%\end{align}
%Next, we take the divergence of the eq.~\eqref{eq:step_int}
%and make use of eq.~\eqref{eq:step4_1sd} to get:
%\begin{align}%\label{eq:q_poissonL_tmp}%\label{eq:q}
%\Delta p'^{,n+1} =  \nabla \cdot \textbf{H}(\tilde{\bu}^{n+1}) - gz \Delta {\rho}^{n+\frac{1}{3}} - \frac{9}{4\Delta t^2} {\rho}^{n+\frac{1}{3}}_i + \frac{3}{2\Delta t} b^{n+1}_{\rho,i}. \label{eq:p_prime}
%\end{align}
The treatment of Step 3 requires careful attention in order to contain the computational cost. We start by plugging eq.~\eqref{eq:step3_2} into eq.~\eqref{eq:step3}. 
The integral form of the resulting equation 
for each volume $\Omega_i$ is given by:
% \begin{align}%\label{eq:mom_if}
% \frac{3}{2\Delta t} \int_{\Omega_i} {\rho}^{n+\frac{1}{3}} \tilde{\bu}^{n+1} d\Omega &+ \int_{\Omega_i} \div \left(\rho^n \bu^n \otimes \tilde{\bu}can be written as^{n+1}\right) d\Omega  + \int_{\Omega_i}\nabla p^{n} d\Omega 
% \cl
% & - \int_{\Omega_i} \nabla \cdot (\mu_a(\nabla \tilde{\bu}^{n+1} + (\nabla \tilde{\bu}^{n+1})^T)) d\Omega
%  = \int_{\Omega_i}\bb^{n+1}_{\bu} d\Omega. \el
% \end{align}
%By applying the Gauss-divergence theorem, the equation above becomes:
\begin{align}
\frac{1}{\Delta t} \int_{\Omega_i} {\rho}^{n+\frac{1}{3}} \bu^{n+1} d\Omega &+\int_{\partial \Omega_i} \left(\rho^n \bu^n \otimes \bu^{n+1}\right) \cdot d\textbf{A}  
+ \int_{\Omega_i}\nabla p'^{,n+1} d\Omega + \int_{\Omega_i} g z \nabla{{\rho}^{n+\frac{2}{3}}} d\Omega \cl
&  - \int_{\partial \Omega_i} 2 \mu_a^n \boldsymbol{\epsilon}(\bu^{n+1}) \cdot d\textbf{A}  + \int_{\partial \Omega_i} \left(\frac{2}{3}\mu_a^{n} \nabla \cdot \bu^{n}\right)  d\textbf{A} = \int_{\Omega_i}{\bb}_\bu^{n+1} d\Omega, \label{eq:step2_if}
\end{align}
after the application of the Gauss-divergence theorem.
Each term in \eqref{eq:step2_if} is approximated like
the corresponding term in \eqref{eq:e2_gauss}. Hence,
the discretized form of \eqref{eq:step2_if}, divided by the control volume 
$\Omega_i$, is given by:
\begin{align}
\frac{1}{\Delta t} {\rho}^{n+\frac{1}{3}}_i \bu^{n+1}_i &+\sum_j^{} \varphi^n_j \bu^{n+1}_{i,j} + \nabla p'^{,n+1}_{i}  + g z_i \grad{{\rho}_i^{n+\frac{2}{3}}} \cl
&- \sum_j^{} 2 \mu_a^n \boldsymbol{\epsilon}(\bu^{n+1}_i)_j \cdot \textbf{A}_j + \sum_j \left(\frac{2}{3}\mu_a^{n} (\nabla \cdot \bu^{n})_j\right) \textbf{A}_j = {\bm b}^{n+1}_{\bu, i}, \label{eq:step2_sd}
\end{align}
where $\bu^{n+1}_i$, $p'^{,n+1}_{i}$ and ${\bm b}^{n+1}_{\bu, i}$ are the average end-of-step velocity, pressure fluctuation, and source term in control volume $\Omega_i$, while
$\bu^{n+1}_{i,j}$ denotes the end-of-step velocity
associated to the centroid of face $j$ normalized by the volume of $\Omega_i$. In \eqref{eq:step2_sd}, $z_i$ is the vertical coordinate of the centriod of cell $\Omega_i$.
%and
%$(\nabla \bu_i^{n+1})_j$ is the gradient of $\bu_i^{n+1}$ at face $j$ approximated in the same way as $(\nabla h_i^{n+1})_j$.
%Such gradient is approximated with second order \anna{(or third order?)} \textcolor{red}{second order!} accuracy.  

Following \cite{jasakphd}, we now write eq.~\eqref{eq:step2_sd} in semi-discretized form, i.e., with some terms in continuous form while all other terms (grouped in $\mathbf{H}$) are in discrete form: %\anna{Michele, controlla i termini continui vs discreti}
\begin{align}
\bu^{n+1} &= \frac{\Delta t}{{\rho}^{n+\frac{1}{3}}} \left(\mathbf{H}(\bu^{n+1}_i) - gz \nabla {\rho}^{n+\frac{2}{3}} - \nabla p'^{,n+1} \right), \label{eq:step3_1_int} \\
\mathbf{H}(\bu_i^{n+1}) &= - \sum_j^{} \varphi^n_j \bu^{n+1}_{i,j}
+ \sum_j^{} 2 \mu_a^n \boldsymbol{\epsilon}(\bu^{n+1}_i)_j \cdot \textbf{A}_j - \sum_j \left(\frac{2}{3}\mu_a^{n} (\nabla \cdot \bu^{n})_j\right) \textbf{A}_j + {\bm b}^{n+1}_{\bu, i}. 
\el
\end{align}
Next, we plug \eqref{eq:step3_1_int} into \eqref{eq:step3_4} to obtain:
\begin{align}
\frac{1}{\Delta t} {\rho}^{n+\frac{2}{3}} %+ \frac{1}{RT^{n+1}}\frac{p_i'^{,n+1} - p_i'^{,n}}{\Delta t} 
 +  \nabla \cdot \left({\rho}^{n+\frac{2}{3}} \left(\frac{\Delta t}{{\rho}^{n+\frac{1}{3}}} \left(\mathbf{H}(\bu^{n+1}) - gz \nabla {\rho}^{n+\frac{2}{3}} - \nabla p'^{,n+1} \right) \right)\right) = b_{\rho}^{n+1}. \label{eq:p_prime}
\end{align}
By integrating eq.~\eqref{eq:p_prime} over the control volume $\Omega_i$, applying the Gauss-divergence theorem, and dividing by the control volume, we get:
\begin{align}
\sum_j {\rho}_j^{n+\frac{2}{3}}(\nabla p'^{,n+1}_i)_j \cdot \textbf{A}_j =  \sum_j  \frac{ {\rho}_j^{n+\frac{2}{3}} \Delta t}{{\rho}_j^{n+\frac{1}{3}}} \left(\mathbf{H}(\bu_i^{n+1})_j  - gz_j (\nabla {\rho}_i^{n+\frac{2}{3}})_j \right) \cdot \textbf{A}_j  - b_{\rho, i}^{n+1} + \dfrac{1}{\Delta t} {\rho}_i^{n+\frac{2}{3}} %\frac{9}{4\Delta t^2} {\rho}^{n+\frac{1}{3}}_i + \frac{3}{2\Delta t} b^{n+1}_{\rho,i}
, \label{eq:p_prime_sd}
\end{align}
which is used to compute the pressure fluctuation.
In eq.~\eqref{eq:p_prime_sd}, $(\nabla p'^{,n+1}_i)_j$ and $(\nabla {\rho}_i^{n+\frac{2}{3}})_j$ are the gradients of $p'^{,n+1}$ and ${\rho}_i^{n+\frac{2}{3}}$ at faces $j$, respectively. %The same scheme is used for $(\nabla p'^{,n}_i)_j$ if a predictor step is used. 
%$(\nabla p'^{,n+1}_i)_j$ is approximated in the same way as $(\nabla\bu^{n+1})_j$ and $(\nabla h^{n+1})_j$. %In
%this work, we choose a partitioned approach to deal with the pressure-velocity coupling. 
In OpenFOAM, there are a few partitioned algorithms that decouple the computation of the pressure from the computation of the velocity, namely SIMPLE \cite{SIMPLE} for steady-state problems, PISO  \cite{PISO}  and PIMPLE \cite{PIMPLE} for time-dependent problems. In this work, we use the PISO algorithm. % without predictor step. %The splitting of operations in the solution of the discretised
%momentum and pressure equations gives rise to a formal order of accuracy of the order of
%powers of ∆t depending on the number of operation-splittings used (see [23] for more details).
%We remark that in the OpenFOAM R© implementation of the PISO solver the mass flux is
%modified through an additional term than can cause artificial dissipation and is not a part
%of the original PISO algorithm in [23]. See [50] for more details

Now that we have computed the pressure fluctuation and the end-of-step velocity, we can get the pressure and the end-of-step kinetic energy density using the discretized forms of \eqref{eq:step3_2} and \eqref{eq:step3_5}:
\begin{align}
p^{n+1}_i &= p'^{,n+1}_i + {\rho}^{n+\frac{2}{3}}_i g z_i, \cl
K^{n+1}_i &= \frac{|\bu_i^{n+1}|^2}{2}. \el
\end{align}
%Next, we compute the end-of-step velocity.
%The space discretization of eq.~\eqref{eq:step4_2} is analogous to that of eq.~\eqref{eq:step2}, i.e.:
%\begin{align}
%\frac{3}{2\Delta t} {\rho}^{n+1}_i {\bu}^{n+1}_i &+\sum_j^{} \varphi^n_j {\bu}^{n+1}_{i,j} + \sum_j^{} p^{n+1}_{i,j} \textbf{A}_j - \sum_j^{} \mu_a ((\nabla{\bu}^{n+1}_i)_j+ (\tilde{\bu}^{n+1}_i)^T_j) \cdot \textbf{A}_j   = {\bm b}^{n+1}_{\bu, i}. \label{eq:step4_2sd}
%\end{align}
Finally, we compute the end-of-step density with the space-discrete version of eq.~\eqref{eq:step3_3}:
\begin{align}
\rho^{n+1}_i = 
\frac{p_i^{n+1}}{R T^{n+1}_i}.\label{eq:step4_4sd}
\end{align}

%\textcolor{blue}{A third-order accurate scheme for the divergence term and a second-order accurate scheme for the Laplacian and gradient terms are used.}
%\anna{Io direi tutto per la priam eq. e poi diciamo che si fa cosi' nel resto delle eq.}

\section{Numerical results}\label{sec:num_res}

We test our OpenFOAM-based solver for the Euler equation with two standard benchmarks that have been widely used to assess atmospheric dynamical cores, i.e., 
the rising thermal bubble and the density current.
Both test cases involve a perturbation of 
a neutrally stratified atmosphere with uniform background potential temperature over a flat terrain. So, before reporting the results for the two benchmarks, in Sec. \ref{sec:HB} we show that an unperturbed stratified atmosphere with uniform background potential temperature over a flat terrain remains unchanged up to a certain tolerance. 
In Sec.~\ref{2dRTBtest} we present our results for the rising thermal bubble benchmark. 
There exist several variations of this benchmark, featuring different geometries and/or initial conditions. We use the settings from \cite{ahmadLindeman2007}. See also \cite{Feng2021} for a recent work using this variation. Our results for the classical density current test  \cite{carpenterDroegemeier1990,strakaWilhelmson1993}  are shown in Sec.~\ref{2dDC}. We would like to point out that neither the rising bubble nor the density current benchmark has an exact solution. Hence, one can only have a relative comparison with other numerical data available in the literature.

\subsection{Hydrostatic atmosphere}\label{sec:HB}
%We conclude by testing our methodology on the two-dimensional equilibrium test.
The goal of this first test case is to verify that an initial resting atmosphere over a flat terrain remains still within a reasonable accuracy for a long time interval.

The computational domain in the $xz$-plane is $\Omega = [0, 16000] \times [0, 800]$ m$^2$. 
In this domain, the hydrostatic atmosphere, initially at rest, is free to evolve until $t$ =  25 days \cite{bottaKlein2004}.
Impenetrable, free-slip boundary conditions are imposed at all the boundaries. We consider a uniform mesh with mesh sizes $h = \Delta x = \Delta z = 250$~m and we set the time step to $\Delta t = 0.1$ s. % \textcolor{red}{o facciamo 0.2? Seguendo esattamente il Botta? o ancora, facciamo dynamically seguendo Marras, che ottiene una media di 0.1?}.
%\subsubsection{Flat terrain}
%The background environment is characterized by the 
%the background pressure is found as:

\begin{figure}[htb]
\centering
 \begin{overpic}[width=0.485\textwidth]{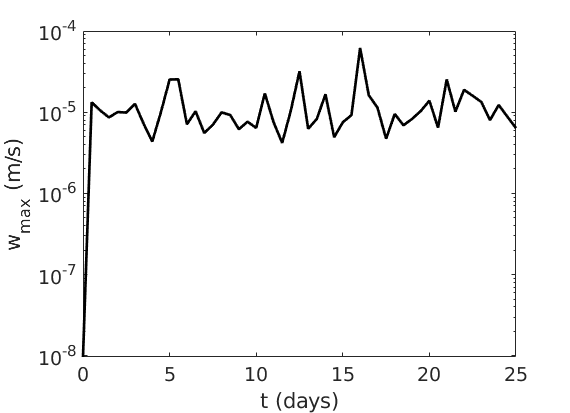}  
        %\put(35,18){FOM}
        %\put(-8,7){$\u$}
      \end{overpic} %\\ \vspace{0.3cm}
       %\begin{overpic}[width=0.8\textwidth]{images/AV75_400m_900s.png}  
      %\end{overpic}
%       \begin{overpic}[width=0.485\textwidth]{images/wmax_HB.png}  
        %\put(35,18){FOM}
        %\put(-8,7){$\u$}
%      \end{overpic} %\\ \vspace{0.3cm}
       %\begin{overpic}[width=0.8\textwidth]{images/AV75_400m_900s.png}  
      %\end{overpic}
\caption{
Hydrostatic atmosphere: time evolution of the maximal vertical velocity $w_{max}$.}
\label{fig:HB}
\end{figure}

Figure \ref{fig:HB} shows the time evolution of the maximal vertical velocity $w_{max}$. We observe that $w_{max}$ has an initial, rather fast, growth but subsequently stabilizes around $1e-5$ m/s. Thus, we conclude that the hydrostatic equilibrium is preserved with reasonably good accuracy, especially in view of the two benchmarks we are going to study next. % and 108 m s1 for single and double precision, respectively%the space-averaged vertical velocity $w_{av}$ \textcolor{red}{Botta fa vedere solo wmax, Marras solo uno snapshot a 56000s della distribuzione della w, ma cmq secondo me va bene introdurre anche wav}, defined as follows
%\begin{equation}\label{eq:w_avg}
%\w_{av} = \dfrac{1}{\Omega} \int_{\Omega} w d\Omega,
%\end{equation}%\label{eq:w_av}
%the maximum vertical velocity $w_{max}$. %The time interval is sampled every 10 s. 
%\textcolor{red}{From Figure \ref{fig:HB} (right), we observe that after an initial transient of about $10000$ s $w_{max}$ stabilizes and does not exceed $1.4e-5$. Figure \ref{fig:HB} (left) shows that $w_{av}$, instead, keeps decreasing and reaches a value of $3.1e-7$ at the end of the time interval under consideration.
%Thus, we conclude that the hydrostatic equilibrium is preserved with a good accuracy.}

%\textcolor{red}{Non sono riuscito a trovare nessuna validazione in letteratura su flat terrain per confronto, ma solo sulla montagna. Ricordiamoci cmq che in questo caso il botta sulla montagna trova dei valori piu' bassi, vicino alla machine precisione (come SM...). Io credo che i valori che otteniamo noi siano ottimi, anche considerando che nel nsotro caso la discretizzazione del termine idrostatico non è forse quella ottimale; guardiamo poi la media che diventa progressivamente più piccola (infatti lo farei notare nei commenti).}

\subsection{Rising thermal bubble in a neutrally stratified atmosphere}
\label{2dRTBtest}

%\subsection{Rising thermal bubble}

%Following \cite{marrasNazarovGiraldo2015}, I'd add $\Delta {\bf x} = 31.25$~m and $\Delta {\bf x} = 15.625$~m to the meshes listed in the table. In \cite{marrasNazarovGiraldo2015}, they report the ``effective resolution'', i.e. mesh size/order of the elements. Please consider the $\Delta {\bf x}$ here as effective resolution. 
%First, I would set the artificial kinematic viscosity to $\nu = 15$ m$^2$/s as in \cite{ahmadLindeman2007} (this amounts to the NSE model) and
%\begin{itemize}
%    \item[-] plot the contour lines of $\Delta \theta$ at $t = 1020$ s. We expect to see that as the mesh is refined the Rayleigh-Taylor instability gets captured.
%    \item[-] extract the maximum $\Delta \theta$ at $t = 1020$, which is expected to be around 1.4 K, as is found in \cite{ahmadLindeman2007} for $\Delta {\bf x} = 125$~m. Actually, let's plot the maximum $\Delta \theta$ over time and see how we compare to Fig.~6 in \cite{ahmadLindeman2007}. 
%\end{itemize}
%If this yields the expected results, then we can think about what to do with the LES models. 
%This last benchmark has been consists of a flow that is triggered by the thermal perturbation of a neutrally stratified
%atmosphere at initially uniform potential temperature θ0 = 300 K and in hydrostatic equilibrium
%such that the pressure decreases with z as:

The computational domain in the $xz$-plane is $\Omega=[0, 5000]\times[0, 10000]$ m$^2$ and the time interval of interest is $(0, 1020]$ s. Impenetrable, free-slip boundary conditions are imposed on all walls.
The initial density is given by \begin{align}
\rho^0 = \frac{p_g}{R \theta_0} \left(\frac{p}{p_g}\right)^{c_{v}/c_p}, \quad p = p_g \left( 1 - \frac{g z}{c_p \theta^0} \right)^{c_p/R}, \label{eq:rho_wb}
\end{align}
with %$p_g = 10^5$ Pa, 
$c_p = R + c_v$, $c_v = 715.5$ J/(Kg K), $R = 287$ J/(Kg K).
In \eqref{eq:rho_wb}, $\theta^0$ is the initial potential temperature, which is defined as:
\begin{equation}
\theta^0 = 300 + 2\left[ 1 - \frac{r}{r_0} \right] ~ \textrm{if $r\leq r_0=2000~\mathrm{m}$}, \quad\theta^0 = 300
~ \textrm{otherwise},
\label{warmEqn1}
\end{equation}
where $r = \sqrt[]{(x-x_{c})^{2} + (z-z_{c})^{2}}$, $(x_c,z_c) = (5000,2000)~\mathrm{m}$ \cite{ahmadLindeman2007,ahmad2018}.
Notice that \eqref{eq:rho_wb}-\eqref{warmEqn1} represents a neutrally stratified atmosphere with uniform background potential temperature of $ 300~\mathrm{K}$ perturbed by a circular bubble of warmer air. 
The initial velocity field is zero everywhere. 
Finally, the initial specific enthalpy is given by:
\begin{align}
h^{0} = c_p \theta^0 \left( \frac{p}{p_g} \right)^{\frac{R}{c_{p}}}.
\label{eq:e0}
\end{align}

% The pressure is:
% \begin{align}
% p = p_g \left( 1 - \frac{g z}{c_p \theta} \right)^{c_p/R}, \label{eq:p_wb}
% \end{align} 
% % while the mean hydrostatic pressure is given by:
% % \begin{align}
% % p_0 = p_g \left( 1 - \frac{g z}{c_p \theta_0} \right)^{c_p/R},  \label{eq:p0_wb}
% % \end{align} 
% with %$p_g = 10^5$ Pa, 
% $c_p = R + c_v$, $c_v = 715.5$ J/(Kg K), $R = 287$ J/(Kg K).

% \begin{rem}
% An important quantity for the benchmarks in Sec.~\ref{sec:num_res} is the potential temperature:
% \begin{align}
% \theta = \frac{T}{\pi}, \quad \pi = \left( \frac{p}{p_g} \right)^{\frac{R}{c_{p}}}, \label{eq:theta}
% \end{align}
% i.e.~the temperature that a parcel of dry air would have if it were expanded or compressed
% adiabatically to standard pressure $p_g = 10^5$ Pa, which is the atmospheric pressure at the ground. 
% Additionally, we split $theta$ into a
% background potential temperature $\theta_0$ and fluctuation $\theta'$:
% \begin{align}
% \theta = \theta_0 + \theta'. \label{eq:theta_wb}
% \end{align} 
% \end{rem}

For this test, we use five different meshes with uniform resolution $h = \Delta x = \Delta z = 250, 125, 62.5$, $31.25, 15.625$~m. The time step is set to $\Delta t = 0.1$ s for all the simulations. 
%We adopt the same pipeline used for the density current benchmark. 
For stabilization, we consider two strategies. 
First, following \cite{ahmadLindeman2007},
we set $\mu_a = 15$ and $Pr = 1$.
Note that while $Pr = 1$ is close to a physically meaningful value as the air Prandtl number is about 0.71 at 20$^\circ$C, $\mu_a = 15$ is an ah-hoc value.
Hereinafter, we refer to this model as AV15, where AV stands for artificial viscosity.
Then, we consider the Smagorinsky model as described in Sec.~\ref{sec:LES}.

\begin{figure}[htb]
\centering
 \begin{overpic}[width=0.3\textwidth, grid=false]{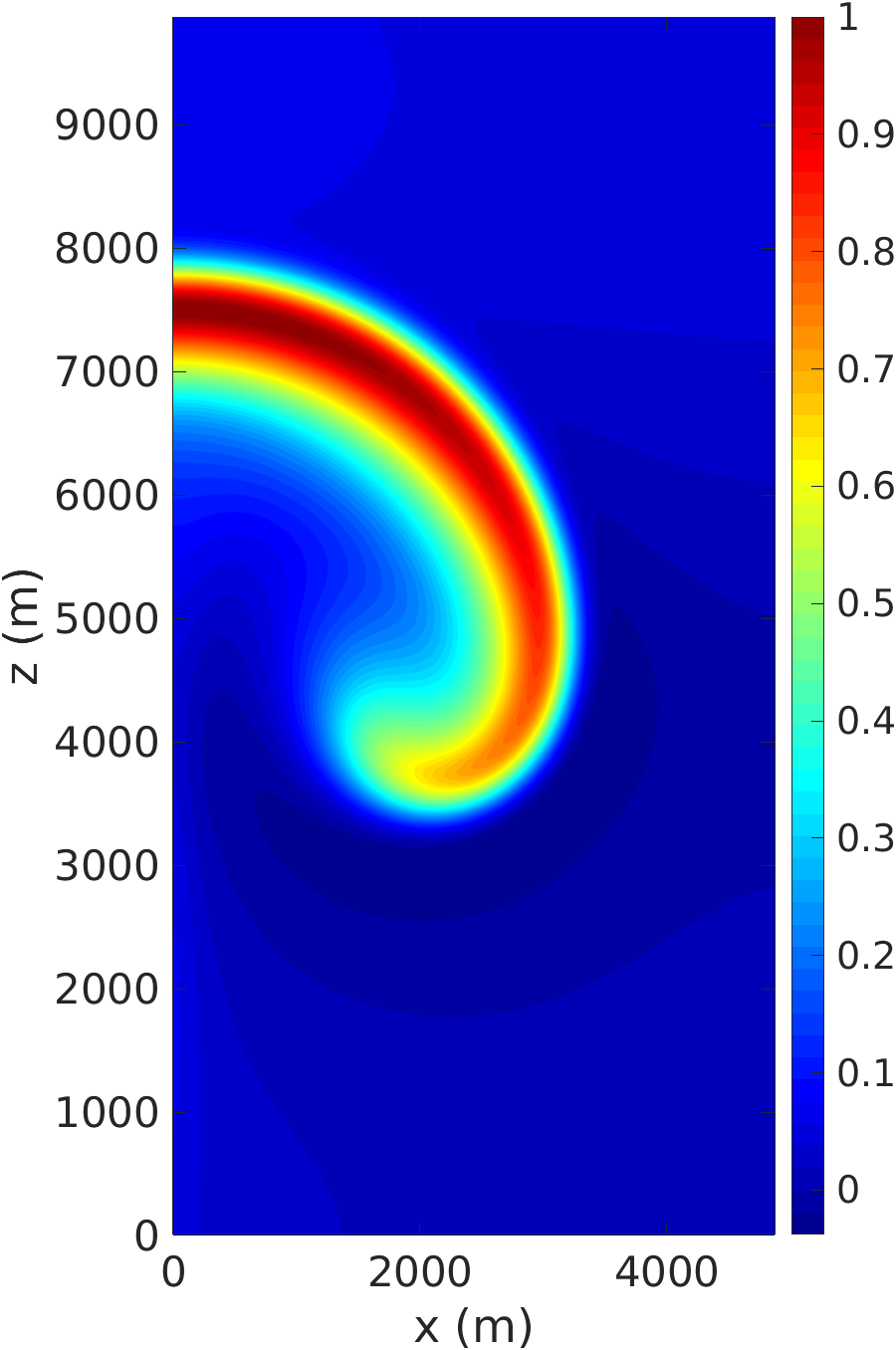}  
        \put(20,90){\textcolor{white}{$h = 15.625$ m}}
      \end{overpic}~ \hspace{0.5cm}
\begin{overpic}[width=0.3\textwidth, grid=false]{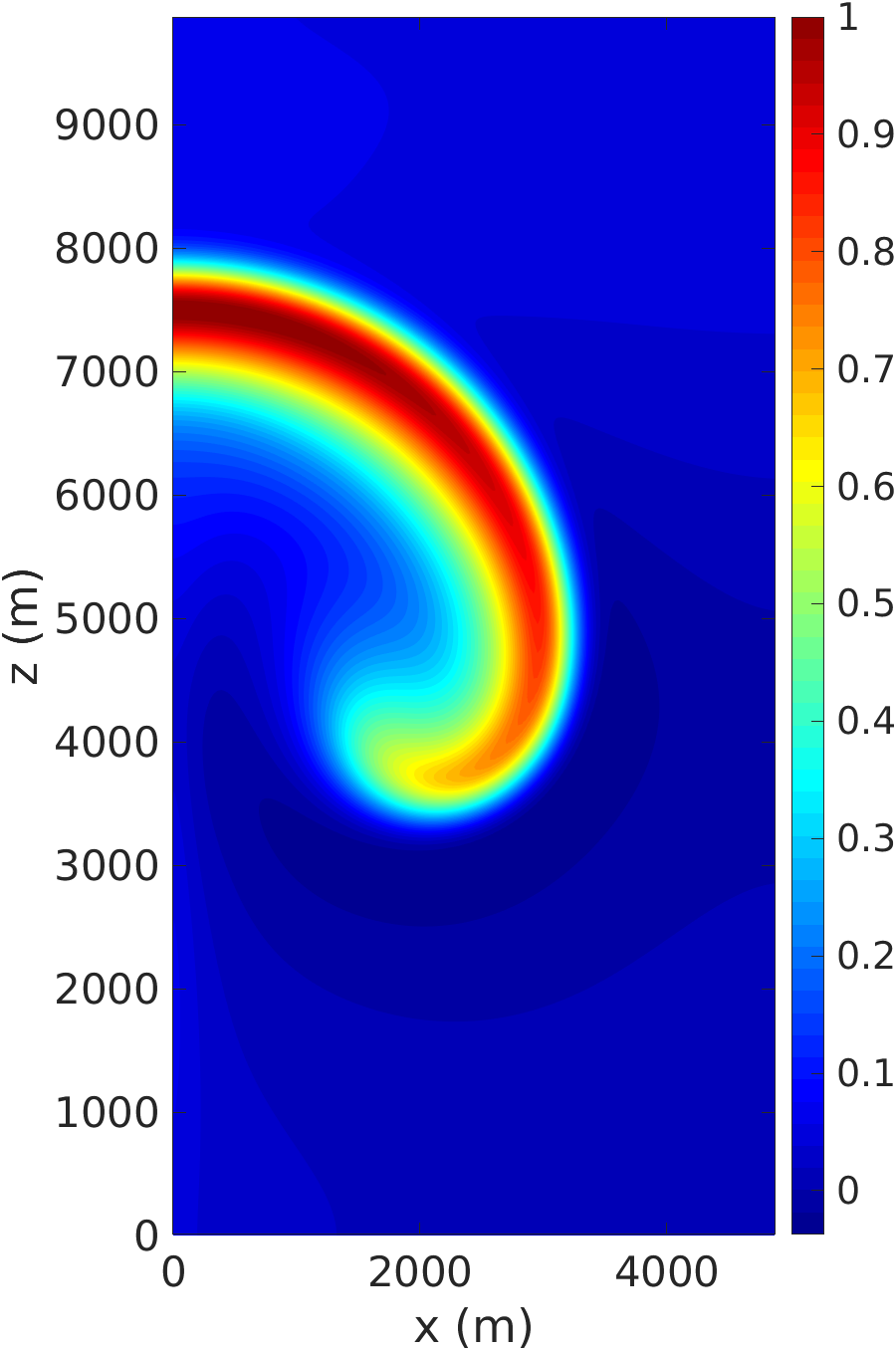} 
        \put(22,90){\textcolor{white}{$h = 32.25$ m}}
      \end{overpic}~ \hspace{0.5cm} 
\begin{overpic}[width=0.3\textwidth, grid=false]{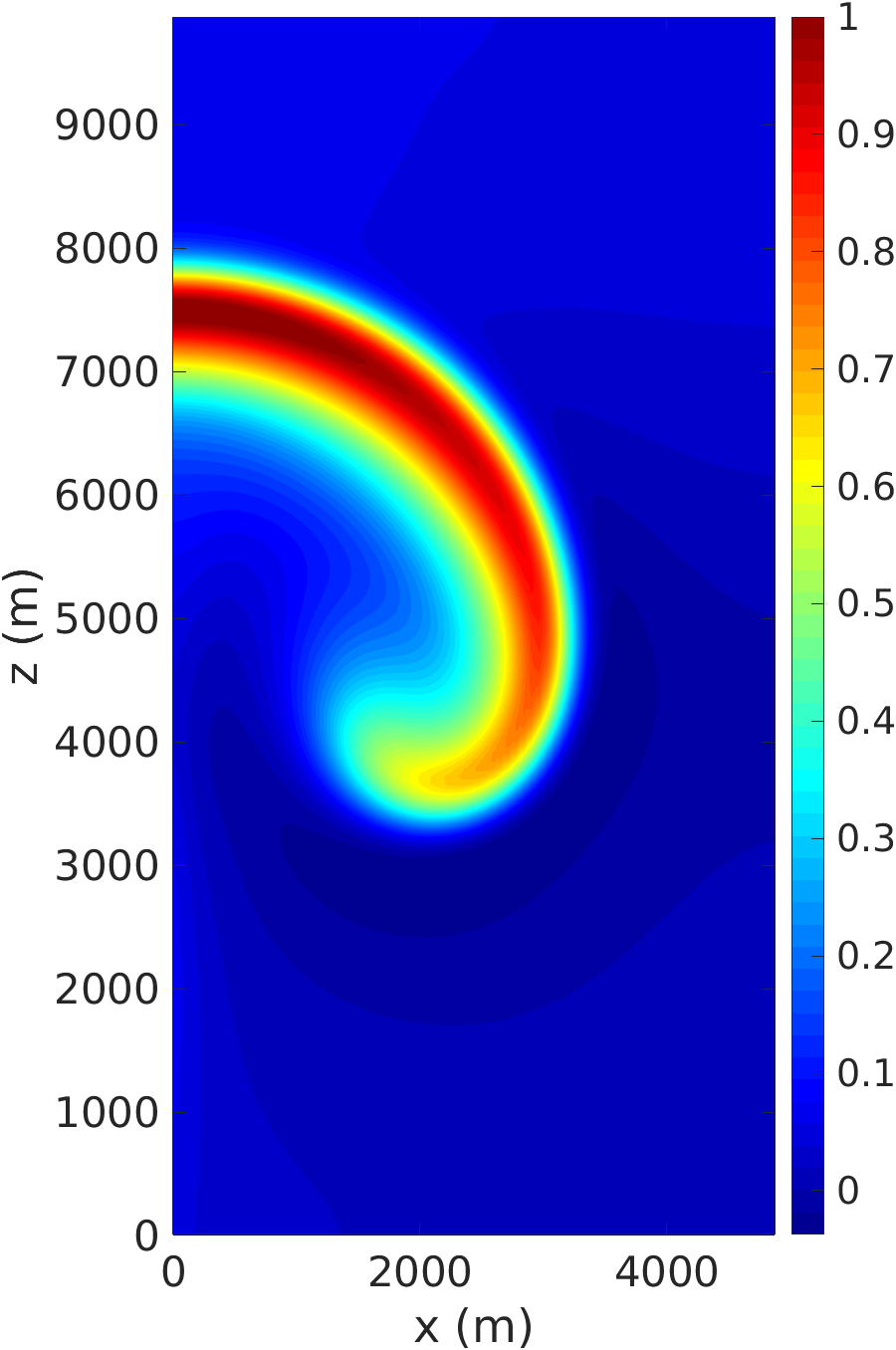} 
\put(23,90){\textcolor{white}{$h = 62.5$ m}}
      \end{overpic}\\ \vspace{0.3cm}
\begin{overpic}[width=0.3\textwidth, grid=false]{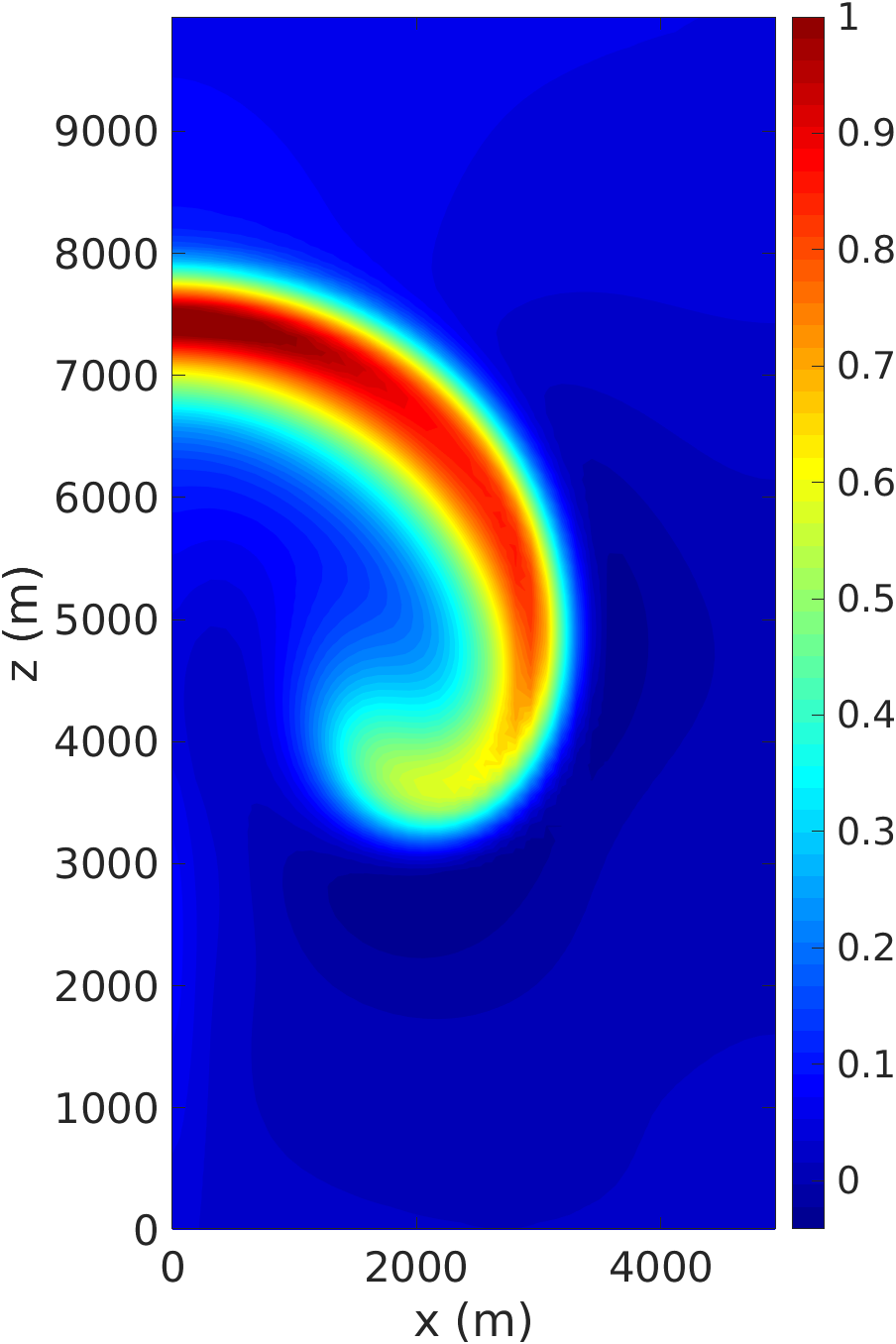} 
\put(23,90){\textcolor{white}{$h = 125$ m}}
      \end{overpic} ~ \hspace{0.5cm}
      \begin{overpic}[width=0.3\textwidth, grid=false]{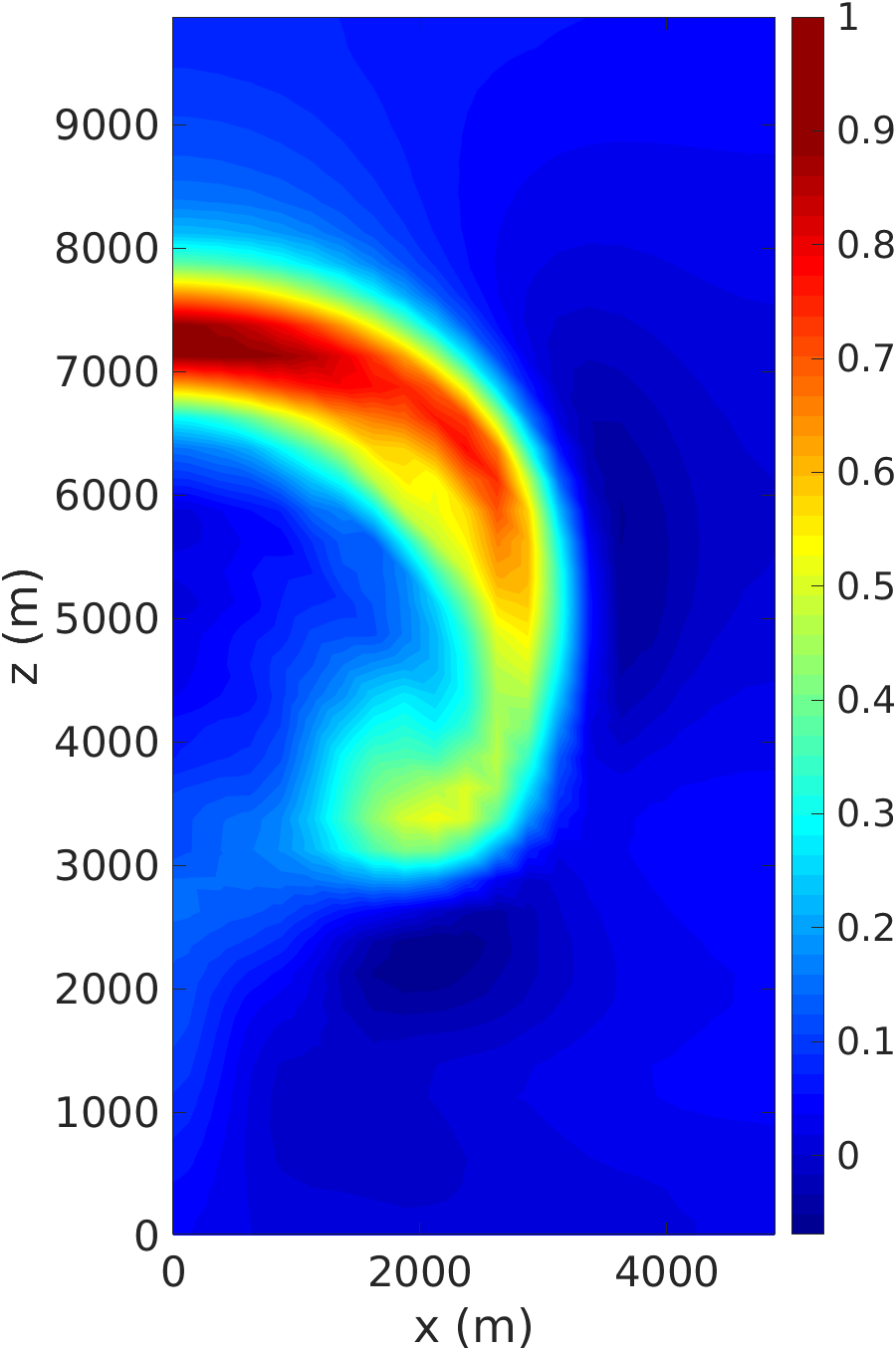} 
\put(23,90){\textcolor{white}{$h = 250$ m}}
      \end{overpic}
    \caption{Rising thermal bubble, AV15 model: perturbation of potential temperature computed with mesh $h = 15.625$ m (top-left), $h = 32.25$ m (top-center), $h = 62.5$ m (top-right), $h = 125$ m (bottom-left), $h = 250$ m (bottom-right)}
\label{fig:RTB1}
\end{figure}

Figure \ref{fig:RTB1} reports the perturbation of potential temperature $\theta'$ at $t = 1020$ s computed by the AV15 model with all the meshes under consideration. By $t = 1020$,  
the air warmer than the ambient has risen due to buoyancy and deformed due to shearing motion. As a result, the bubble has evolved into a mushroom shape. From 
Figure \ref{fig:RTB1}, 
%The different resolutions were used to analyze the behavior of the method as the grid is refined %although no proper convergence study is made.
we observe that the no substantial change
in the computed $\theta'$ when the mesh 
is refined past $h = 125$ m. We remark that, to facilitate the comparison of the panels in Figure \ref{fig:RTB1}, we have forced the colorbar to range from 0 to 1.
Qualitatively, these results are in very good agreement
with those reported in the literature. See, e.g., \cite{ahmadLindeman2007,ahmad2018,marrasNazarovGiraldo2015}.
%\anna{Michele, per favore aggiungi tu altre referenze in cui la bolla ha le stesse condizioni nostre (non vorrei citare una delle mille variazioni).}

\begin{figure}[htb]
\centering
 \begin{overpic}[width=0.22\textwidth,grid=false]{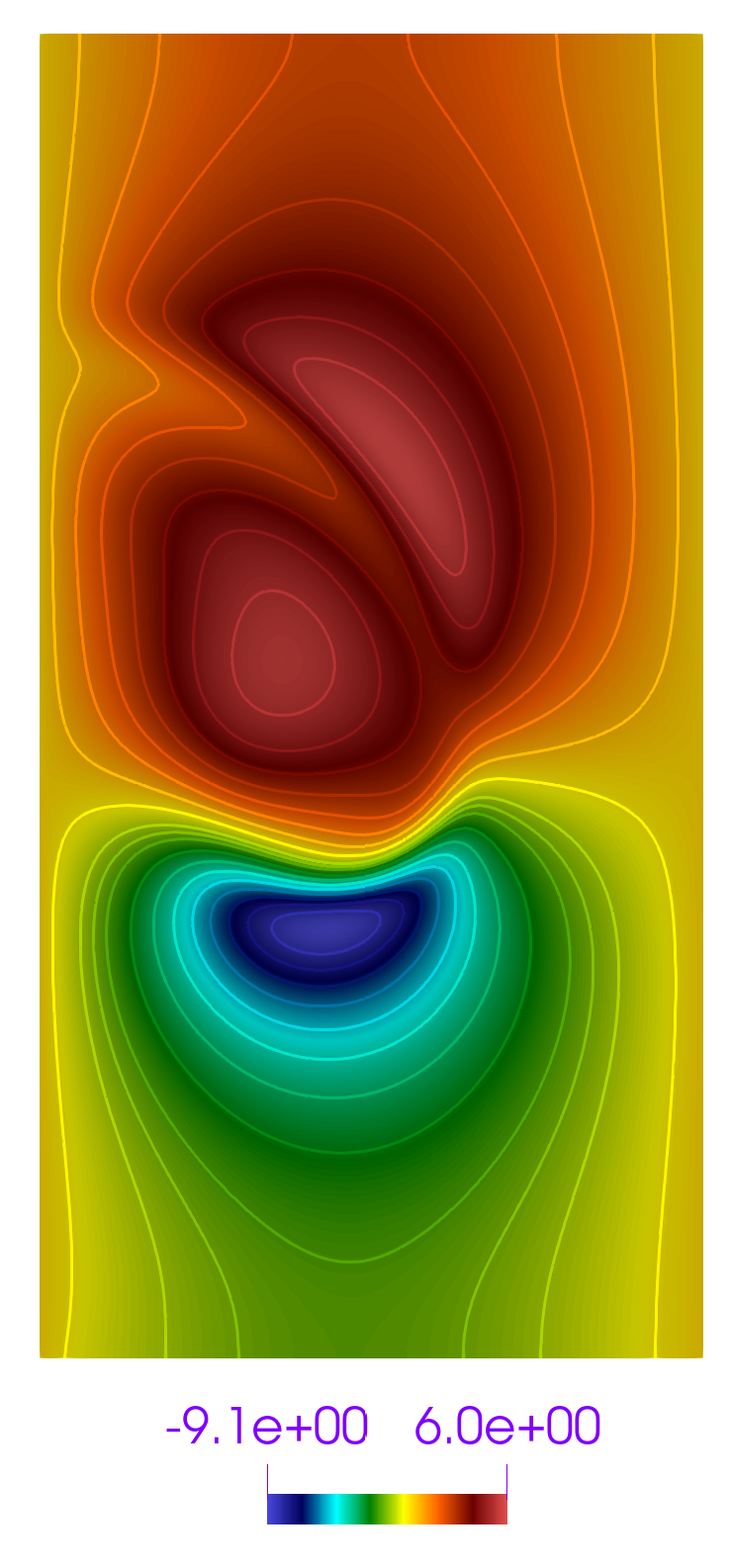}  
        \put(17,100){$u$ (m/s)}
      \end{overpic}~
       \begin{overpic}[width=0.22\textwidth]{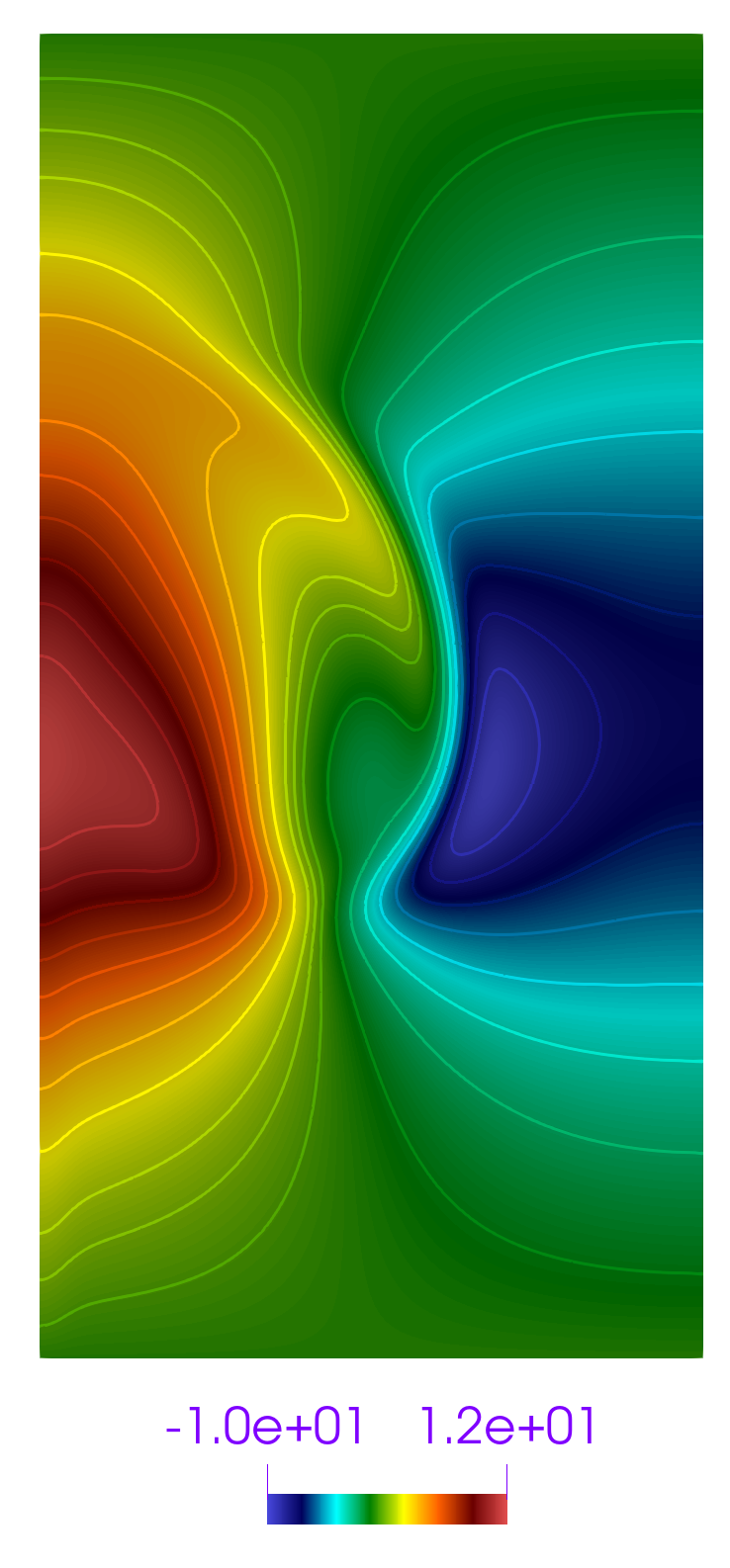}  
        \put(15,100){$w$ (m/s)}
      \end{overpic}
\caption{Rising thermal bubble, AV15 model: contour plots of the horizontal velocity component $u$ (left) and the vertical velocity component $w$ (right) at $t = 1020$ s computed with mesh $h = $ 125 m.}
\label{fig:RTB3}
\end{figure}

For further qualitative assessment,
Figure \ref{fig:RTB3} displays velocity components $u$ and $w$ at $t = 1020$ s computed by the AV15 model with mesh $h = $ 125 m. These contour plots are in very good agreement with those reported in the literature. See, e.g., Figure 7 in \cite{ahmadLindeman2007}.

\begin{figure}[htb]
\centering
 \begin{overpic}[width=0.485\textwidth]{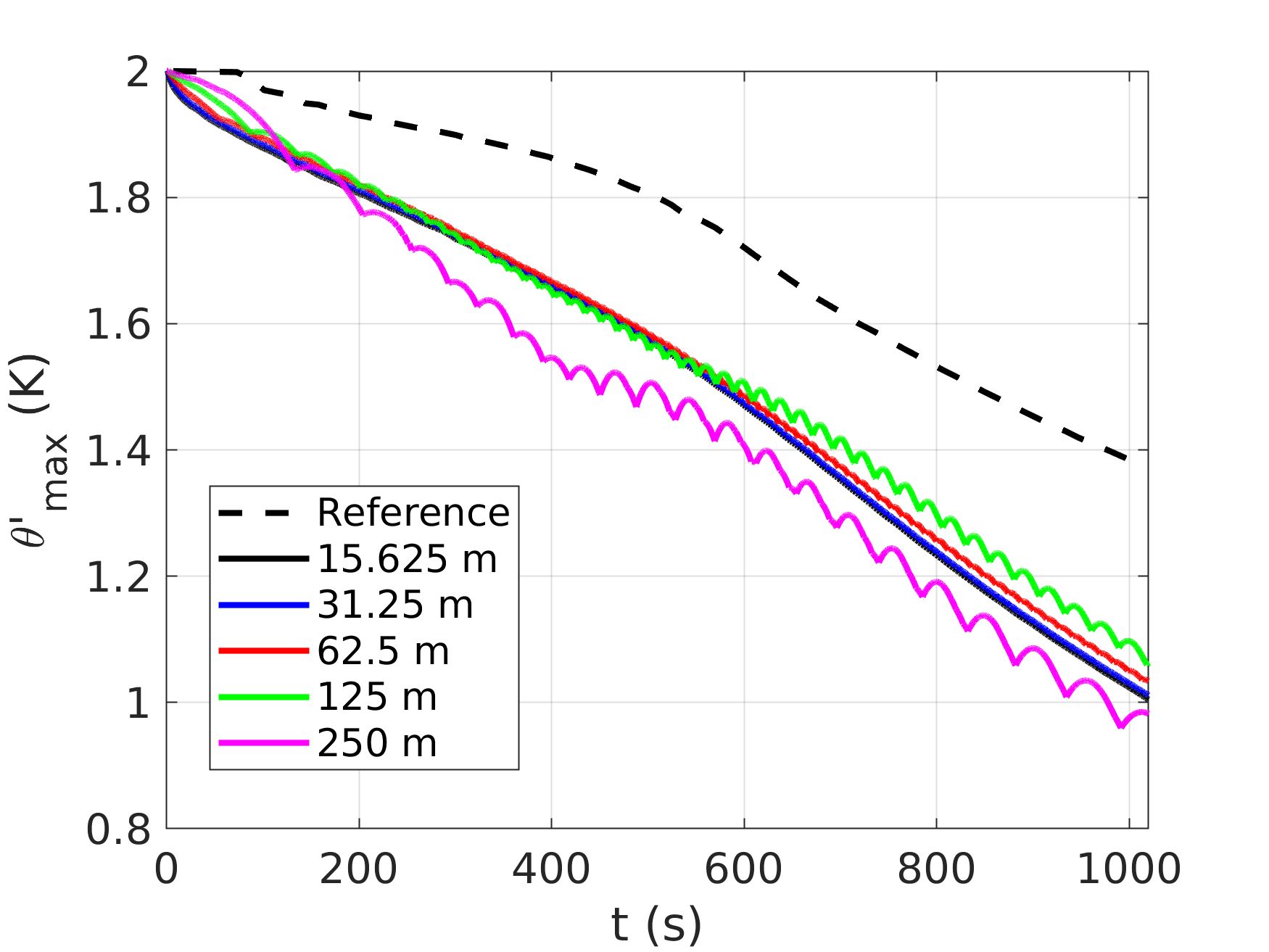}  
        %\put(35,18){FOM}
        %\put(-8,7){$\u$}
      \end{overpic}~
       \begin{overpic}[width=0.485\textwidth]{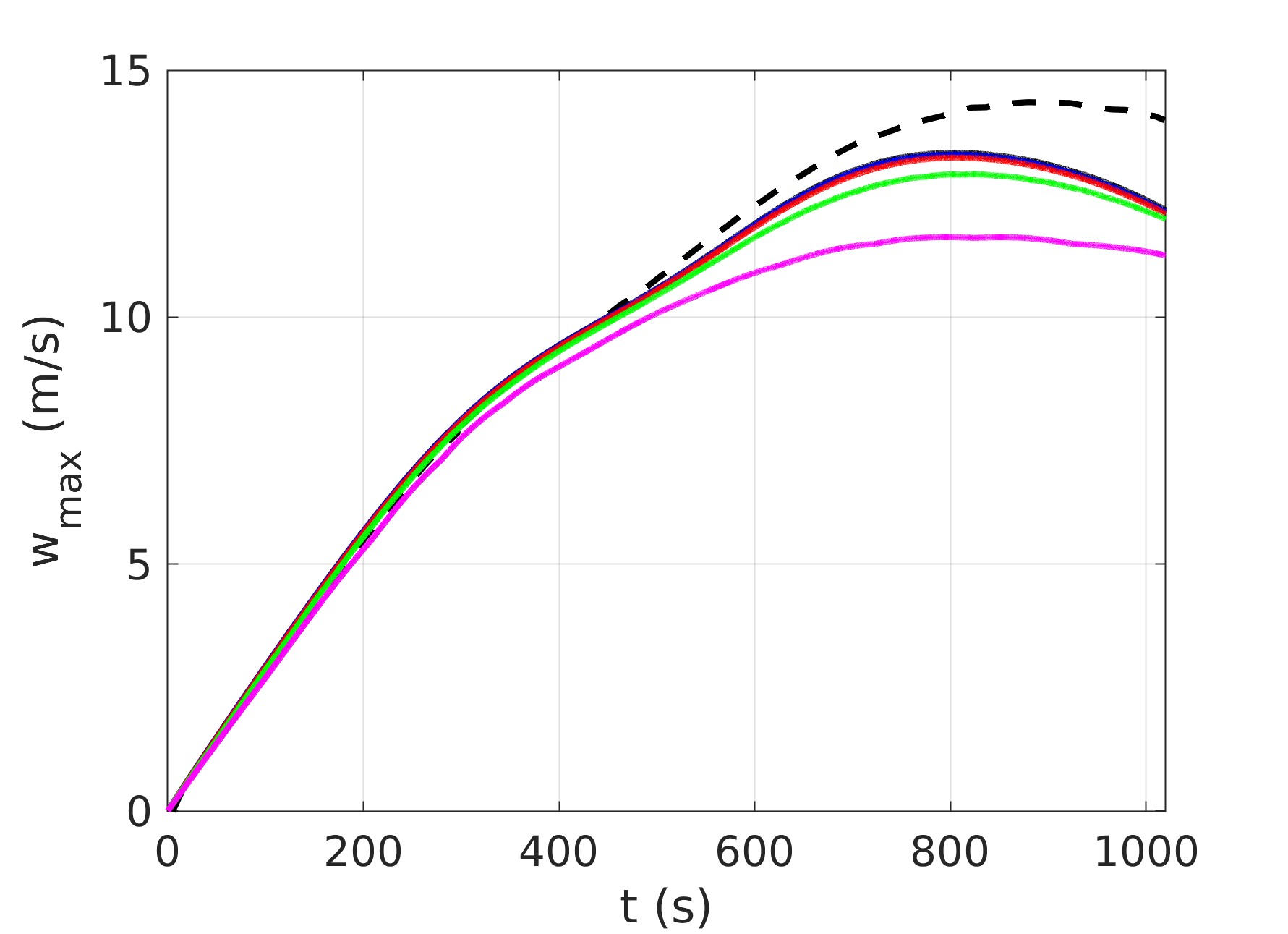}  
        %\put(35,18){FOM}
        %\put(-8,7){$\u$}
      \end{overpic}
\caption{Rising thermal bubble, AV15 model: time evolution of the maximum perturbation of potential temperature $\theta'_{max}$ (left) and the maximum vertical component of the velocity $w_{max}$ (right) computed with all the meshes under consideration. The reference values are taken from \cite{ahmadLindeman2007} and refer to resolution 125 m.}
\label{fig:RTB2}
\end{figure}

For a more quantitative comparison, in Figure \ref{fig:RTB2} we show
the time evolution of the maximum perturbation of potential temperature $\theta'_{max}$ and maximum vertical component of the velocity $w_{max}$ computed 
by the AV15 model with all the meshes, together with 
the corresponding results from \cite{ahmadLindeman2007}. We observe that the evolution of $\theta'_{max}$ computed with meshes $h = 250$ m and $h = 125$ m is affected by spurious oscillations. 
Such oscillation disappear at higher resolution and they are not present at all in the evolution of $w_{max}$. We see that $\theta'_{max}$ and $w_{max}$ computed with meshes $h = 31.25$ m and $h = 15.625$ m are practically overlapped for the entire time interval, indicating that we are close to convergence. The ``converged'' $w_{max}$ overlaps with the reference value till about $t = 500$ s and it remains close to it till about $t = 800$ s. The agreement of the ``converged'' $\theta'_{max}$ with the results from \cite{ahmadLindeman2007} is not as good.
These trends are confirmed from
Table \ref{tab:2}, which reports the extrema for the vertical velocity $w$ and potential temperature perturbation $\theta'$ at $t = 1020$ s obtained with the AV15 model, together with the values extracted from the figures in \cite{ahmadLindeman2007}. 
The results from \cite{ahmadLindeman2007} are obtained with mesh resolution of 125 m and a density-based approach developed from a Godunov-type scheme that employs flux-based wave decompositions for the solution of Riemann problem. The authors of \cite{ahmadLindeman2007} start from the Euler equations written in density, velocity, and potential temperature. Other differences with our methodology include the orders of space and time discretizations and the different treatment of the hydrostatic term. Given all these differences, we believe that our results cannot be considered off with respect to the reference. In addition, other references, such as \cite{marrasNazarovGiraldo2015} (see Fig.~1, top left panel), report a $\theta'_{max}$ at $t = 1020$ s computed with resolution 125 m  closer to  1K (like in our case) than to 1.4 K (like in \cite{ahmadLindeman2007}).

%However we see that there is a significant difference between our converged solution and the reference one.
%\textcolor{red}{io attribuirei la differenza all'ordine degli schemi spaziale (anche se S nel suo paper dice che Lindemann usa FV 2 ordine?!? Non credo ci sia da fidarsi naturalmente...), allo schema in tempo (non ricordo cosa Lind usi ma di certo non BDF1) nonchè al fatto che Lindemann usa un Godunov like scheme associato ad un density based solver e un tipo di discretizzazione diversa del termine idrostatico (questa cosa forse riusciamo a mostrarla nell'altro paper in corso, infatti nelle prove che ho fatto ottengo 1.2 K che è piu' vicino al valore di Lind ;)). Altra speculazione potrebbe essere il fatto che noi risolviamo rispetto all'energia e theta è una variabile calcolata, non risolta direttamente (che si contrappone a quella che succede con $w_{max}$ dove abbiamo un comportamento meno critico appunto).} On the other hand, in Figure \ref{fig:RTB2} (right) we depict the time evolution of the $w_{max}$ by the AV15 model with all the meshes mentioned above. Unlike what observed for $\theta'_{max}$, the solution is not affected by spurious oscillations for the coarser meshes and the convergence is closer to the reference.

\begin{table}[htb]
\begin{tabular}{|c|c|c|c|c|c|} \hline
Model & Resolution [m] & $w_{min}$ (m/s) & $w_{max}$ (m/s) & $\theta'_{min}$ (K) & $\theta'_{max}$ (K) \\
 \hline
 AV15 & 15.625 & -10.02 & 12.17 & -0.038 & 1\\
 AV15 & 31.25  & -10.05 & 12.13 & -0.037 & 1.01\\
 AV15 & 62.5  & -10.08 & 12.09 & -0.037 & 1.03\\
 AV15 & 125 & -10.02 & 11.97 & -0.042 & 1.06\\
 AV15 & 250 & -9.59 & 11.23 & -0.069 & 0.89\\
  Ref.~\cite{ahmadLindeman2007} & 125 & -7.75 & 13.95 & -0.01 & 1.4 \\
%Smagorisnky (LES) & 15.625 & -12.24 & 15.42 & -0.18 & 2.04\\
%Smagorisnky (LES) & 31.25 & -11.54 & 15.04 & -0.072 & 1.89\\
 \hline  
\end{tabular}
\caption{Rising thermal bubble, AV15 model: minimum and maximum vertical velocity $w$ and potential temperature perturbation $\theta'$ at $t = 1020$ s 
compared with the values extracted from the figures in \cite{ahmadLindeman2007}.
%\textcolor{red}{CITARE. Con chi ci confrontiamo? Lindemann direi e magari la soluzione WRF?. Giraldo e Restell, in teoria, non sono utilizzabili perchè hanno una IC diversa dalla nostra ma guarda la tabella 1 di: \url{https://congress.cimne.com/iacm-eccomas2014/admin/files/filePaper/p1448.pdf}. Qui secondo me fanno un casino...usano anche loro la IC alla Giraldo e Restelli ma in tabella mettono pure Lindemann...che calderone!}.
%For reference \cite{strakaWilhelmson1993}, we reported the range of mesh sizes and front location values obtained with different methods.
%For reference \cite{marrasNazarovGiraldo2015}, we report only the front location computed with the finest resolution.
}\label{tab:2}
\end{table}

\begin{figure}[htb]
\centering
 \begin{overpic}[width=0.335\textwidth]{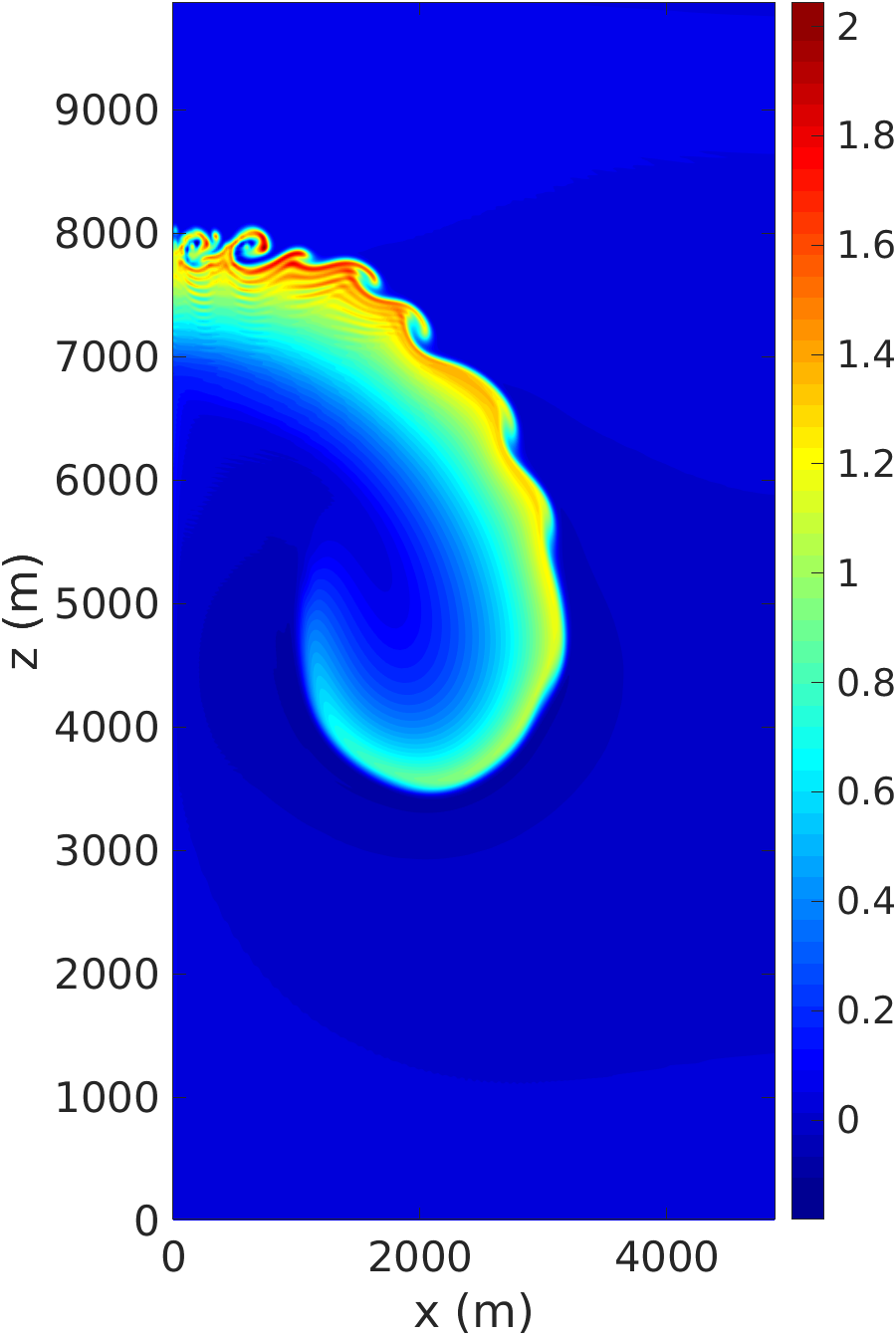}  
\put(22,90){\textcolor{white}{$h = 15.625$ m}}
      \end{overpic}~ \hspace{0.5cm}
       \begin{overpic}[width=0.335\textwidth]{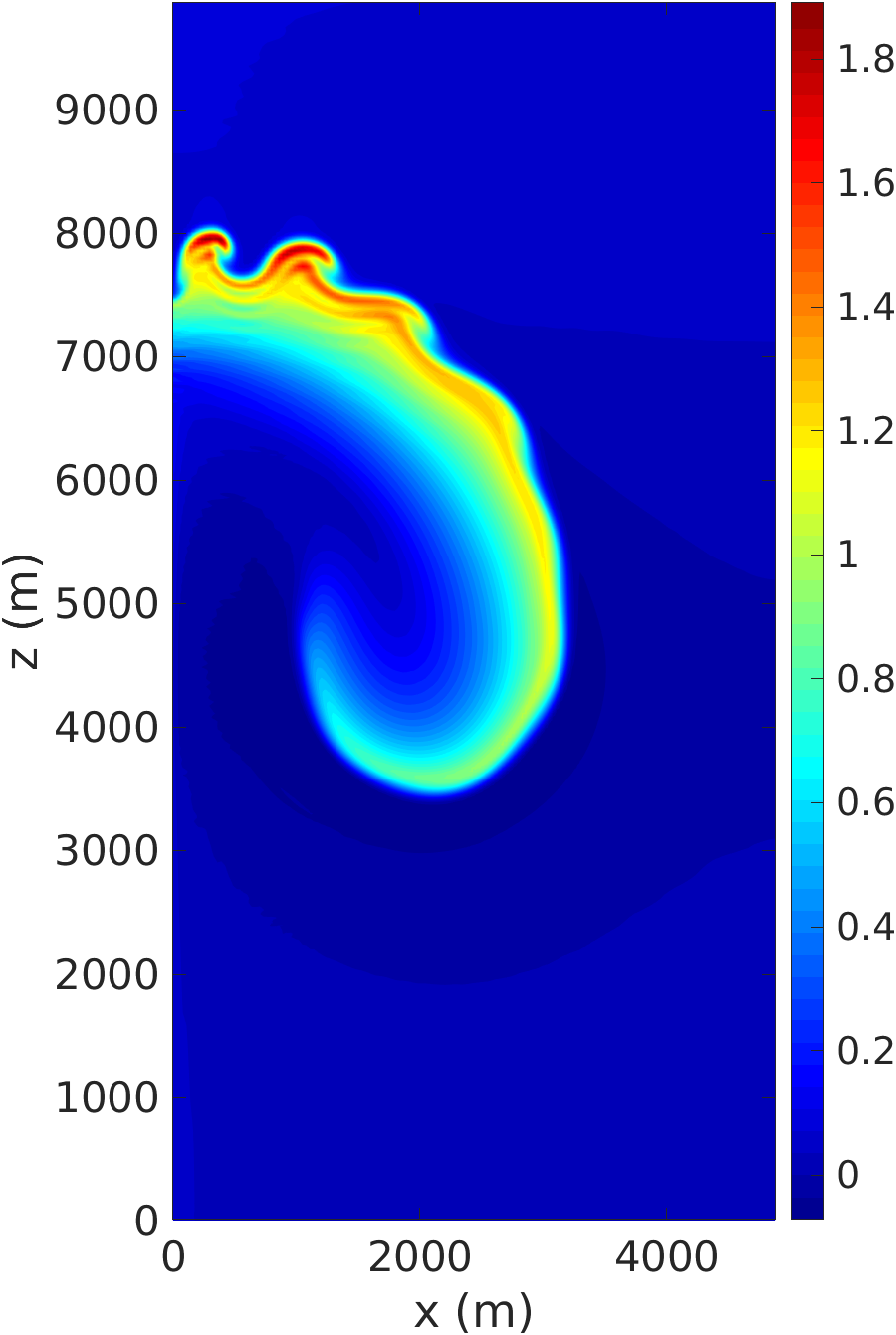}  
\put(23,90){\textcolor{white}{$h = 32.25$ m}}
      \end{overpic}
\caption{Rising thermal bubble, Smagorinsky model: perturbation of potential temperature computed with mesh $h = 15.625$ m (left) and $h = 32.25$ m (right).}
\label{fig:RTB4}
\end{figure}

Finally, we run this test with the Smagorinsky model. We set the parameter $C_s = 0.094$, with $C_k = 0.21$ and $C_\epsilon = 1.048$. Details on the choice of these parameters are reported in Sec.~\ref{2dDC}. 
%Like in the case of the AV15 model, we start from a qualitative comparison. 
Figure \ref{fig:RTB4} depicts the spatial distribution of $\theta'$ at $t = 1020$ s computed with meshes $15.625$ m and $32.25$ m. As expected, with these fine meshes the Smagorinsky model is able to capture a larger amount of vortical structures than the AV15 model. The beautiful vortex strip created by the Rayleigh-Taylor instability at the edge of the bubble shown in Figure \ref{fig:RTB4} (left) matches well with that shown in \cite{marrasNazarovGiraldo2015} (see Fig.~1, bottom right panel). Table \ref{tab:3} reports the extrema for the vertical velocity $w$ and potential temperature perturbation $\theta'$ at $t = 1020$ s obtained with the Smagorinsky model.
Since we could not find any data obtained with a LES model for the exact setting of \cite{ahmadLindeman2007}, Table \ref{tab:3} does contain reference data for comparison.

\begin{table}[htb]
\begin{tabular}{|c|c|c|c|c|c|} \hline
Model & Resolution [m] & $w_{min}$ (m/s) & $w_{max}$ (m/s) & $\theta'_{min}$ (K) & $\theta'_{max}$ (K) \\
 \hline
Smagorisnky (LES) & 15.625 & -12.24 & 15.42 & -0.18 & 2.04\\
Smagorisnky (LES) & 31.25 & -11.54 & 15.04 & -0.072 & 1.89\\
 \hline  
\end{tabular}
\caption{Rising thermal bubble, Smagorinsky model: minimum and maximum vertical velocity $w$ and potential temperature perturbation $\theta'$ at $t = 1020$ s.
}\label{tab:3}
\end{table}

%\begin{center}
%\end{center}

\subsection{Density current}\label{2dDC}

\begin{figure}[htb]
\centering
 \begin{overpic}[width=0.6\textwidth]{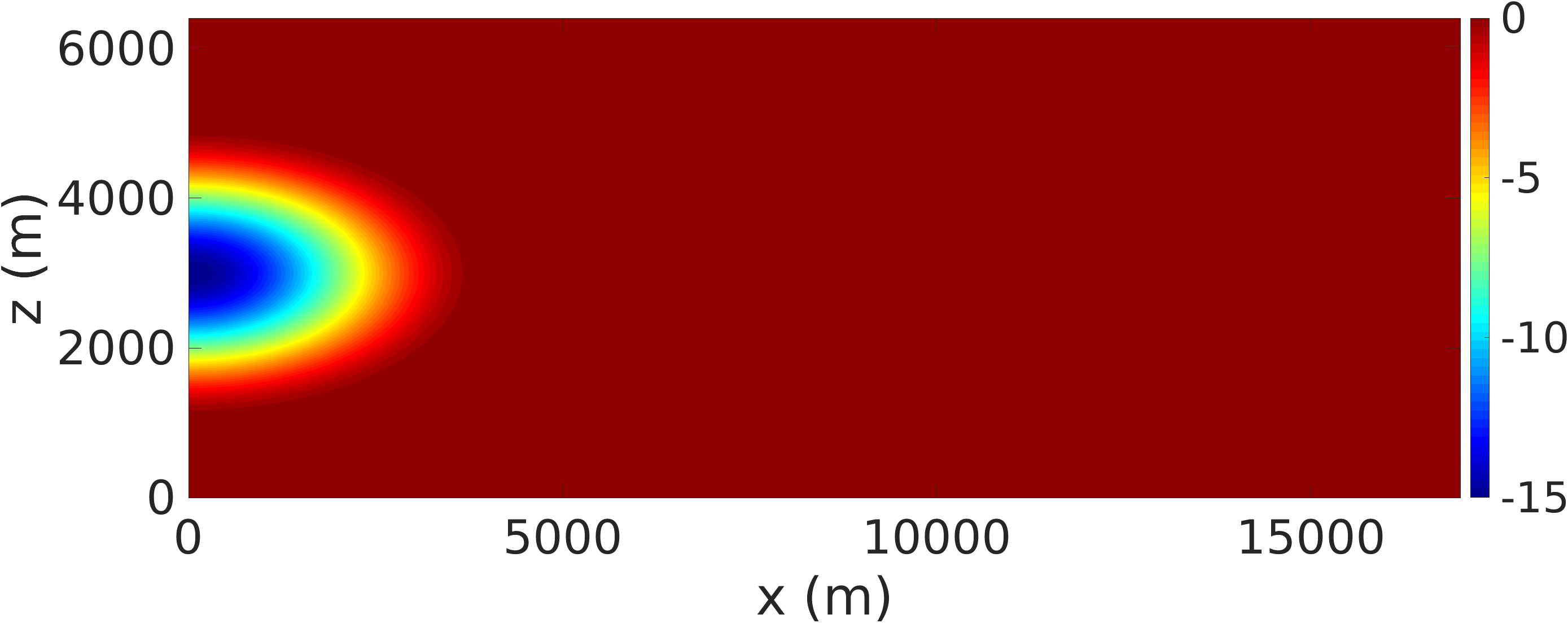}  
        %\put(35,18){FOM}
        %\put(-8,7){$\u$}
      \end{overpic} %\\ \vspace{0.3cm}
       %\begin{overpic}[width=0.8\textwidth]{images/AV75_400m_900s.png}  
      %\end{overpic}
\caption{
Density current: initial potential temperature fluctuation $\theta'$ on part of the computational domain.}
\label{fig:IC_DC}
\end{figure}

The computational domain in the $xz$-plane is $\Omega=[0,25600]\times[0,6400]~\mathrm{m}^2$ and the time interval of interest is $(0,900]$ s. 
Impenetrable, free-slip boundary conditions are imposed on all the walls. 
The initial density is given by \eqref{eq:rho_wb} with initial potential temperature:
\begin{equation}
\theta^0 = 300 - \frac{15}{2}\left[  1 + \cos(\pi r)\right] ~ \textrm{if $r\leq 1$},\quad\theta^0 = 300
~ \textrm{otherwise},
\label{dcEqn1}
\end{equation}
where $r = \sqrt[]{\left(\frac{x-x_{c}}{x_r}\right)^{2} + \left(\frac{z-z_{c}}{z_r}\right)^{2}}$, with $(x_r,z_r)=(4000, 2000)~{\rm m}$ and $(x_c,z_c) = (0,3000)~\mathrm{m}$.
The initial potential temperature fluctuation \eqref{eq:theta_split} on part of the domain $\Omega$ is shown in Fig.~\ref{fig:IC_DC}.
Notice that in this case the initial bubble is cold, while the bubble in \eqref{warmEqn1} is warm. 
The initial velocity field is zero everywhere and the initial specific enthalpy is given by \eqref{eq:e0}.

We first consider uniform meshes with commonly used \cite{strakaWilhelmson1993,ahmadLindeman2007,giraldoRestelli2008a,marrasEtAl2013a,marrasNazarovGiraldo2015} mesh sizes $h = \Delta x = \Delta z = [400, 200, 100, 50, 25]$~m. 
The time step is set to $\Delta t = 0.1$ s. 
Since this benchmark features more complex vortical structures than the rising thermal bubble, we consider three stabilization strategies.
%It is well known that the simulation of this benchmark problem with
%the Euler equations becomes unstable and numerical stabilization is needed. See, e.g., \cite{strakaWilhelmson1993,ahmadLindeman2007,giraldoRestelli2008a,marrasEtAl2013a,marrasNazarovGiraldo2015}.
%We consider two stabilization strategies.
Following \cite{strakaWilhelmson1993,ahmadLindeman2007} we first set $\mu_a = 75$ %m$^2$/s.
in \eqref{eq:mom_LES}-\eqref{eq:ent_LES}
and $Pr = 1$ in \eqref{eq:ent_LES}.
%Note that while $Pr = 1$ is close to a physically meaningful value as the air Prandtl number is about 0.71 at 20$^\circ$C, $\mu_a = 75$ is an ah-hoc value. 
We will refer to this model as AV75, since 75 is an ad-hoc artificial value. 
Then, we consider the Smagorinsky and the kEqn models as described in Sec.~\ref{sec:LES}. 

%We are going to compare the results obtained by using the NSE model \textcolor{red}{introduciamo degli acronimi come nostro solito?} and  the Smagorinsky model. %(already adopted in \textcolor{red}{citazioni}) and the WALE model (that, to the best of our knowledge, it is used here for the first time for this benchmark). 

\begin{figure}
\centering
 \begin{overpic}[width=0.49\textwidth, grid=false]{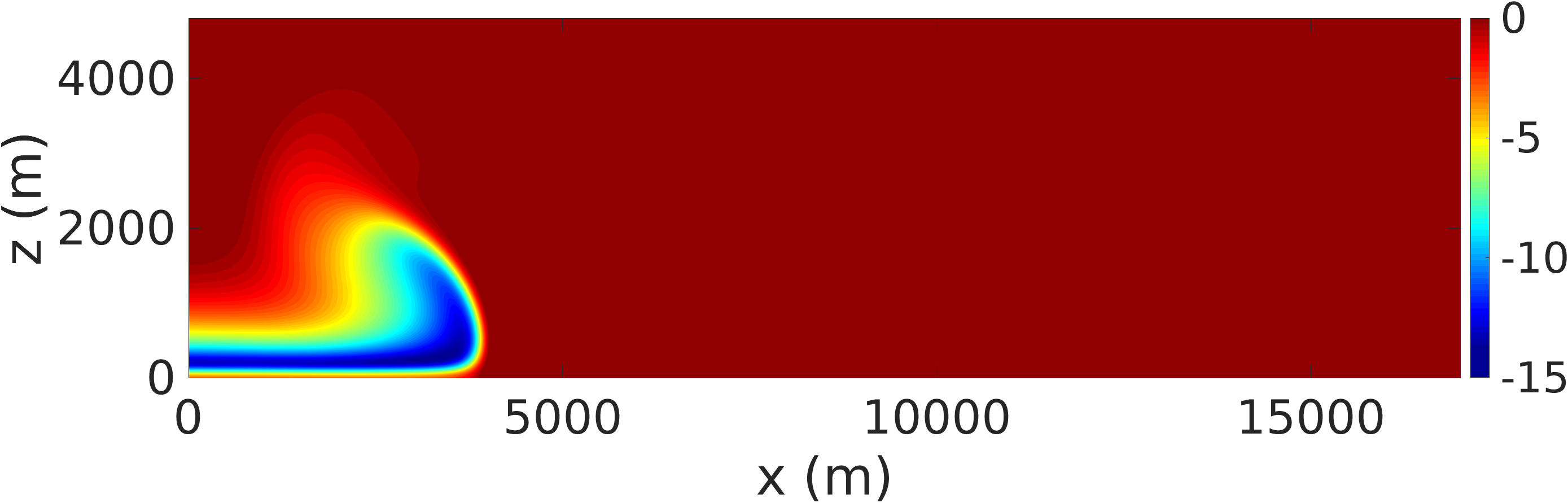}  
        \put(70,25){\textcolor{white}{$t = 300$ s}}
        %\put(-8,7){$\u$}
      \end{overpic}~
 \begin{overpic}[width=0.49\textwidth]{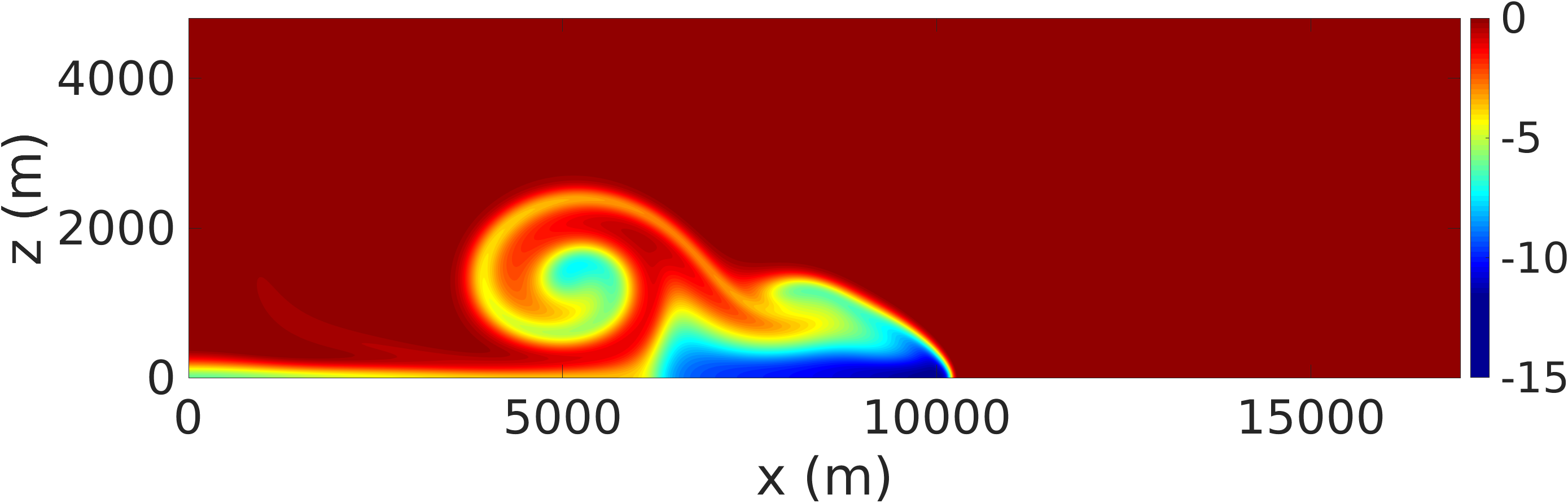}
 \put(70,25){\textcolor{white}{$t = 600$ s}}
      \end{overpic} \\ \vspace{0.3cm}
 \begin{overpic}[width=0.49\textwidth]{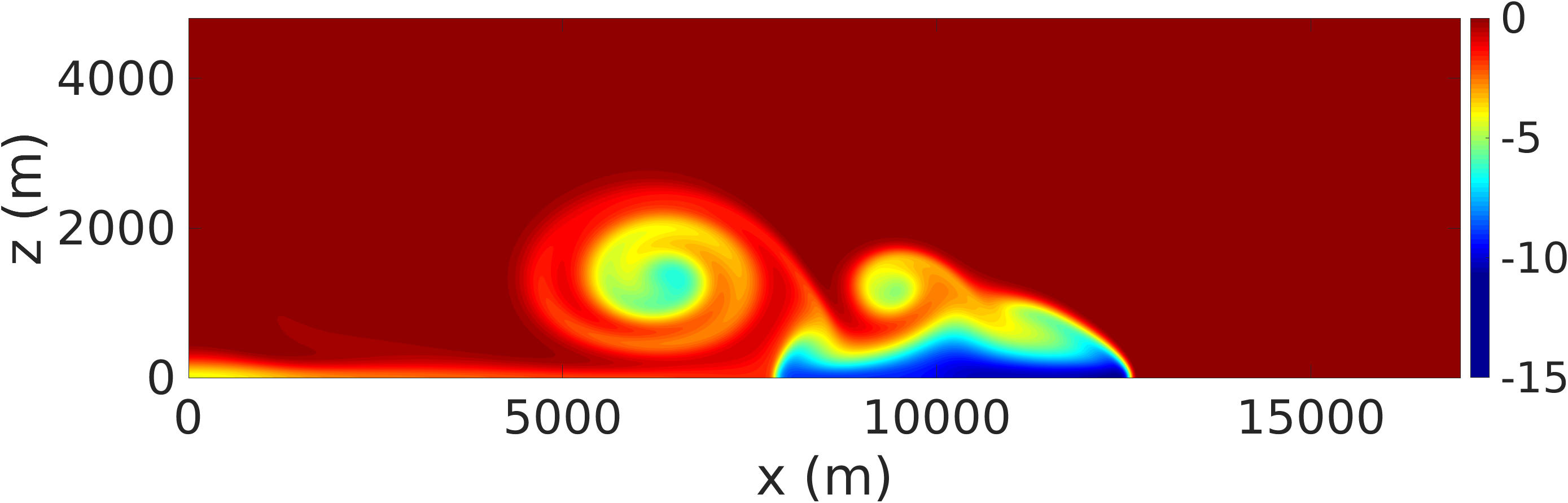}
 \put(70,25){\textcolor{white}{$t = 750$ s}}
      \end{overpic}~
       \begin{overpic}[width=0.49\textwidth]{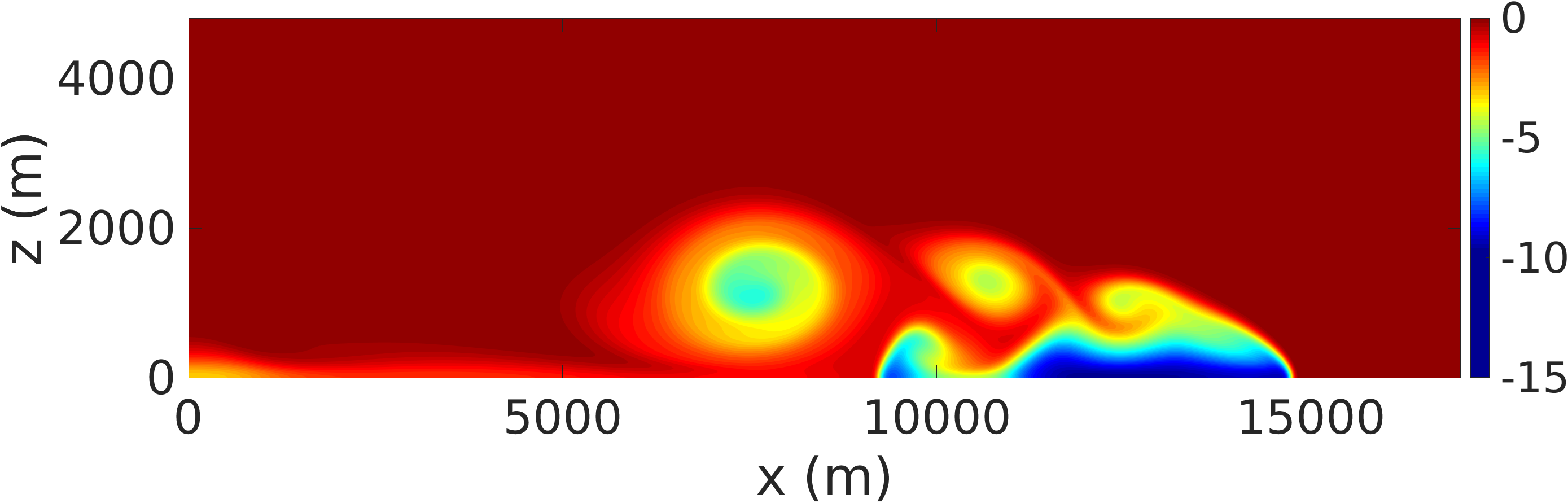} 
       \put(70,25){\textcolor{white}{$t = 900$ s}}
      \end{overpic}
\caption{%\sm{[SM] These plots need to have identical formatting style. Use the formatting of the top-left figure with large fonts.} 
Density current, AV75 model: time evolution of potential temperature fluctuation $\theta'$ computed with mesh $h = 25$ m.}
\label{fig:DC1}
\end{figure}

Let us start with the AV75 model and a qualitative illustration of the flow evolution. 
Figure \ref{fig:DC1} shows $\theta'$ computed with the AV75 model and mesh 
 $h = 25$ m (i.e., the finest mesh among those considered) %resolutions $\Delta x = \Delta z = 400$ m, $\Delta x = \Delta z = 200$ m, $\Delta x = \Delta z = 100$ m, $\Delta x = \Delta z = 50$ m and $\Delta x = \Delta z = 25$ m \textcolor{red}{decidere se fare anche 12 metri} 
at $t = 300, 600, 750, 900$ s. 
We observe very good agreement with the results reported in Fig.~1 of \cite{strakaWilhelmson1993}, which were obtained with the same resolution. Indeed, we see that, as expected, the cold air descends due to negative buoyancy and
strong downdrafts develop at the center of the cold bubble. When the cold air reaches the ground, it rolls up and forms a front. As this front propagates, shear is generated at its top boundary with a resulting Kelvin-Helmholtz type instability that leads to a three-rotor structure at $t = 900$ for the resolution under consideration. 

\begin{figure}[htb]
\centering
 \begin{overpic}[width=0.6\textwidth,grid=false]{images/AV75_25m_900s.png}  
        \put(75,27){\textcolor{white}{$h = 25$ m}}
        %\put(-8,7){$\u$}
      \end{overpic} \\ \vspace{0.3cm}
 \begin{overpic}[width=0.6\textwidth]{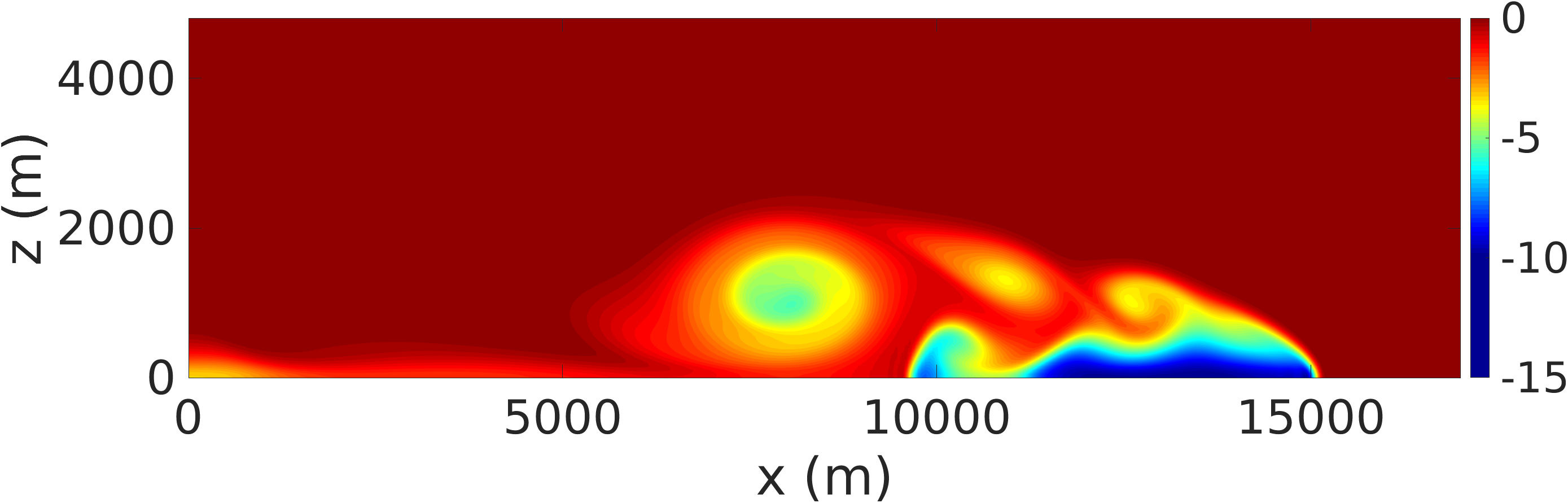} 
 \put(75,27){\textcolor{white}{$h = 50$ m}}
      \end{overpic} \\ \vspace{0.3cm}
 \begin{overpic}[width=0.6\textwidth]{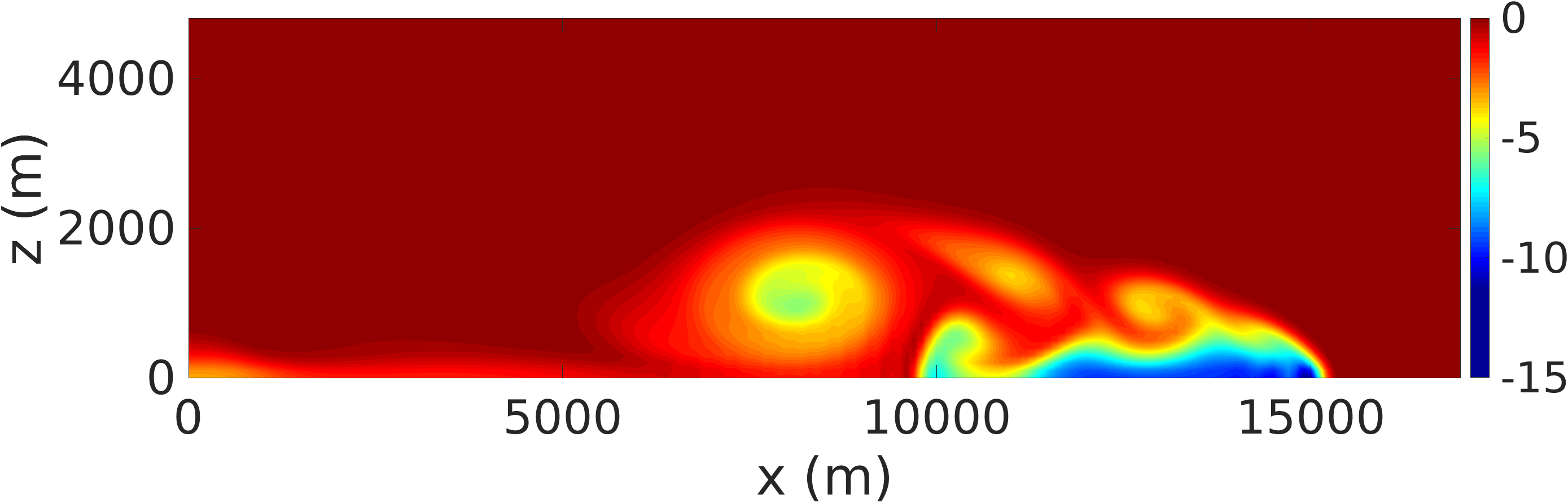} 
 \put(75,27){\textcolor{white}{$h = 100$ m}}
      \end{overpic}
       \begin{overpic}[width=0.6\textwidth]{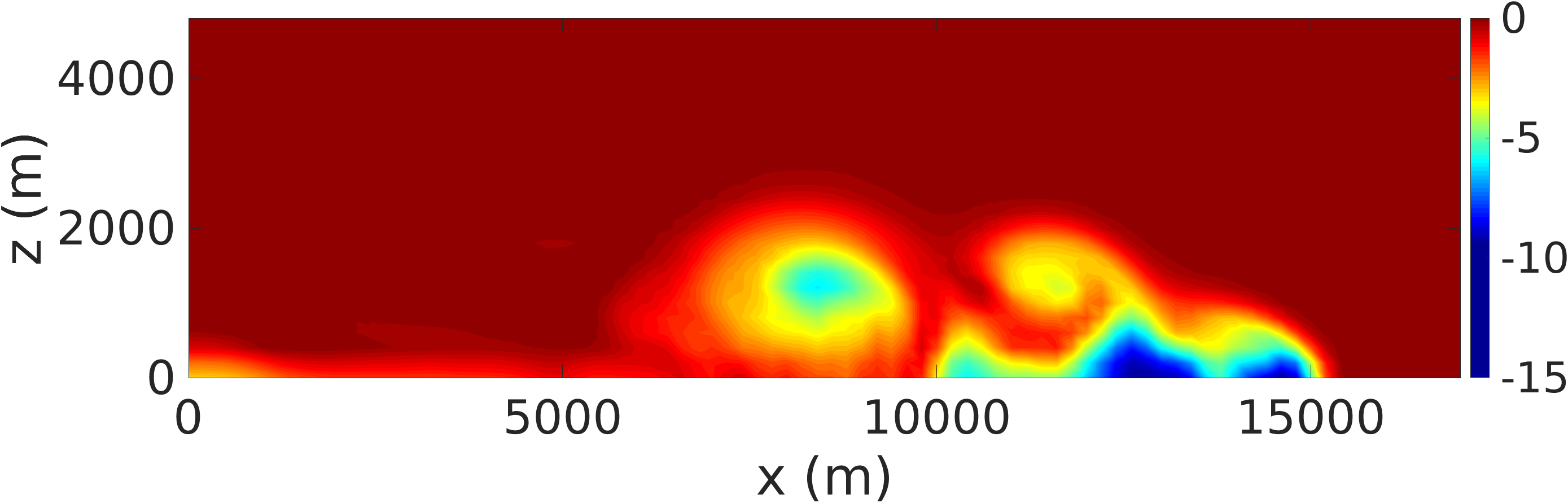}  
       \put(75,27){\textcolor{white}{$h = 200$ m}}
      \end{overpic}
       \begin{overpic}[width=0.6\textwidth]{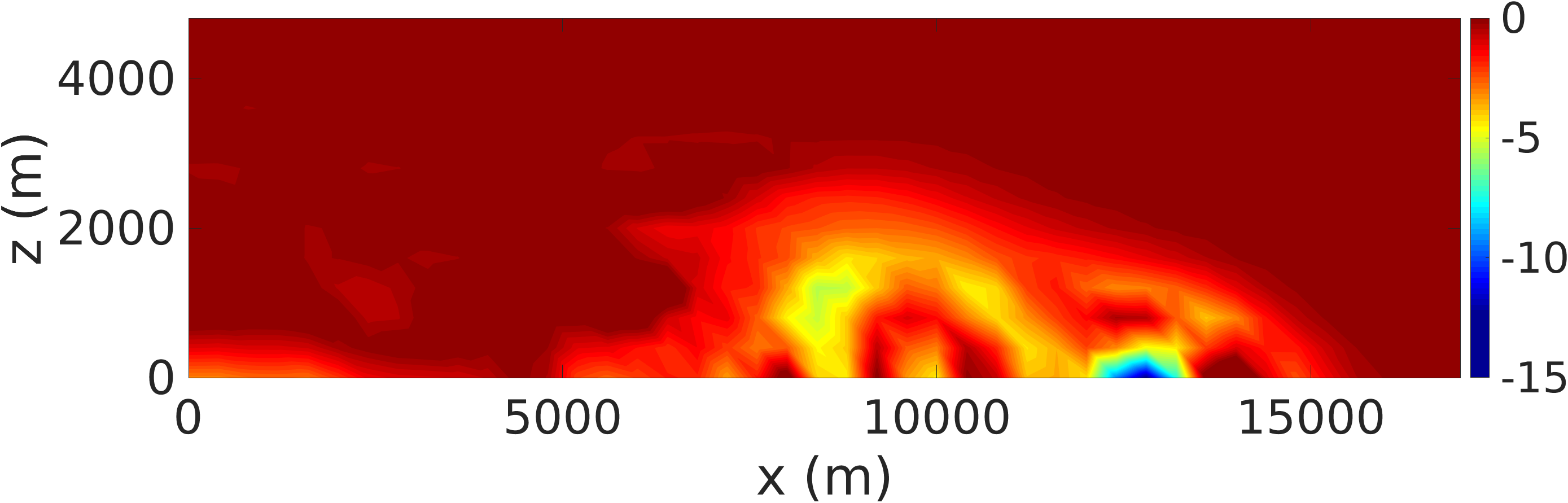}  
       \put(75,27){\textcolor{white}{$h = 400$ m}}
      \end{overpic}
\caption{Density current, AV75 model: potential temperature fluctuation $\theta'$ computed at $t = 900$ s with meshes $h = 25, 50, 100, 200, 400~$m. The mesh size is increasing from top to bottom.}
\label{fig:DC2}
\end{figure}

In Figure \ref{fig:DC2}, we report $\theta'$ computed at $t = 900$ s by the AV75 model with all the meshes mentioned above. 
We observe the emergence of a clear three-rotor structure when the resolution
is equal to or smaller than $h = 100$ m, with more definition when the mesh is finer. Also the results in Figure \ref{fig:DC2} are in very good agreement with those reported in the literature. See, e.g., \cite{strakaWilhelmson1993,ahmadLindeman2007,giraldoRestelli2008a,marrasEtAl2013a,marrasNazarovGiraldo2015}.
%The formation and the propagation of the front and the development of these
%rotors can be seen in Figure 8. 
%The final front location is also indicated in Figure 8, which compares well with the
%solutions given in Straka et al. (1993). The front location (in terms of potential temperature) at 14975m corresponds
%to the cell with a potential temperature value of 299.99 K – the value of potential temperature in the cell next to it (at
%15025m) is 300.0K. The WRF 2nd order simulation (not shown here) became unstable for this test case also – but
%the addition of explicit diffusion stabilized the solution. %\textcolor{blue}{qui scriverei semplicemente che la time evolution è quella che ci si aspetta, confrontabile con altre (WRF, ecc...) e che a partire da 100m raggiungiamo la convergenza!}.
%By considering the accepted use of the Smagorinsky model [37; 47] for this benchmark, %in numerical weather prediction, %to
%support our hypothesis that a properly designed SGS model can serve as a stabilization method,
%we compare the results of the current model with the constant coefficient Smagorinsky at 25 m and
%50 m resolutions. 

\begin{figure}[htb]
\centering
 \begin{overpic}[width=0.49\textwidth]{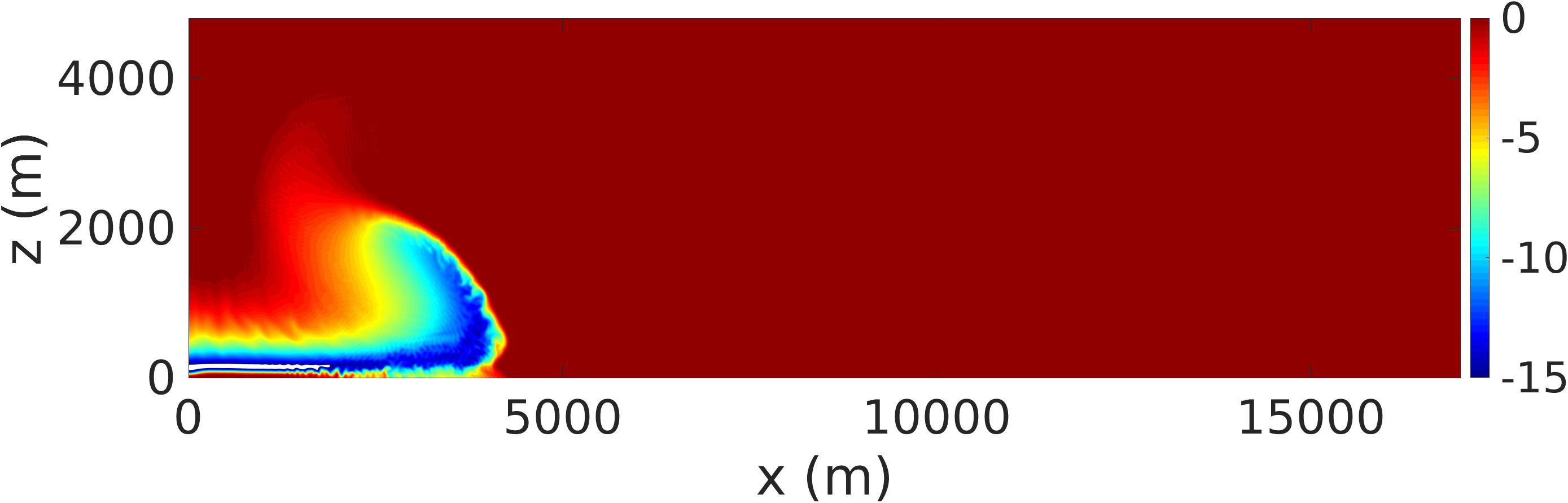}  
        \put(70,25){\textcolor{white}{$t = 300$ s}}
      \end{overpic}~
 \begin{overpic}[width=0.49\textwidth]{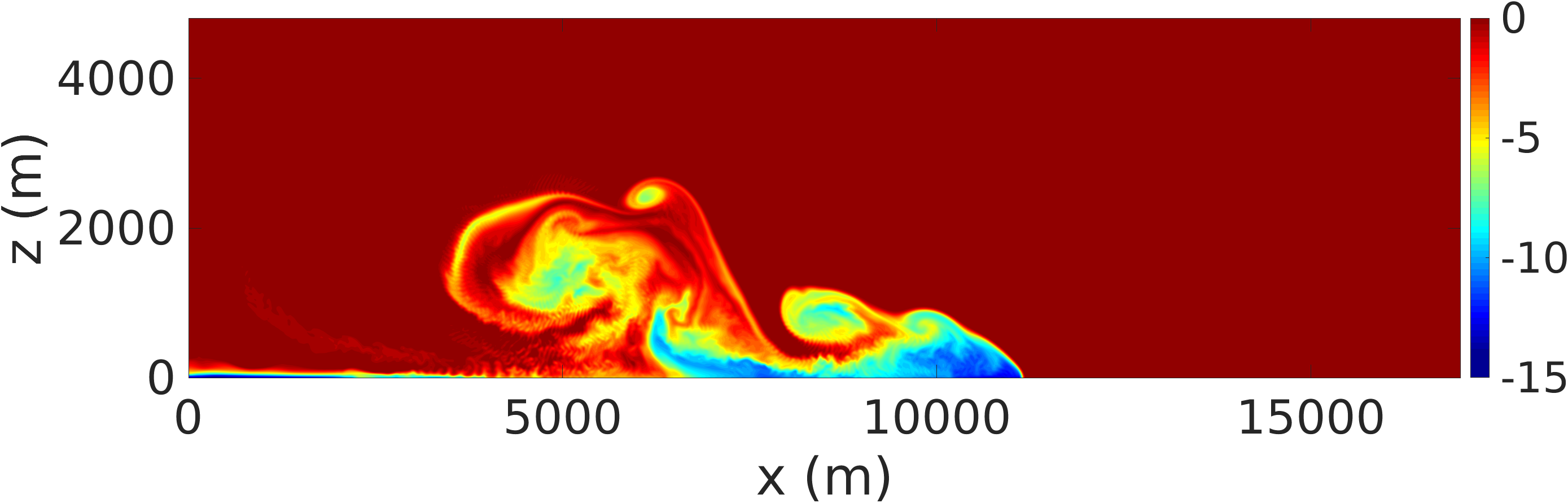}
         \put(70,25){\textcolor{white}{$t = 600$ s}}
      \end{overpic} \\
 \begin{overpic}[width=0.49\textwidth]{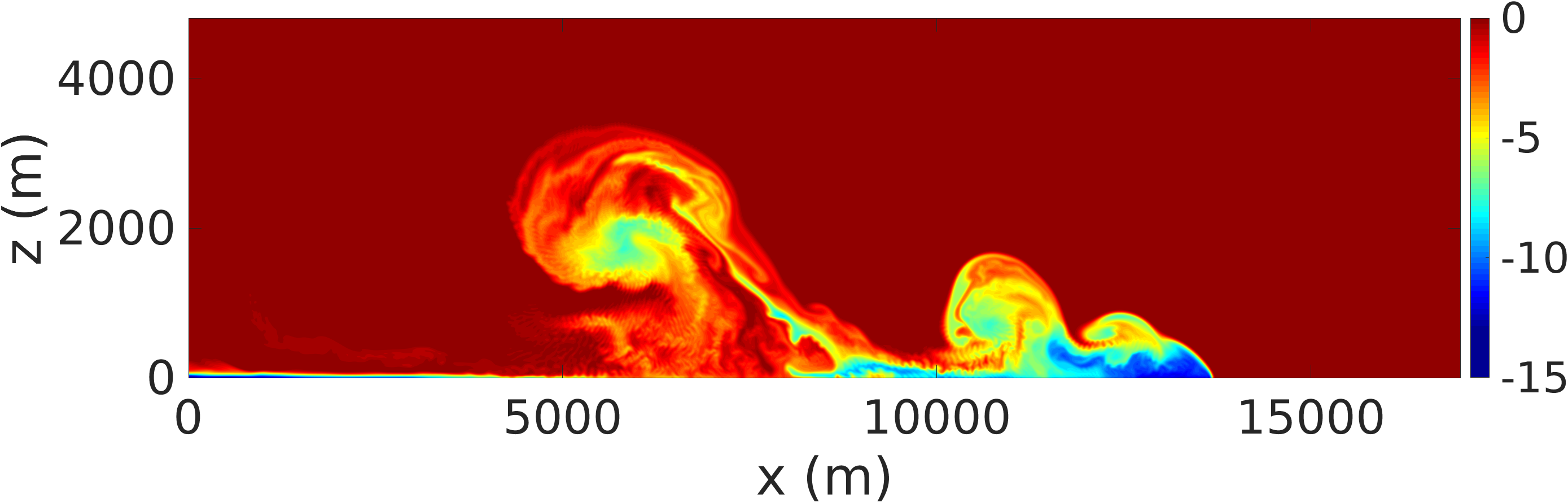}  
        \put(70,25){\textcolor{white}{$t = 750$ s}}
      \end{overpic}~
 \begin{overpic}[width=0.49\textwidth]{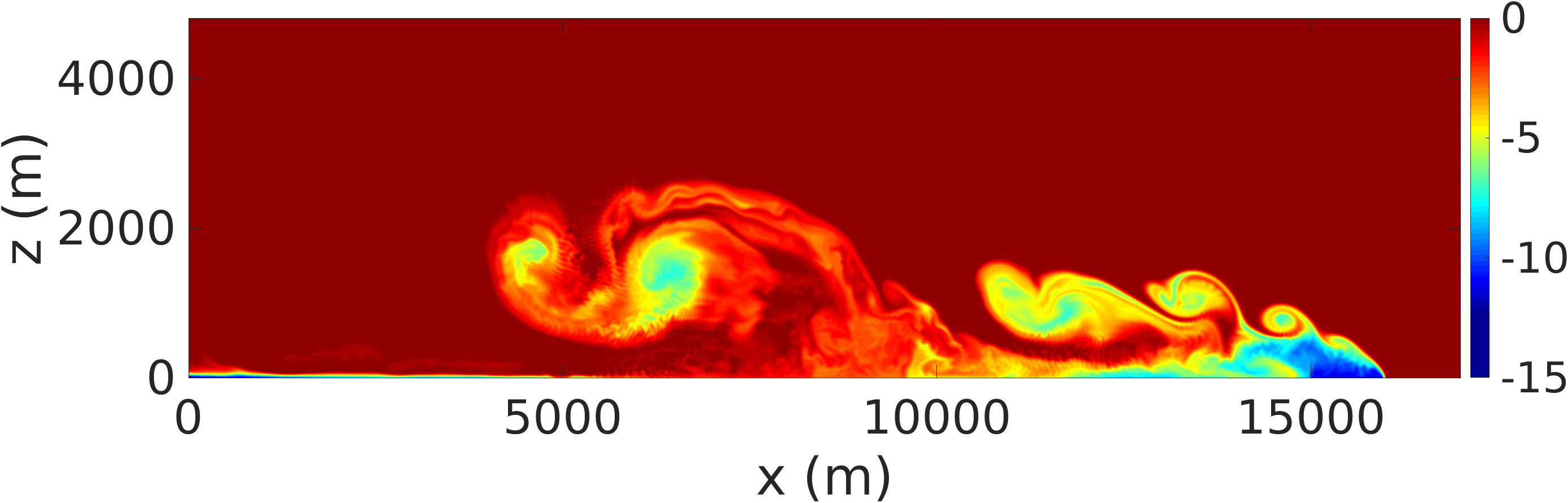}  
        \put(70,25){\textcolor{white}{$t = 900$ s}}
      \end{overpic}
\caption{
Density current, Smagorinsky model:
time evolution of potential temperature fluctuation $\theta'$ computed with mesh $h = 12.5$ m.
%\anna{Linea bianca a $t = 300$}
}
\label{fig:DC3}
\end{figure}

\begin{figure}[htb]
\centering
 \begin{overpic}[width=0.485\textwidth]{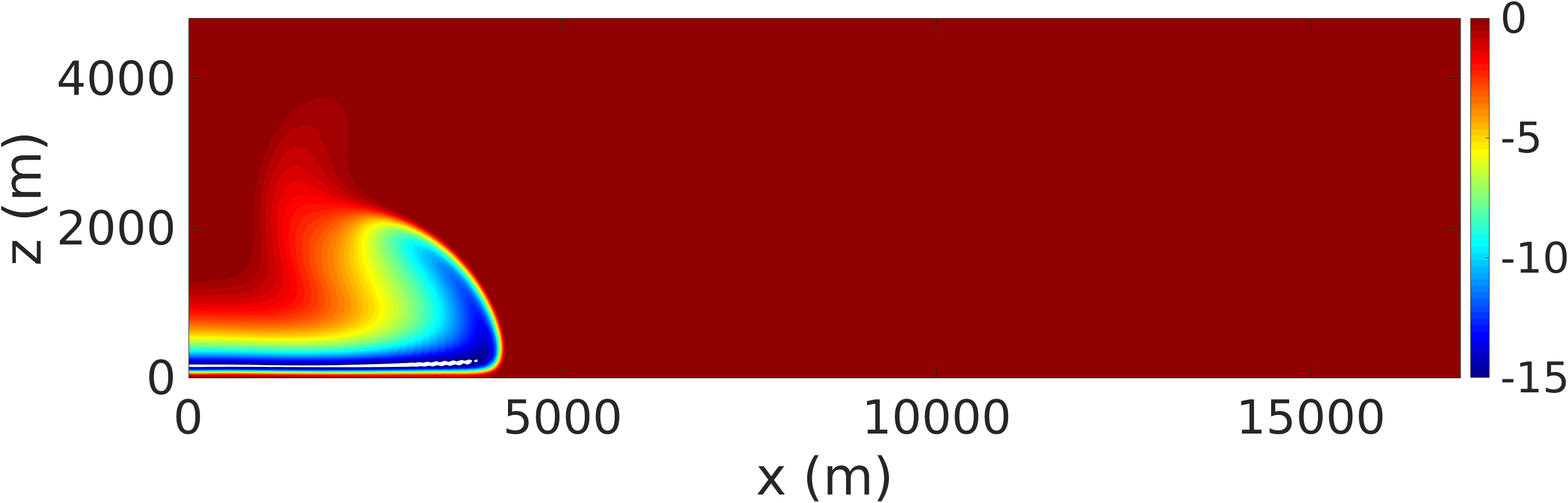}  
        \put(70,25){\textcolor{white}{$t = 300$ s}}
      \end{overpic} ~
 \begin{overpic}[width=0.485\textwidth]{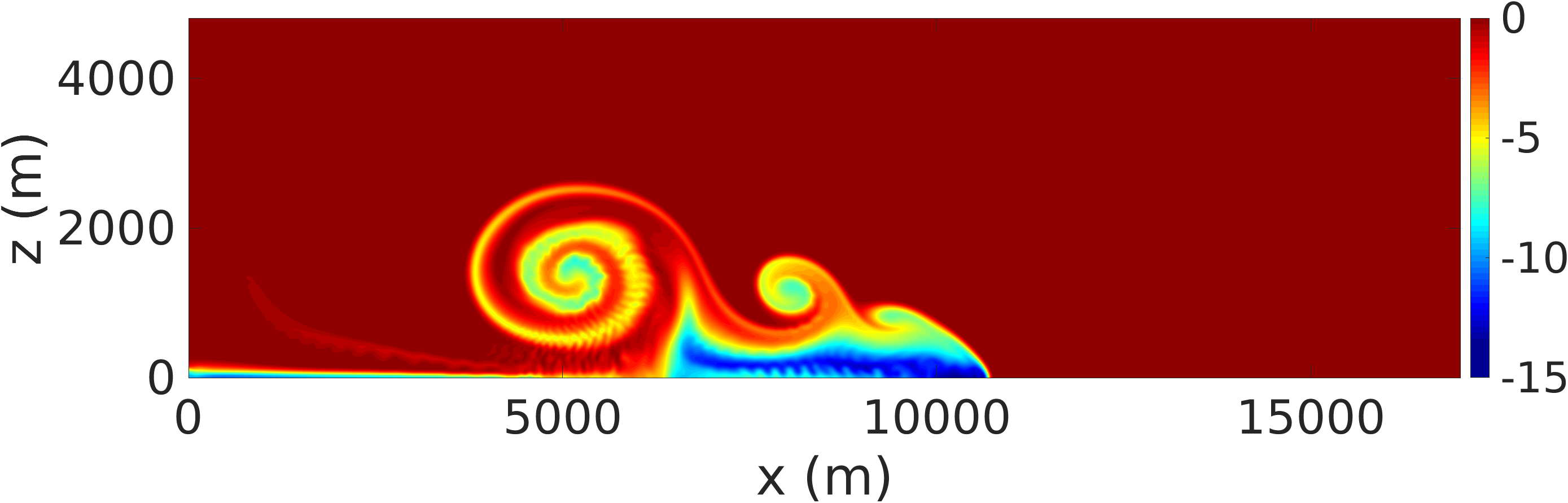}  
        \put(70,25){\textcolor{white}{$t = 600$ s}}
      \end{overpic} \\
 \begin{overpic}[width=0.485\textwidth]{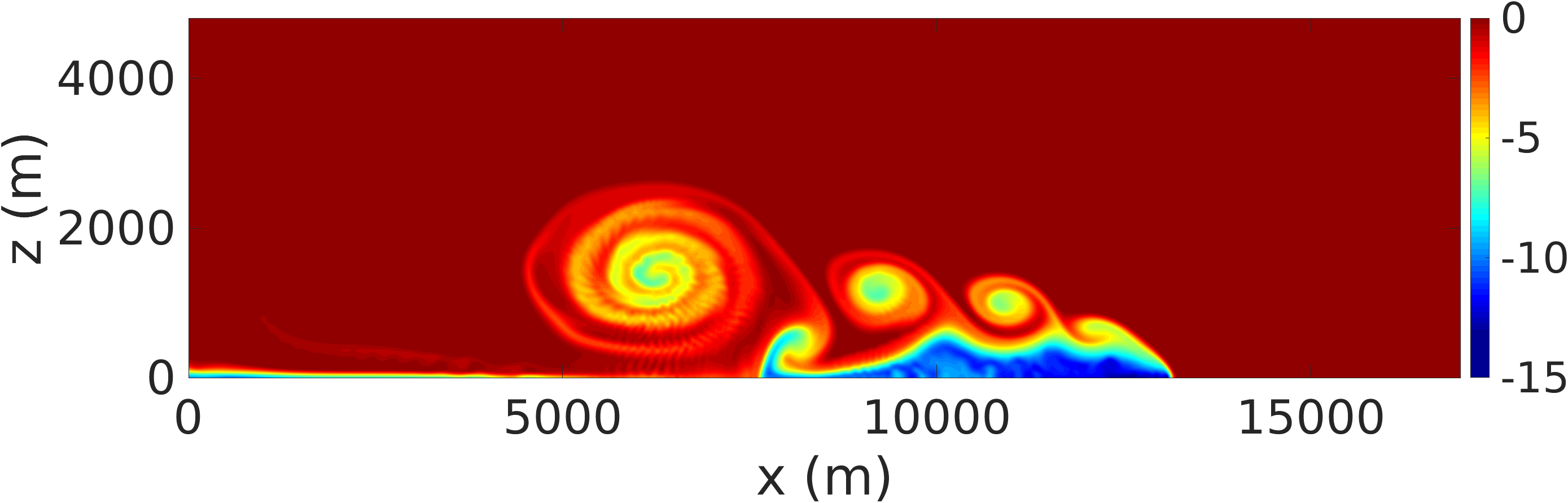}  
        \put(70,25){\textcolor{white}{$t = 750$ s}}
      \end{overpic} ~
 \begin{overpic}[width=0.485\textwidth]{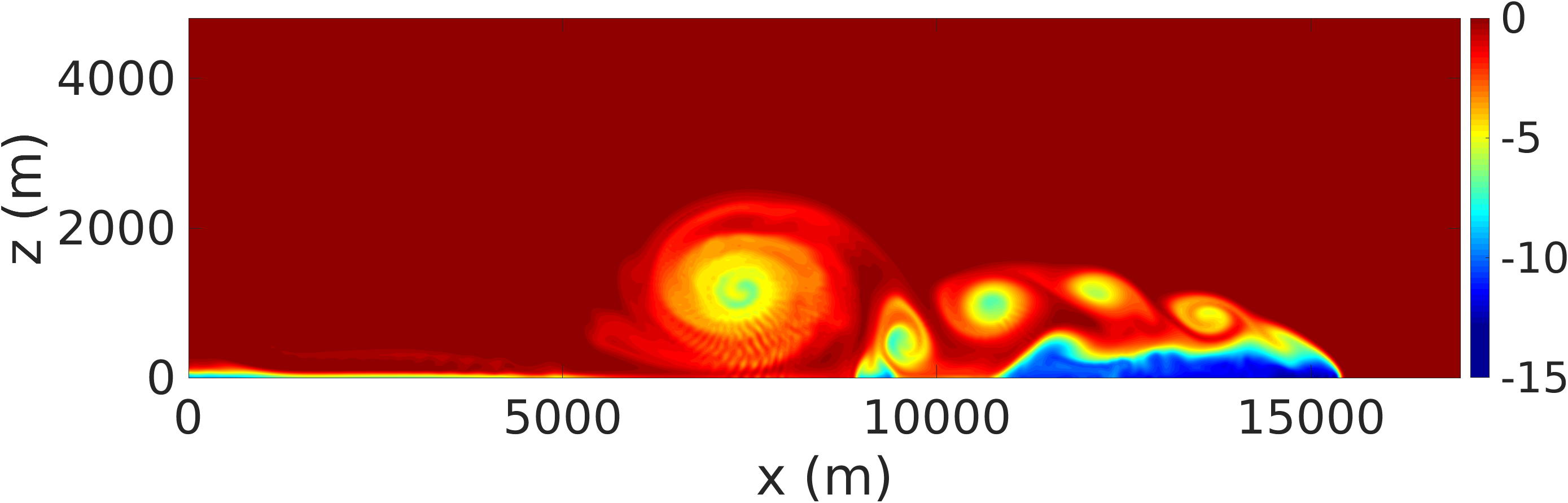}  
        \put(70,25){\textcolor{white}{$t = 900$ s}}
      \end{overpic}
\caption{Density current, Smagorinsky model:
time evolution of potential temperature fluctuation $\theta'$ computed with mesh $h = 25$ m.
%\anna{Linea bianca a $t = 300$}
}
\label{fig:DC4}
\end{figure}

\begin{figure}[htb]
\centering
 \begin{overpic}[width=0.485\textwidth]{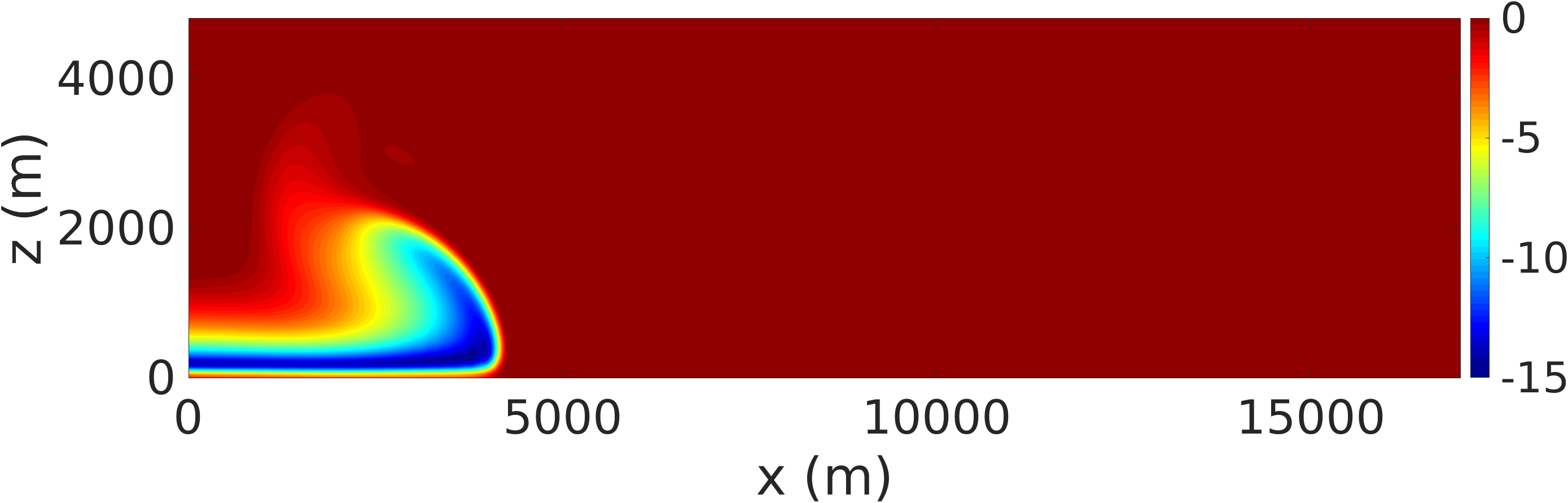}  
        \put(70,25){\textcolor{white}{$t = 300$ s}}
      \end{overpic} ~
 \begin{overpic}[width=0.485\textwidth]{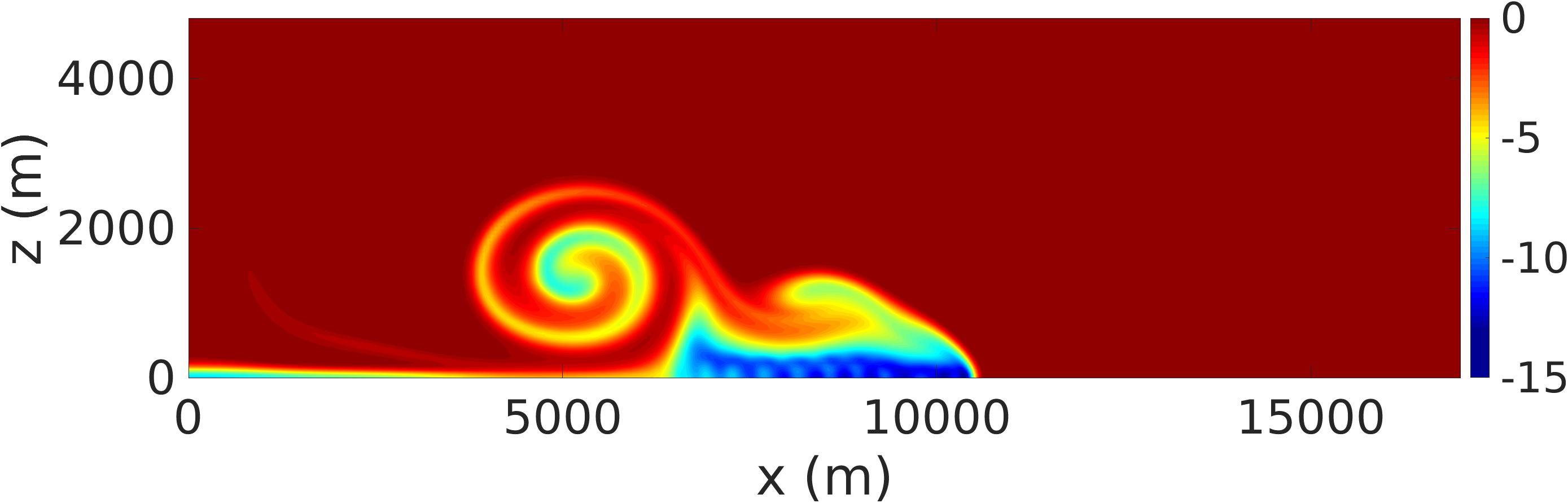}  
        \put(70,25){\textcolor{white}{$t = 600$ s}}
      \end{overpic} \\
 \begin{overpic}[width=0.485\textwidth]{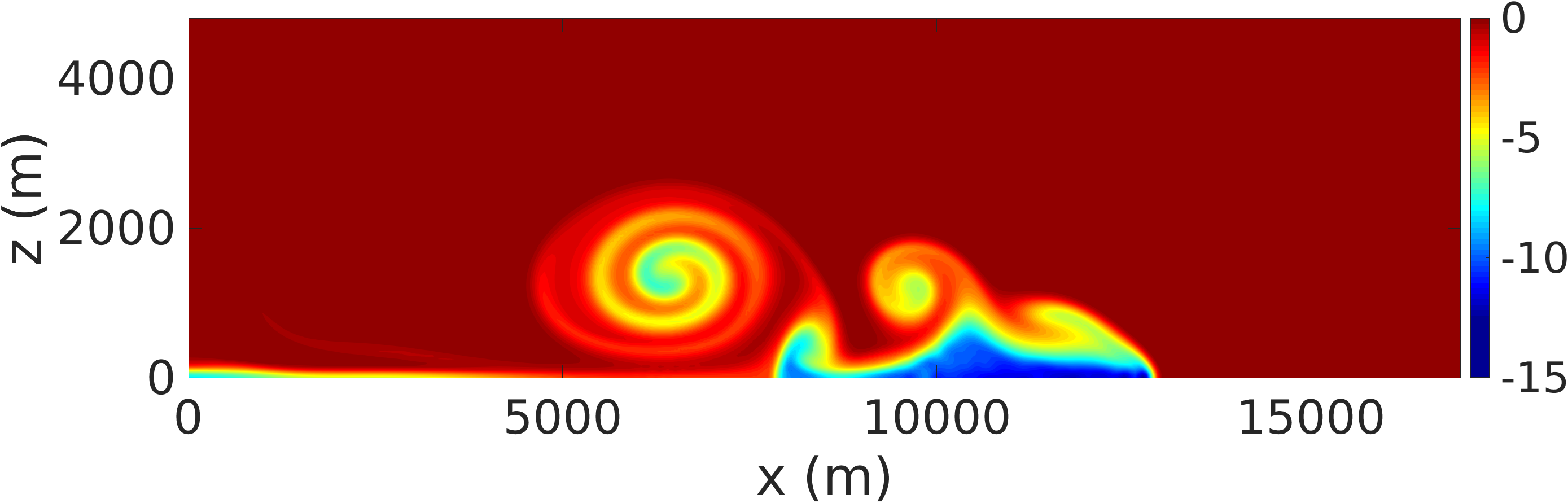}  
        \put(70,25){\textcolor{white}{$t = 750$ s}}
      \end{overpic}~
 \begin{overpic}[width=0.485\textwidth]{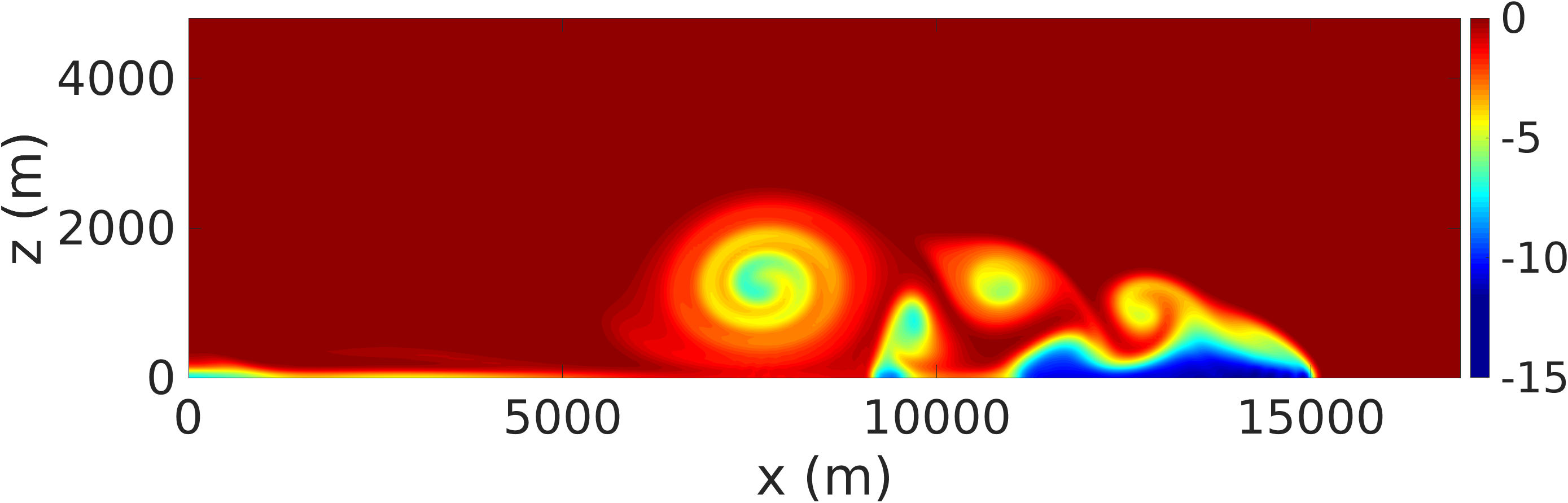}  
        \put(70,25){\textcolor{white}{$t = 900$ s}}
      \end{overpic}
\caption{
Density current, Smagorinsky model:
time evolution of potential temperature fluctuation $\theta'$ computed with mesh $h = 50$ m.}
\label{fig:DC5}
\end{figure}

Next, we focus on the Smagorinsky model. 
Given the poor quality of the solutions computed with the AV75 model and meshes coarser than $h = 100$ m, we will use only the meshes finer than $h = 100$ m. Moreover, following \cite{marrasNazarovGiraldo2015} for the LES models we will also consider a very fine mesh with size $h = 12.5$ m. 
Figures \ref{fig:DC3}, \ref{fig:DC4}, and \ref{fig:DC5} display the time evolution of the potential temperature fluctuation computed with the Smagorinsky model and meshes $h = 12.5, 25, 50$ m, respectively.  For all these simulations, the parameter $C_s $ \eqref{eq:smago} was set to 0.454, with $C_k = 0.6$ and $C_\epsilon = 1.048$.
We note that the ratio of the values of $C_s$ used for this benchmark and for the rising thermal bubble is about 5, which is the ratio of the ad-hoc artificial viscosities.
%\textcolor{red}{Scrivere che il rapporto tra  di Straka e quello della bolla è vicino a 5 che il rapporto tra le viscosità dinamiche 75 e 15 del modello AV}
%\anna{add value} \textcolor{red}{aggiunto! :)}. 
From Figures \ref{fig:DC3}-\ref{fig:DC5}, we clearly see that more vortical structures appear as the mesh size decreases. %, as one would expect. 
Indeed, we can see what while with $h = 50$ m there is a three-rotor structure at $t = 900$ s (see Fig.~\ref{fig:DC5}, bottom right panel), with $h = 25$ m a quadri-rotor structure is present (see Fig.~\ref{fig:DC4}, bottom right panel). The largest recirculation in the bottom right panels of \ref{fig:DC5} and \ref{fig:DC4} is broken into two recirculations when mesh $h = 12.5$ m is used. See Fig.~\ref{fig:DC3}, bottom right panel.

\begin{figure}[htb]
\centering
 \begin{overpic}[width=0.485\textwidth]{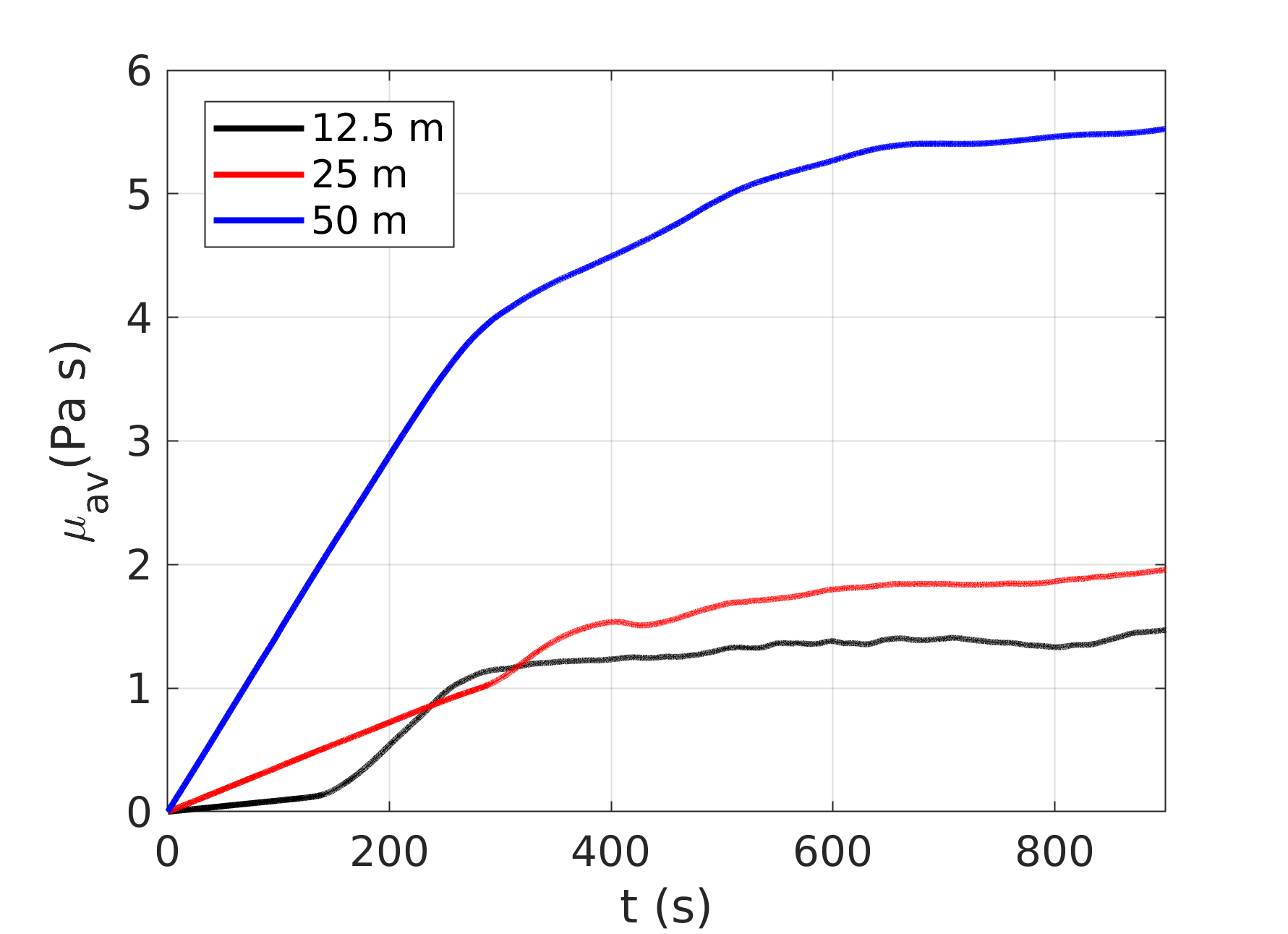}  
        %\put(35,18){FOM}
        %\put(-8,7){$\u$}
      \end{overpic}
\caption{Density current, Smagorinsky model: time evolution of the average eddy viscosity \eqref{eq:mu_av} for meshes $h = 12.5, 25, 50$ m.}
\label{fig:DC10}
\end{figure}

To further justify our choice of $C_s$,
we report in Figure \ref{fig:DC10} the time evolution of the space-averaged eddy viscosity
\begin{equation}\label{eq:mu_av}
\mu_{av} = \dfrac{1}{\Omega} \int_{\Omega} \mu_a d \Omega
\end{equation}
for meshes $h = 12.5, 25, 50$ m.
We see that $\mu_{av}$ decreases with increasing resolution over most of the time interval. This is expected since the eddy viscosity \eqref{eq:smago} is a quadratic function of the filter width, which is related to the mesh size. However, even  with the coarsest mesh (i.e., $h = 50$ m), the space-averaged eddy viscosity is much smaller than 75, i.e., the value used by the AV75 model.
Values of $C_s$ smaller than 0.454 do not provide enough artificial dissipation to stabilize the solution, leading to under-resolved regions and a significant amount of noise above the recirculations. 
The value $C_s = 0.454$ ensures a good compromise between accuracy and stability and allows us
to obtain a very good qualitative agreement with the solutions reported in \cite{strakaWilhelmson1993,ahmadLindeman2007,giraldoRestelli2008a,marrasEtAl2013a,marrasNazarovGiraldo2015}.

%Moreover, we see that, at a given resolution,  at increasing of the time $\nu_t$ increases. This could be due to the fact that more and more vortical structures occur \textcolor{red}{(dire meglio!)}.

%\anna{Io farei un paragrafo a parte su $C_s$. Lo metterei qui e lo giustificherei con i grafici della viscosita'.} \textcolor{red}{Ottima idea Anna! Concordo in pieno, così contestualizziamo meglio il comparison tra 75 e la vicosità media turbolenta!}

\begin{figure}[htb]
\centering
 \begin{overpic}[width=0.485\textwidth]{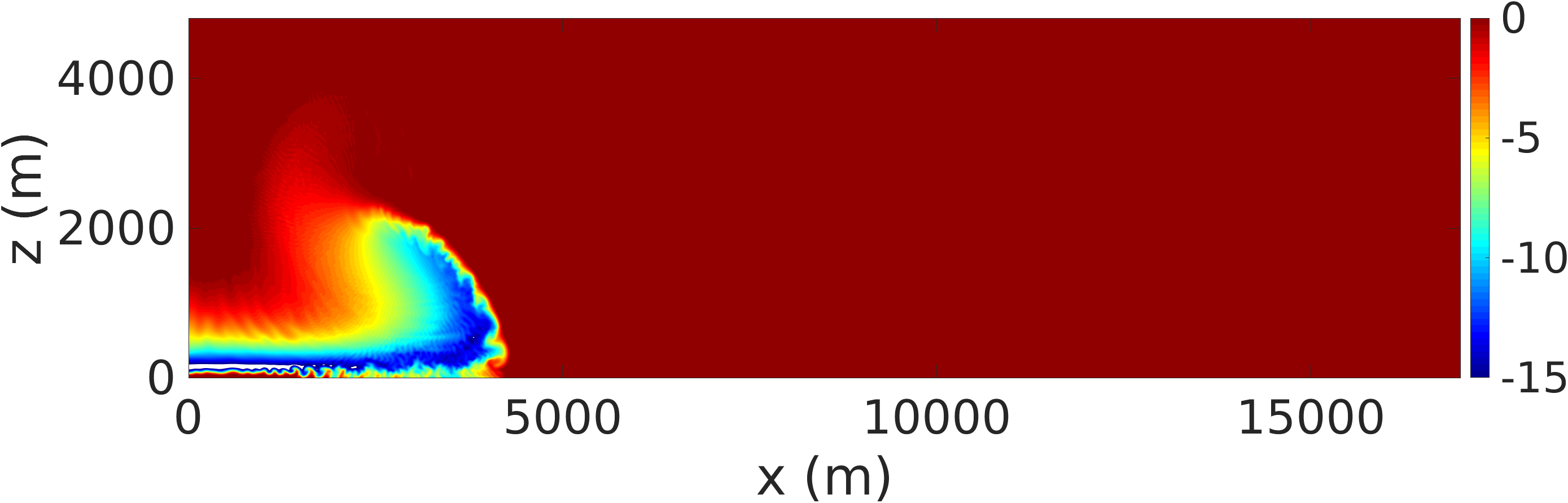}  
        \put(70,25){\textcolor{white}{$t = 300$ s}}
      \end{overpic} ~
 \begin{overpic}[width=0.485\textwidth]{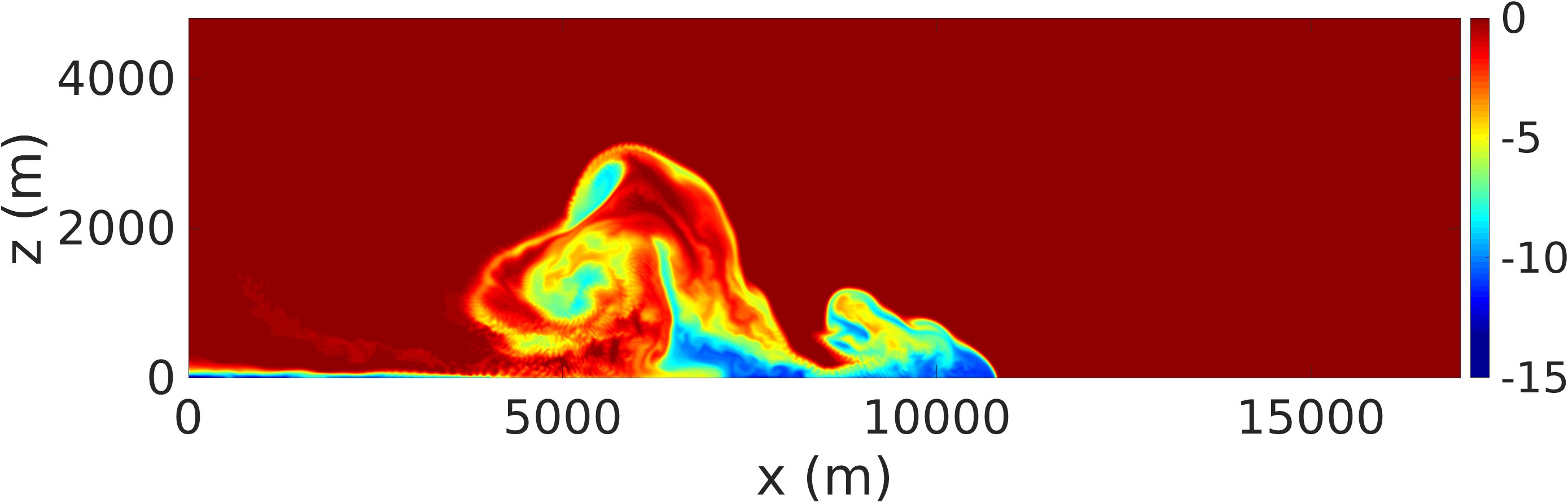}  
        \put(70,25){\textcolor{white}{$t = 600$ s}}
      \end{overpic} \\
 \begin{overpic}[width=0.485\textwidth]{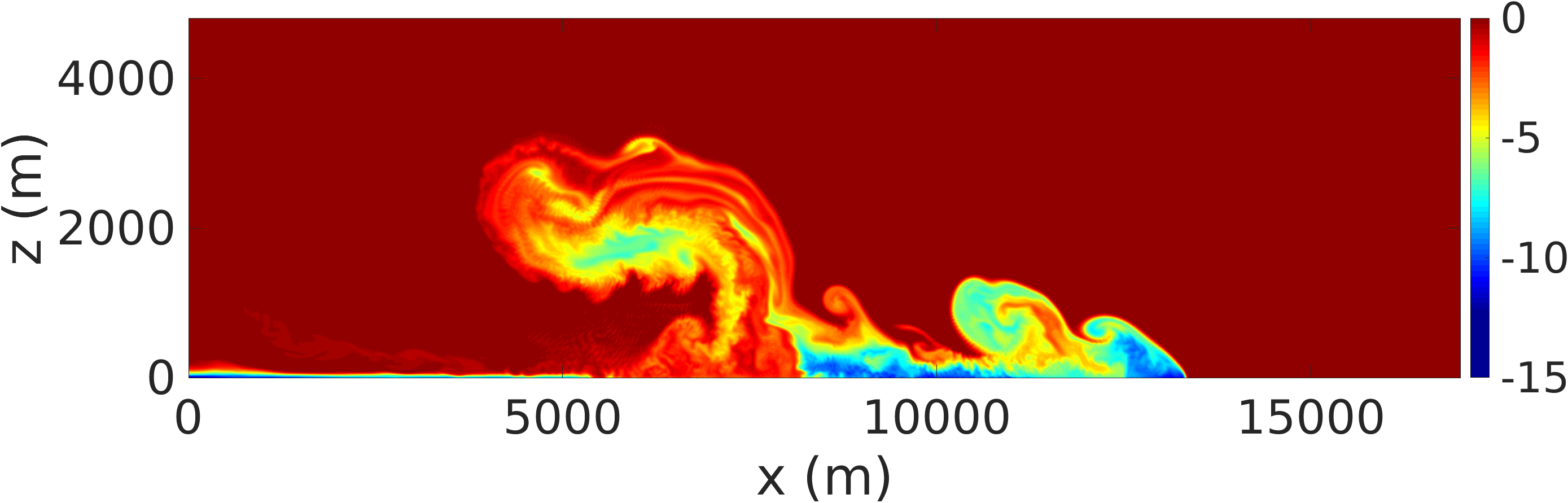}  
        \put(70,25){\textcolor{white}{$t = 750$ s}}
      \end{overpic} ~
 \begin{overpic}[width=0.485\textwidth]{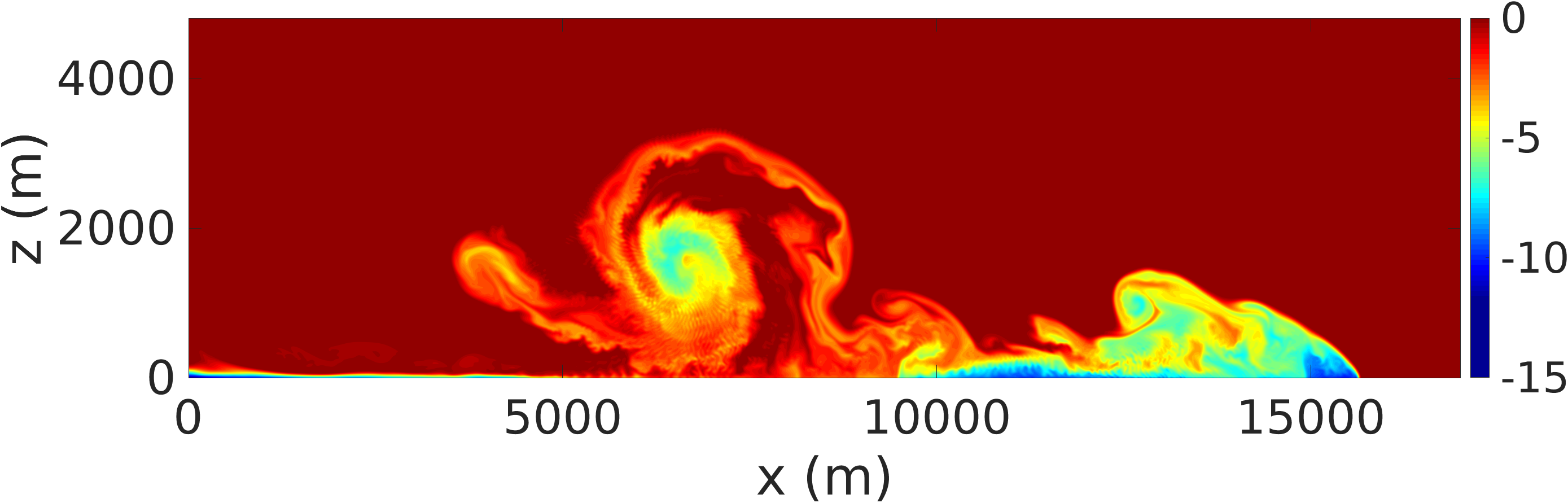}  
        \put(70,25){\textcolor{white}{$t = 900$ s}}
      \end{overpic}
\caption{Density current, kEqn model:
time evolution of potential temperature fluctuation $\theta'$ computed with mesh $h = 12.5$ m.
%\anna{Linea bianca a $t = 300$}
}
\label{fig:DC6}
\end{figure}

\begin{figure}[htb]
\centering
 \begin{overpic}[width=0.485\textwidth]{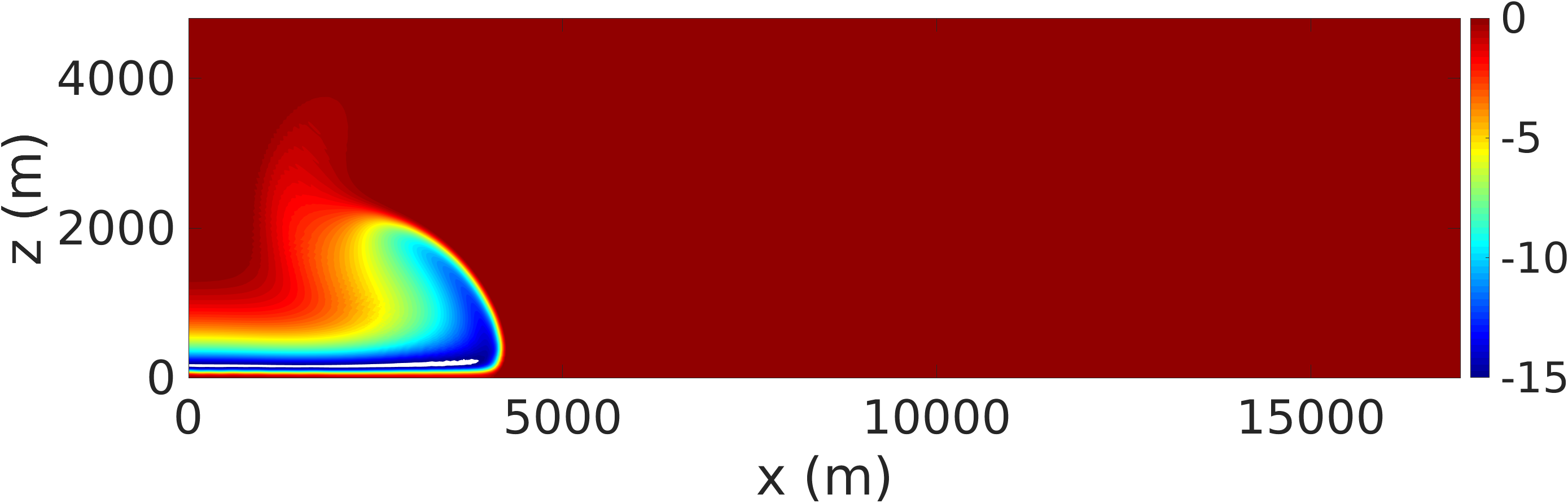}  
        \put(70,25){\textcolor{white}{$t = 300$ s}}
      \end{overpic} ~
 \begin{overpic}[width=0.485\textwidth]{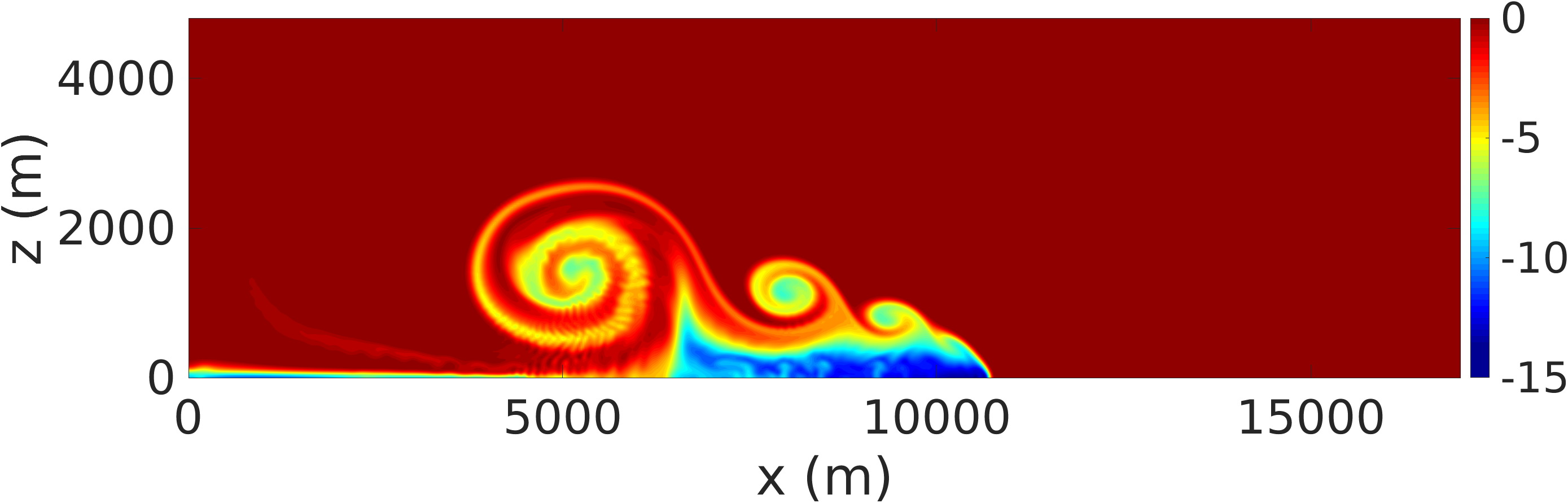}  
        \put(70,25){\textcolor{white}{$t = 600$ s}}
      \end{overpic} \\
 \begin{overpic}[width=0.485\textwidth]{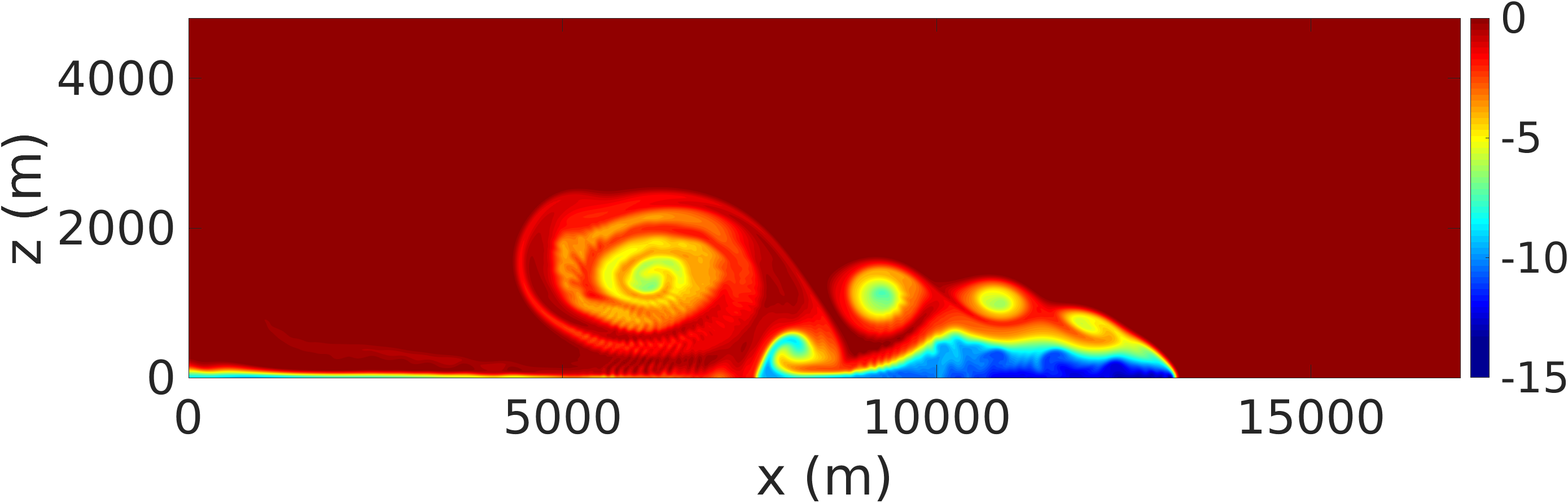}  
        \put(70,25){\textcolor{white}{$t = 750$ s}}
      \end{overpic} ~
 \begin{overpic}[width=0.485\textwidth]{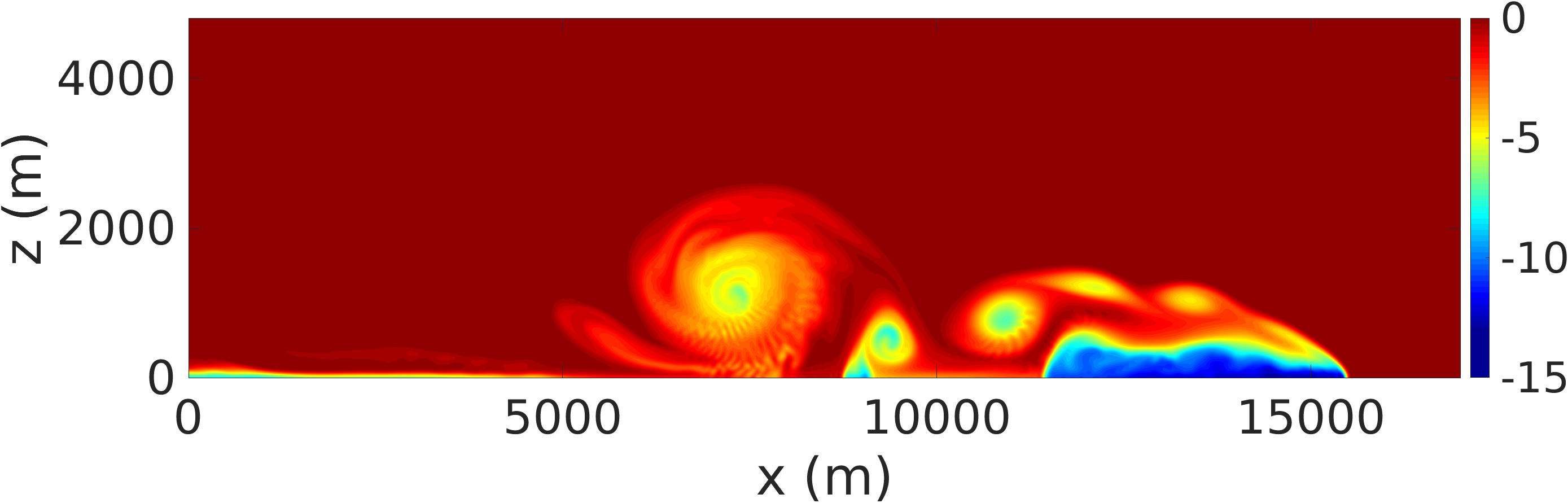}  
        \put(70,25){\textcolor{white}{$t = 900$ s}}
      \end{overpic}
\caption{Density current, kEqn model:
time evolution of potential temperature fluctuation $\theta'$ computed with mesh $h = 25$ m. %\anna{Linea bianca a $t = 300$}
}
\label{fig:DC7}
\end{figure}

\begin{figure}[htb]
\centering
 \begin{overpic}[width=0.485\textwidth]{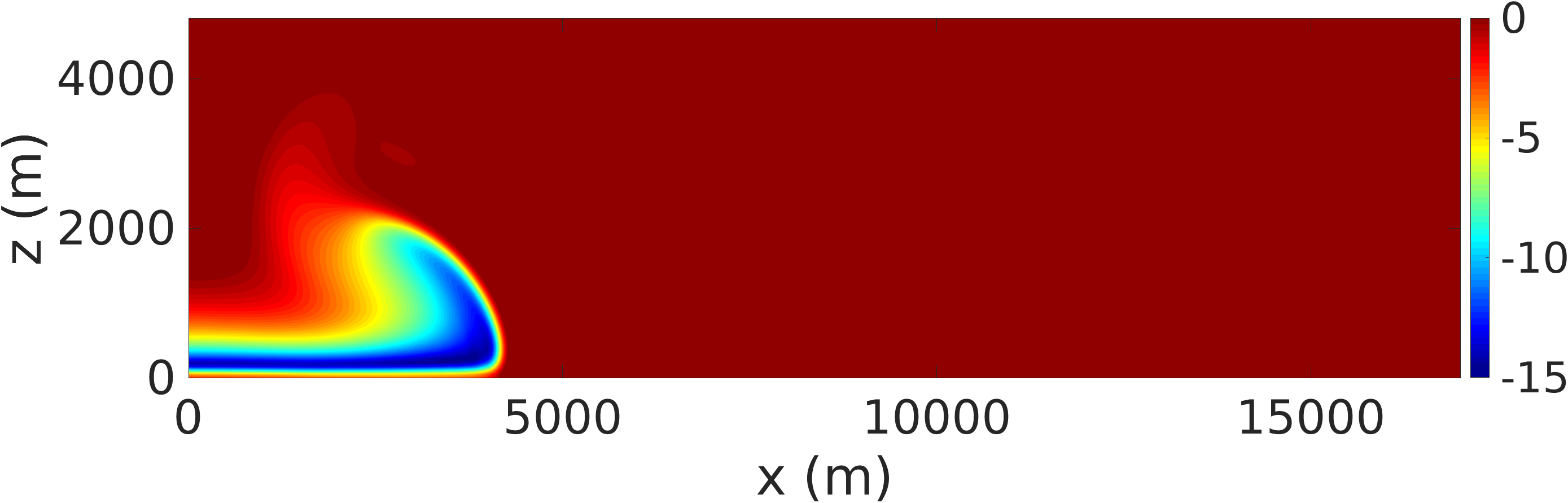}  
        \put(70,25){\textcolor{white}{$t = 300$ s}}
      \end{overpic} ~
 \begin{overpic}[width=0.485\textwidth]{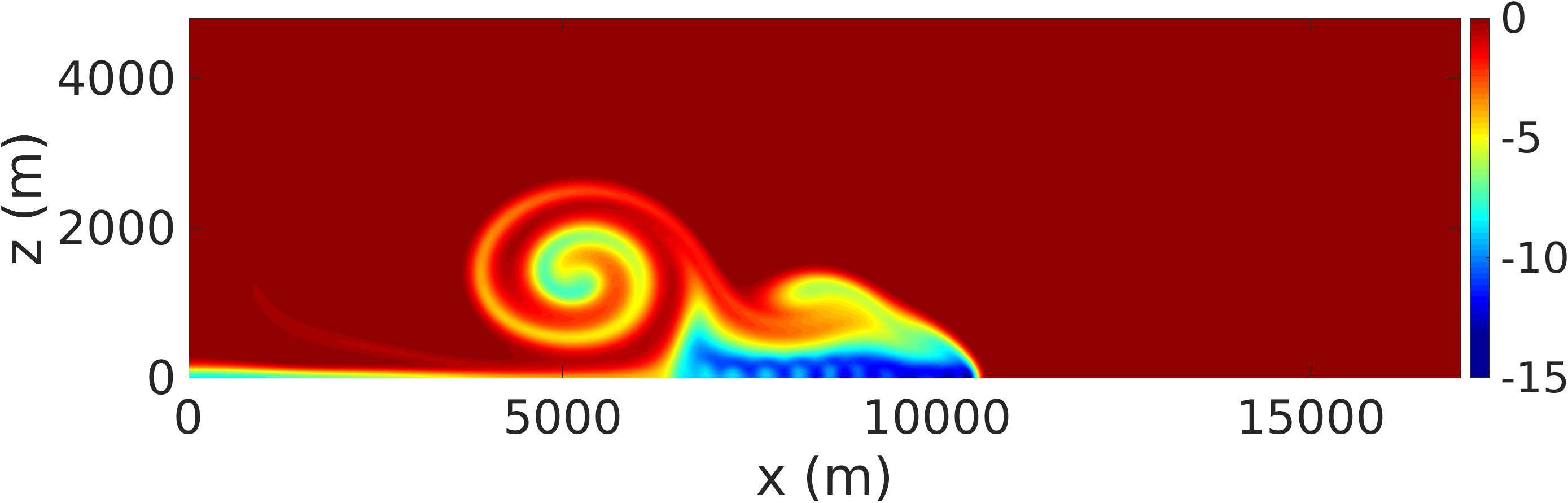}  
        \put(70,25){\textcolor{white}{$t = 600$ s}}
      \end{overpic} \\
 \begin{overpic}[width=0.485\textwidth]{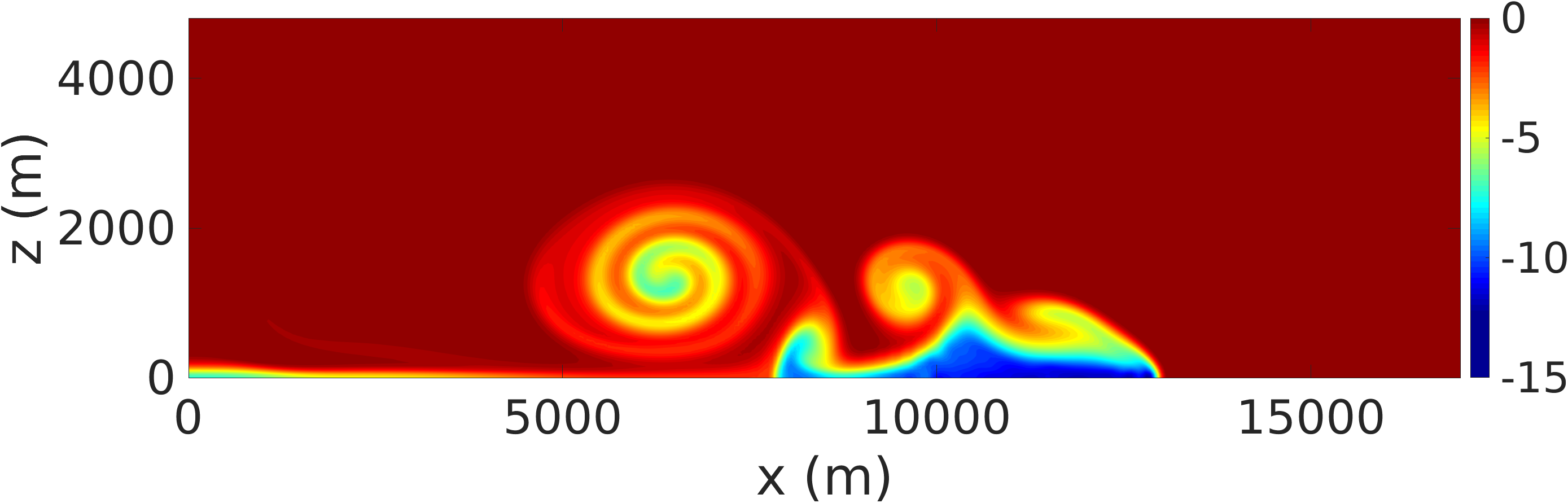}  
        \put(70,25){\textcolor{white}{$t = 750$ s}}
      \end{overpic} ~
 \begin{overpic}[width=0.485\textwidth]{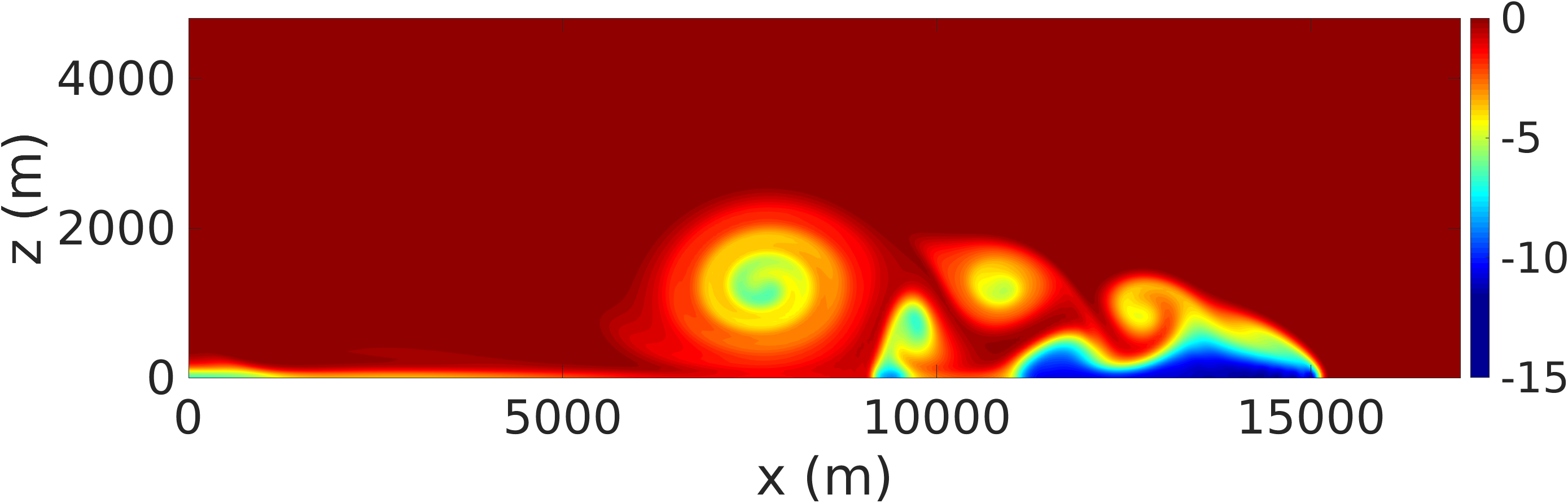}  
        \put(70,25){\textcolor{white}{$t = 900$ s}}
      \end{overpic}
\caption{Density current, kEqn model:
time evolution of potential temperature fluctuation $\theta'$ computed with mesh $h = 50$ m.}
\label{fig:DC8}
\end{figure}

We now switch to the kEqn model.
Figures \ref{fig:DC6}, \ref{fig:DC7}, and \ref{fig:DC8} display the time evolution of the potential temperature fluctuation computed with the kEqn model and meshes $h = 12.5, 25, 50$ m, respectively.
We observe little difference between the solutions computed by the Smagorinsky model and the kEqn model with mesh $h = 50$ m for the entire duration of the time interval. 
Compare Figure \ref{fig:DC5} with Figure \ref{fig:DC8}.
The differences in the solutions given by the two LES models with mesh $h = 25$ m become visible around $t = 750$ s, when the largest recirculation given by the kEqn model has a more flattened top boundary. In addition, the quadri-rotor structure at $t = 900$ s given by the Smagorinsky model is more defined. Compare the bottom panels in  \ref{fig:DC4} with Figure \ref{fig:DC7}. The differences in the solutions given the two LES models with mesh $h = 12.5$ m are pretty remarkable already at $t = 600$ s. Compare Figure \ref{fig:DC3} with Figure \ref{fig:DC6}.

\begin{figure}[htb]
\centering
%  \begin{overpic}[width=0.485\textwidth]{images/turbulent_visc.png}  
%         %\put(35,18){FOM}
%         %\put(-8,7){$\u$}
%       \end{overpic}
       \begin{overpic}[width=0.485\textwidth]{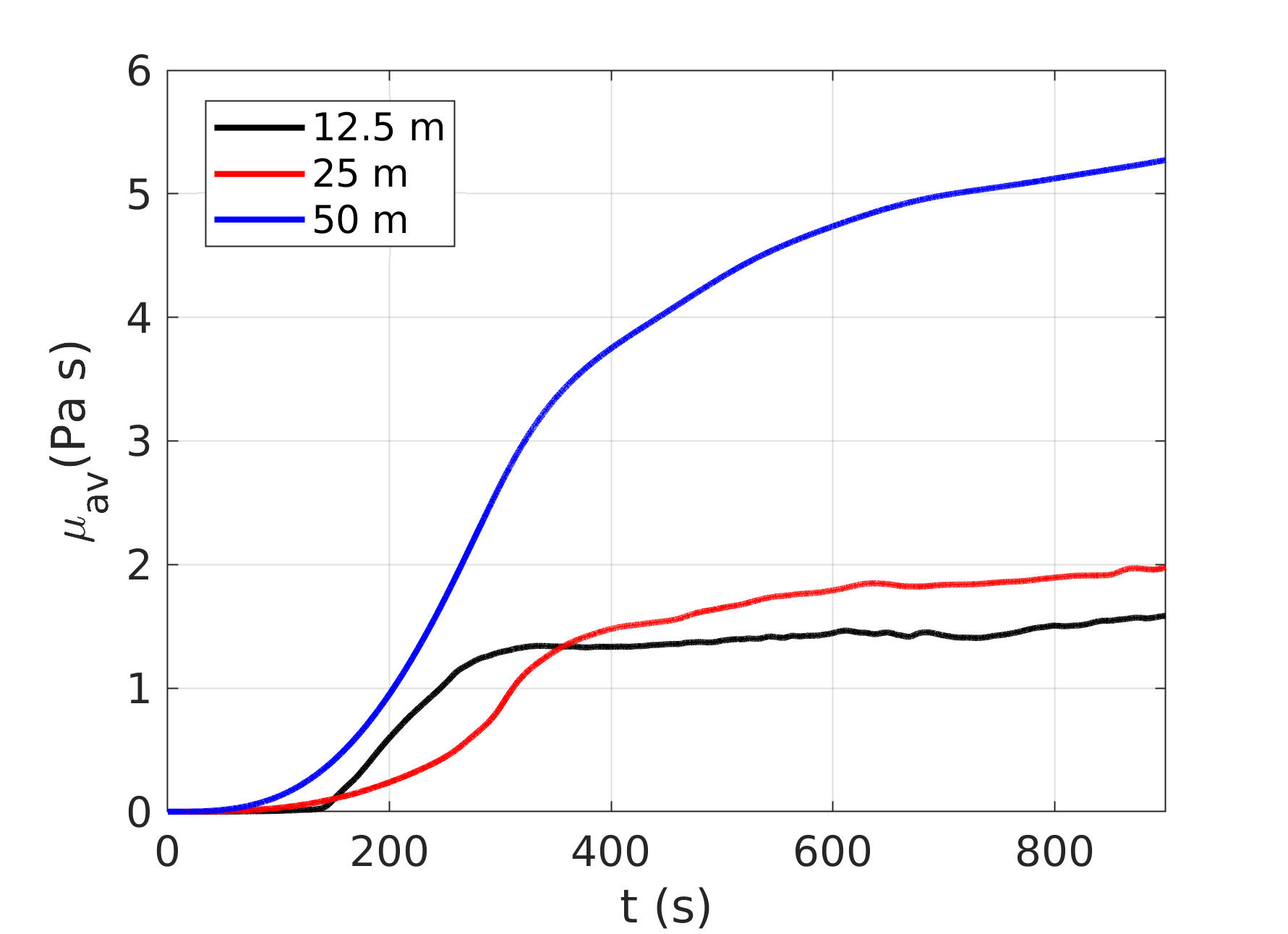}  
        %\put(35,18){FOM}
        %\put(-8,7){$\u$}
      \end{overpic}
\caption{Density current, kEqn model: time evolution of the average eddy viscosity \eqref{eq:mu_av} for meshes $h = 12.5, 25, 50$ m.}
\label{fig:DC11}
\end{figure}

Figure \ref{fig:DC11} shows the time evolution of the space-averaged eddy viscosity \eqref{eq:mu_av} given by the kEqn model for meshes $h = 12.5, 25, 50$ m. By comparing Figure \ref{fig:DC11} with Figure \ref{fig:DC10}, we observe that
the two LES models have a very similar evolution of $\mu_{av}$
for mesh $h = 12.5$ m. For the other two meshes, the Smagorinsky model ramps up the value of $\mu_{av}$ faster than the kEqn model. Finally, we notice that the $\mu_{av}$ introduced by both models keeps increasing over the time interval $[0, 900]$ s since more and more vortical structures develop. 

For a more quantitative comparison, we consider the front location, defined as the location on the ground where 
$\theta' = 1$ K, at $t = 900$ s. 
In Table \ref{tab:front_loc}, we report the space interval that contains the front location for our implementations of the AV75, Smagorinsky, and kEqn models. %\anna{(Michele, dobbiamo spiegare perche' nella tabella c'e' un intervallo per la front location invece che un valore secco)} \textcolor{red}{Di fatto io prendo l'intervallo campionato in paraview, le due stazioni che mi danno valore maggiore e minore di -1. Gli altri probabilmente fanno un'interpolazione tra i valori a cavallo di -1 e calcolano mediante interpolante la stazione? Credo sia più giusto dare un intervallo tra l'altro.}
We compare our values with the data in \cite{strakaWilhelmson1993} (Table 4), which refer to the AV75 model and 14 different approaches for the numerical solution of this benchmark problem. We note that our results fall well within the values reported in
\cite{strakaWilhelmson1993}. Furthermore, from Table \ref{tab:front_loc} we see that with our implementation of the Smagorinsky and kEqn models the front becomes faster as the mesh is refined. This is more evident with the Smagorinsky model than with the kEqn model, while the AV75 model shows the opposite trend. The front locations obtained with the AV75 model and different meshes are within about 500 m from each other. The same is true for the kEqn model, while the Smagorinsky model gives a larger variation in front location (about 900 m) as the mesh size varies. Since in \cite{strakaWilhelmson1993} no results are reported for mesh $h = 12.5$ m, in Table \ref{tab:front_loc} we list the front location found with a LES model and resolution 12.5 m in \cite{marrasNazarovGiraldo2015}. Such front location is within about 500 m of the value we obtain with the kEqn model and about 900 m of the value we get with the Smagorinsky model.

%\begin{sidewaystable}
\begin{table}[htb]
\begin{tabular}{|c|c|c|c|} \hline
Model & Resolution [m] & Front Location [m] \\
 \hline
 AV75 &  100 &  [15269, 15295]\\
 AV75 &  50 &  [15141, 15167]\\
 AV75 &  25 &  [14783, 14808]\\
 Smagorisnky (LES) &  50  & [15104, 15130]\\
 Smagorisnky (LES) &  25 & [15411, 15437]\\
 Smagorisnky (LES) &  12.5  & [16000, 16026]\\
 kEqn (LES)  & 50  & [15181, 15206]\\
 kEqn (LES)  & 25  & [15514, 15539]\\
 kEqn (LES)  & 12.5  & [15642, 15667]\\
 Ref.~\cite{strakaWilhelmson1993} & (25, 200) & (14533,17070)
 \\
 Ref.~\cite{marrasNazarovGiraldo2015} & 12.5 & 15056 \\
 \hline  
\end{tabular}
\caption{Density current: our results for the front location at $t = 900$ s obtained with the AV75, Smagorinsky, and kEqn models and different meshes compared against results reported in 
\cite{strakaWilhelmson1993,marrasNazarovGiraldo2015}.
For reference \cite{strakaWilhelmson1993}, we reported the range of mesh sizes and front location values obtained with different methods.
For reference \cite{marrasNazarovGiraldo2015}, we report only the front location computed with the finest resolution.
} 
\label{tab:front_loc}
\end{table}
%\end{sidewaystable}

\begin{figure}[htb]
\centering
 \begin{overpic}[width=0.485\textwidth]{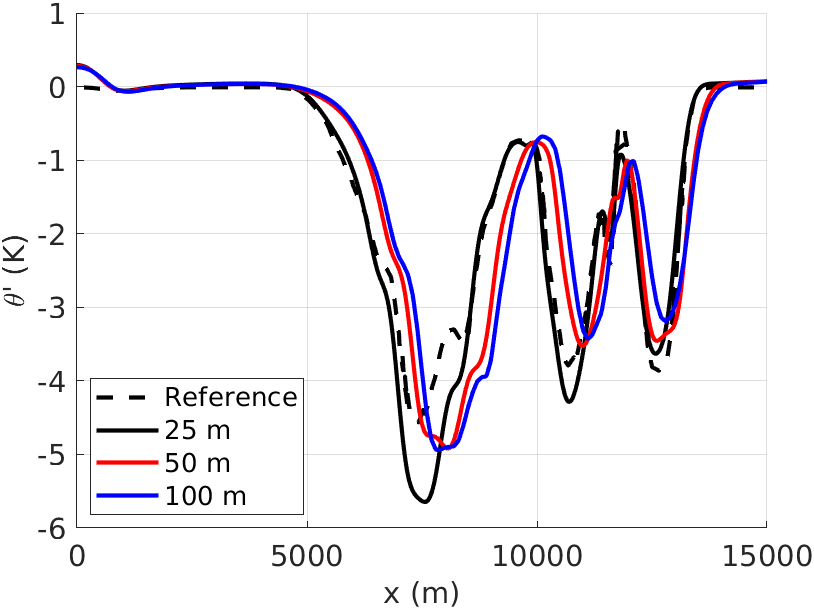}  
        %\put(35,18){FOM}
        %\put(-8,7){$\u$}
      \end{overpic}
\caption{Density current: potential temperature perturbation $\theta'$ at $t = 900$ s
along the horizontal direction at a height of $z = 1200$ m given the AV75 model 
on meshes $h = 100, 50, 25$ m
and compared with data from \cite{giraldo_2008} and resolution 25 m (denoted as ``Reference'').}
\label{fig:DC9}
\end{figure}

For a further quantitative comparison, we consider the 
potential temperature perturbation $\theta'$ at $t = 900$ s
along the horizontal direction at a height of $z = 1200$ m.
Figure \ref{fig:DC9} displays a comparison between the results given the AV75 model 
on meshes $h = 100, 50, 25$ m
and the results obtained with the same model and resolution 25 m in \cite{giraldo_2008}. Two numerical methods were used in \cite{giraldo_2008} to approximate the solution: a spectral element method and a discontinuous Galerkin method, which provide curves that are superimposed in Figure \ref{fig:DC9} and are denoted as ``Reference''. 
We observe that our results are slightly out of phase for meshes $h = 50, 100$ m with respect to the data in \cite{giraldo_2008}. The potential temperature fluctuation obtained with mesh $h = 25$ m is in phase but has larger negative peaks, especially in correspondence of the first recirculation. However, given all the differences between our approach and the one in \cite{giraldo_2008}, the comparison is satisfactory.%\anna{Michele, secondo te a cosa e' dovuto questo?} \textcolor{red}{non ho una risposta purtroppo Anna...ti posso pero' dire che feci una prova a 12.5 metri con AV75 e i risultati cominciavano a cambiare un po', credo tra l'altro di avertelo già detto...si vede che una vera convergenza per questo caso si ottiene a CFL fissato, d'altronde non ha molto senso infittire la mesh e mantenere costante il dt, bisogna risolvere anche le scale temporali piccole...quindi, probabilmente, una possibile spiegazione potrebbe essere cercata in questo contesto. In ogni caso in letteratura la gente non va oltre 25 m per questo caso, addirittura se ben ricordo la soluzione di riferimento su WRF è a 50 metri di risoluzione, e c'è chi lavora a dt fisso, CFL fisso, chi non riporta proprio questi dati ecc... Tornassi indietro non seguirei nessuna indicazione da parte di S e da parte dei vari Lindemann ecc...ma prenderei a riferimento solo Straka. In futuro comunque si potrebbe investigare questa cosa in dettaglio, quando avremo piena consapevolezza del sistema :) Sappiamo bene che su questi problemi c'è ancora tantissimo da dire in fatto di convergenza}
%\anna{Sono indecisa se scrivere un commento o lasciare perdere e aspettare che il reviewer chieda spiegaazioni. Tu cosa dici?}

%\textcolor{red}{direi di aspettare. Teniamo conto che come scrivevo su probabilmente questo comportamento è dovuto al fatto che continuando ad infittire tenendo pero' dt fisso potrebbe determinare un cambiamento nella soluzione. Cmq per fortuna direi che il confronto con la letteratura è ottimo visto tutte le differenze del caso!}
%Referring to 100, 50, or 25 $m$ resolutions, one can see that the associated curves are directly on top of each other, which demostrates that our NSE model has converged at 100 $m$ resolution.
%\anna{Non sono sicura che capisco la frase precedente} 

\section{Concluding remarks}\label{sec:concl}

The goal of this paper is to show that OpenFOAM provides a reliable and accurate framework for the simulation of non-hydrostatic atmospheric flows. To achieve this goal, we developed a pressure-based solver for the Euler equations written in conservative form using density, momentum, and total energy as variables and tested it against numerical data available in the literature for two well-known benchmarks, namely the rising thermal bubble and the density current. For the rising bubble and the density current tests, we obtain good qualitative and quantitative comparisons when the classical Smagorinsky model and/or the one equation eddy-viscosity
model are adopted for stabilization.
The code created for this paper is available within open source package GEA (Geophysical and Environmental Applications) \cite{GEA}.

%\anna{Do we want to add perspectives? If so, which?}

%\textcolor{red}{Anna, qui possiamo dire che abbiamo intenzione di mettere su sempre in openfoam un density based per poterci confrontare in maniera diretta con tutti gli altri autori? E di migliorare il modello con l'inserimento di uno sponge layer magari per simulare HB su orografia, gravity waves ecc...? Implementare schemi di ordine piu' elevato in OF? il density based di extended che dovrebbe essere ok? DensityBasedTurbo }

\section*{Acknowledgements}
%\textcolor{red}{ringraziamo SM e Asma? -.-} \anna{yes}
We thank Dr.~S.~Marras for fruitful conversations.
We acknowledge the support provided by the European Research Council Executive Agency by the Consolidator Grant project AROMA-CFD ``Advanced Reduced Order Methods with Applications in Computational Fluid Dynamics" - GA 681447, H2020-ERC CoG 2015 AROMA-CFD, PI G. Rozza, and INdAM-GNCS 2019-2020 projects.
This work was also partially supported by US National Science Foundation through grant DMS-1953535 (PI A.~Quaini). A.~Quaini acknowledges 
support from the Radcliffe Institute for Advanced Study at Harvard University where she has been the 2021-2022 William and Flora Hewlett Foundation Fellow.

\bibliographystyle{plain}
\bibliography{bibliography}
\end{document}